\documentclass[fleqn,usenatbib,usedcolumn]{mnras}
\usepackage[british]{babel}
\usepackage{graphicx}
\usepackage{color,soul}
\usepackage{mathtools}
\usepackage{newtxtext}
\usepackage[slantedGreek]{newtxmath}
\usepackage[T1]{fontenc}
\usepackage{bm}
\usepackage{amsmath}
\usepackage{hyperref}
\usepackage{booktabs}
\usepackage[inline]{enumitem}
\usepackage{relsize}
\usepackage{todonotes}
\usepackage{cuted}

\addto\extrasbritish{%
}

\let\mr\mathrm
\DeclarePairedDelimiter\abs{\lvert}{\rvert}%
\graphicspath{{Figures/}}

\newcommand{\sPhi}{\bm{\upPhi}}     
\newcommand{\sSigma}{\bm{\upSigma}} 
\newcommand{\sLambda}{\upLambda}    

\begin{document}
\title[High fidelity Bayesian calibration II]{A Bayesian approach to high fidelity interferometric calibration II: demonstration with simulated data}
\author[Sims et al.]{Peter H. Sims,$^{1,2}$\thanks{E-mail: peter.sims@mail.mcgill.ca} Jonathan C. Pober,$^3$ and Jonathan L. Sievers$^{1,2}$ \\
$^1$McGill Space Institute, McGill University, 3550 University Street, Montreal, QC H3A 2A7, Canada \\
$^2$Department of Physics, McGill University, 3600 University Street, Montreal, QC H3A 2T8, Canada \\
$^3$Department of Physics, Brown University, Providence, RI 02912, USA \\ 
}

\maketitle
\label{firstpage}
\begin{abstract}
In a companion paper, we presented \textsc{BayesCal}, a mathematical formalism for mitigating sky-model incompleteness in interferometric calibration. In this paper, we demonstrate the use of \textsc{BayesCal} to calibrate the degenerate gain parameters of full-Stokes simulated observations with a HERA-like hexagonal close-packed redundant array, for three assumed levels of completeness of the a priori known component of the calibration sky model. We compare the \textsc{BayesCal} calibration solutions to those recovered by calibrating the degenerate gain parameters with only the a priori known component of the calibration sky model both with and without imposing physically motivated priors on the gain amplitude solutions and for two choices of baseline length range over which to calibrate. We find that \textsc{BayesCal} provides calibration solutions with up to four orders of magnitude lower power in spurious gain amplitude fluctuations than the calibration solutions derived for the same data set with the alternate approaches, and between $\sim10^7$ and $\sim10^{10}$ times smaller than in the mean degenerate gain amplitude on the full range of spectral scales accessible in the data. Additionally, we find that in the scenarios modelled only \textsc{BayesCal} has sufficiently high fidelity calibration solutions for unbiased recovery of the 21 cm power spectrum on large spectral scales ($k_\parallel \lesssim 0.15~h\mathrm{Mpc}^{-1}$). In all other cases, in the completeness regimes studied, those scales are contaminated.
\end{abstract}


\section{Introduction}
\label{Sec:Introduction}

High fidelity radio interferometric data calibration that minimises spurious spectral structure in the calibrated data is essential in astrophysical applications, such as 21 cm cosmology, which rely on knowledge of the relative spectral smoothness of distinct astrophysical emission components to separate the signal of interest from foreground emission (see e.g. \citealt{2020PASP..132f2001L} for a review of signal separation techniques in 21 cm data analysis). 
 
Existing approaches to determining the antenna dependent complex gains during interferometric calibration rely on either \begin{enumerate*}\item fitting a forward model of the sky to the data, or \item the presence of redundancy in the baseline layout of the array to solve for the majority of the calibration degrees of freedom, followed by fitting a forward model of the sky to the data to solve for the remaining calibration terms \end{enumerate*}. However, both approaches produce biased results if the sky model used to calibrate the data is incomplete (e.g. \citealt{2016MNRAS.461.3135B} and \citealt{2017MNRAS.470.1849E}, and \citealt{2019ApJ...875...70B}, respectively) and, without mitigation, sky-model-incompleteness-sourced biases in the gain parameters are sufficiently large to produce calibration-error-induced foreground systematics at a level that can overwhelm the high-redshift 21 cm signal and preclude unbiased recovery of modes of the 21 cm power spectrum.

In a companion paper (hereafter, paper I), we present \textsc{BayesCal}, a Bayesian framework for interferometric calibration designed to recover high fidelity gain solutions by mitigating sky-model incompleteness. \textsc{BayesCal} addresses the sky-model incompleteness problem by supplementing the a priori known component of the forward model of the sky with a statistical model for the a priori unknown component of the sky brightness distribution associated with \begin{enumerate*} \item sources below the completeness level of catalogues in the field of interest and \item with the residual diffuse emission deriving from the difference between the true brightness distribution of diffuse emission on the sky and our model for it.\end{enumerate*}

The \textsc{BayesCal} framework presented in Paper I addresses the calibration of visibilities corresponding to a single instrumental polarisation, deriving from full-Stokes sky emission, while accounting for the dominant component of sky-model incompleteness associated with missing and mis-modelled emission in Stokes $I$. The \textsc{BayesCal} calibration model is comprised of \begin{enumerate*}\item a simulated visibility model, $\mathbfit{V}^\mathrm{sim}$, which describes the expected contribution to the observed visibilities associated with a priori known sky emission (this model component is standard in sky-referenced calibration) and \item a fitted visibility model, $\mathbfit{V}^\mathrm{fit}$, that models the contribution to the observed visibilities from emission not included in $\mathbfit{V}^\mathrm{sim}$ due to the incompleteness and uncertainties associated with our a priori knowledge of the brightness distribution of the sky.\end{enumerate*} When calibrating data with \textsc{BayesCal}, one jointly fits for instrumental gains, constrained by available priors on their spectral structure, and the most probable sky-model parameter values for $\mathbfit{V}^\mathrm{fit}$, constrained by a prior on their total power. 
 
In this paper, we demonstrate the extent to which the \textsc{BayesCal} framework mitigates these sky-model-incompleteness-sourced biases in the gain parameters for varying levels of completeness of the simulated calibration sky model, in the context of absolute calibration of visibilities deriving from simulated interferometric observations, accounting for both diffuse and point source emission, with a HERA-like hexagonal close-packed redundant array. Additionally, we compare the fidelity of \textsc{BayesCal} calibration solutions to those recovered with two alternate absolute calibration algorithms:
\begin{enumerate}
\item  standard\footnote{Here, we define standard absolute calibration as absolute calibration of redundantly calibrated visibilities without imposing informative priors on the degenerate gain parameters (as considered in \citealt{2019ApJ...875...70B}). We defer to future work, a comparison to versions of absolute calibration with spectrally smooth gain solutions enforced through the gain parametrisation (e.g. \citealt{2016MNRAS.461.3135B}) or filtering of the gain solutions (e.g. \citealt{2020ApJ...890..122K}).} absolute calibration with an incomplete sky model,
\item absolute calibration with an incomplete sky model when the spectral structure in the gain solutions is constrained by a physically motivated prior.
\end{enumerate} 
The first of these approaches provides a fiducial calibration fidelity that calibration solutions derived with \textsc{BayesCal} can be compared against. The second represents an intermediate step between standard absolute calibration and \textsc{BayesCal} and illustrates the limited efficacy of constraining the gain solutions with a physical prior\footnote{Here, by physical prior, we mean a soft prior with finite uncertainty. In contrast, applying a hard (zero-uncertainty) prior on the gain solutions, if accurate, can limit sky-model-incompleteness-sourced bias introduced into the calibration solutions on particular spectral scales (e.g. \citealt{2016MNRAS.461.3135B, 2020ApJ...890..122K}). However, this approach implies hard (zero-uncertainty) knowledge about structure in the gains which, if imperfect, will bias covariant propagation of uncertainties on the calibration parameters through to the calibrated data and create an additional avenue for introducing spurious spectral structure into the gain solutions, and correspondingly into the calibrated data, if the gains are overly constrained and are unable to fit real instrumental structure (see paper I for details).
} when using an incomplete calibration model.

Through this comparison, we will illustrate that \textsc{BayesCal} enables recovery of significantly higher fidelity  calibration solutions, with in an improvement of 1--4 orders of magnitude additional suppression of power in spurious gain amplitude fluctuations in the calibrated data on 3, 4.5 and 9 MHz spectral scales relevant to constraining the astrophysics of CD and the EoR with 21 cm cosmology ($k_{\parallel} \simeq 0.17, 0.11$ and $0.06~h\mathrm{Mpc}^{-1}$, respectively, for redshifted 21 cm emission in the frequency range of our simulated observed data) relative to the alternate calibration approaches. 

In the context of 21 cm cosmology applications, we will show that, in the scenarios modelled, only \textsc{BayesCal} enables recovery of the 21 cm signal at $k_\parallel \lesssim 0.15~h\mathrm{Mpc}^{-1}$.  For all the other techniques in all completeness regimes, that scale is contaminated.

A number of ongoing and upcoming interferometric 21 cm experiments have redundancy in their array configurations and, thus, can take advantage of the \textsc{BayesCal} approach we demonstrate here. These include the Canadian Hydrogen Intensity Mapping Experiment (CHIME; \citealt{2014SPIE.9145E..4VN}), the Hydrogen Epoch of Reionization Array (HERA; \citealt{2017PASP..129d5001D}), the Hydrogen Intensity and Real-time Analysis eXperiment (HIRAX; \citealt{2016SPIE.9906E..5XN}), Phase II of the Murchison Widefield Array (MWA; \citealt{2018PASA...35...33W}) and Tianlai (\citealt{2020SCPMA..6329862L, 2021MNRAS.506.3455W}). For the simulated observations calibrated in this paper we use a HERA-like hexagonal close-packed redundant array.

The generalisation of \textsc{BayesCal} to non-redundant and semi-redundant interferometric array configurations discussed in paper I is applicable to 21 cm experiments including the Giant Metrewave Radio Telescope (GMRT; \citealt{2013MNRAS.433..639P}), the LOw Frequency ARray (LOFAR; \citealt{2013A&A...556A...2V}) and the Square Kilometre Array (SKA; \citealt{2013ExA....36..235M}). The application of \textsc{BayesCal} in this regime will be illustrated in future work.

The remainder of the paper is organised as follows. In \autoref{Sec:SummaryOfPaperI}, we briefly summarize the results of Paper I relevant to estimation of absolute calibration solutions for redundantly calibrated visibilities using standard absolute calibration absolute calibration supplemented by priors on the gain solutions and \textsc{BayesCal}. In \autoref{Sec:InstrumentModel}, we describe the instrument model we use when constructing the simulated observed visibility data which we use to demonstrate the \textsc{BayesCal} algorithm. In \autoref{Sec:SkyModel}, we describe the sky models used in the simulated observations. In \autoref{Sec:Results}, we demonstrate the result of applying \textsc{BayesCal} to derive absolute calibration solutions for the simulated observed data from a HERA-like hexagonal close-packed redundant array and compare them to standard calibration solutions for the same data sets. We discuss our conclusions in \autoref{Sec:SummaryAndConclusions}.

\section{Summary of paper I}
\label{Sec:SummaryOfPaperI}

In this section, we summarize the results of Paper I relevant to estimation of absolute calibration solutions for redundantly calibrated visibilities with the three calibration approaches considered in \autoref{Sec:Results}. \autoref{Sec:CalibrationFormalism} introduces per-instrumental-polarisation direction-independent interferometric calibration. Sections \ref{Sec:RedundantCalibration} and \ref{Sec:ImperfectArrayRedundany} briefly describes the relative and absolute calibration steps that the calibration of redundant array can be subdivided into and the impact of low-level non-redundancy in the baselines of an assumed redundant array. \autoref{Sec:StandardAbscal} describes the Gaussian likelihood for standard absolute calibration of redundantly calibrated visibilities, which we maximise to derive `absolute calibration solutions'. Building on this, \autoref{Sec:AbscalWithGainPriors} describes the posterior probability distribution for absolute calibration of redundantly calibrated visibilities supplemented by priors on the spectral structure of the degenerate amplitude parameters, which we maximise to derive `absolute calibration plus gain prior solutions'. \autoref{Sec:BayesCal} describes the posterior probability distribution for absolute calibration of redundantly calibrated visibilities with \textsc{BayesCal}, which we sample from to derive `\textsc{BayesCal} solutions'. Finally, \autoref{Sec:TemporalModel} describes our model for the temporal evolution of the gain solutions.

\subsection{Calibration formalism}
\label{Sec:CalibrationFormalism}

The polarised interferometric visibilities corresponding to the cross-correlation of the voltages $v_{p}^{a}$ and $v_{q}^{b}$ induced in feeds $a$ and $b$, respectively, of two dual-polarisation antennas, $p$ and $q$, by the electric field due to a continuous sky brightness distribution, expressed in an equatorial basis, can be written in the frame of the instrument, in which the brightness distribution on the celestial sphere is rotating overhead, using the Radio Interferometer Measurement Equation (RIME; \citealt{1996A&AS..117..137H, 2011A&A...527A.106S}; hereafter, HBS96 and Sm11, respectively) as,
\begin{multline}
\label{Eq:MEqSingePol}
V^{ab}_{pq}(\mathbfit{u}_{pq},\nu, t) = n(\nu, t) + (g_{p}^{a}(\nu, t))(g_{q}^{b}(\nu, t))^{*} \\ 
\times \iint\limits_{lm}\frac{\mr{d}l\mr{d}m}{n}\,  
[P_{{ab},I}(\mathbfit{l}, \nu) I(\mathbfit{l}, \nu, t) + P_{{ab},Q}(\mathbfit{l}, \nu) Q(\mathbfit{l}, \nu, t) \\
+ P_{{ab},U}(\mathbfit{l}, \nu) U(\mathbfit{l}, \nu, t) + P_{{ab},V}(\mathbfit{l}, \nu) V(\mathbfit{l}, \nu, t)] \mr{e}^{ -2\pi i \mathbfit{u}_{pq}^{T}(\nu) \mathbfit{l}(\nu) } \ .
\end{multline}
Here, $\mathbfit{u}_{pq} = \mathbfit{b}_{pq}/\lambda = (u_{pq},v_{pq},w_{pq})$ are time-stationary $uvw$-coordinates of the baseline in the frame of the instrument; $n(\nu, t)$ describes the noise on $V^{ab}_{pq}(\mathbfit{u}_{pq},\nu, t)$; $P_{{ab},j}(\mathbfit{l}, \nu)$, where $j \in [I,Q,U,V]$, are the elements of the polarised primary beam matrix that map the polarised sky expressed in an equatorial basis in terms of Stokes parameters $I$, $Q$, $U$, and $V$ onto the apparent sky viewed in a topocentric basis of the antennas, in which the basis vectors are aligned with feeds $a$ and $b$ (see \autoref{Sec:PrimaryBeamModel}); $g_{p}$ and $g_{q}$ are complex direction-independent gains of antennas $p$ and $q$ and we have assumed that the antenna configuration is sufficiently compact and the antennas have a sufficiently narrow FoV that direction-dependent sky-based signal propagation effects can be neglected. 

Going forward, we consider antennas $p$ and $q$ each with north-south and east-west oriented orthogonal dual feeds, n and e, respectively. The four cross-correlations of the voltages $v_{p}^{n}$, $v_{p}^{e}$, $v_{q}^{n}$ and $v_{q}^{e}$ induced in the feeds of antennas $p$ and $q$ yield the interferometric visibilites $V_{pq}^\mathrm{nn}=\langle (v_{p}^\mathrm{n})(v_{q}^{\mathrm{n}})^{*} \rangle$, $V_{pq}^\mathrm{ne}=\langle (v_{p}^\mathrm{n})(v_{q}^{\mathrm{e}})^{*} \rangle$, $V_{pq}^\mathrm{en}=\langle (v_{p}^\mathrm{e})(v_{q}^{\mathrm{n}})^{*} \rangle$ and $V_{pq}^\mathrm{ee}=\langle (v_{p}^\mathrm{e})(v_{q}^{\mathrm{e}})^{*} \rangle$, where, $\langle . \rangle$ denotes an average over a small interval in time and frequency. In this work, we demonstrate calibration in the context of $V_{pq}^\mathrm{nn}$ visibilities; thus, all calculated visibilities refer to correlations of pairs of north-south oriented feeds so we omit the `$\mathrm{n}$' superscripts for brevity. However, each of the remaining three instrumental visibility correlations can be calibrated in an equivalent manner upon substitution of the corresponding terms of the polarised primary beam matrix and complex gains for the visibity correlation to be calibrated.

In the following sections, it will be helpful to write a discretised version of \autoref{Eq:MEqSingePol} for a set of baselines as,
\begin{equation}
\label{Eq:MEqSingePolShorthand}
\mathbfit{V}^\mathrm{obs} = \mathbfit{n} + \mathbfss{G}\mathbfit{V}^\mathrm{true} \ ,
\end{equation}
where $\mathbfit{V}^\mathrm{obs}$, $\mathbfit{n}$ and $\mathbfit{V}^\mathrm{true}$ are given by the concatenation over baselines of the observed visibilities, noise on the data and `true visibilities' (corresponding to the discretised version of the integral in \autoref{Eq:MEqSingePol}), respectively, each vectorised over a discrete set of frequencies and times. $\mathbfss{G}$ is a diagonal matrix encoding the antenna, frequency and time dependent instrumental gains, and, for a single frequency and time, it has elements $G_{ij} = \delta_{ij}g_{p}g_{q}^{*}$, with $p$ and $q$ the antennas associated with the $i$th visibility, such that $\mathbfit{V}^\mathrm{obs}$ has elements,
\begin{equation}
\label{Eq:MEqSingePolShorthandElement}
V^\mathrm{obs}_{i} = n_{pq} + g_{p}g_{q}^{*} V^\mathrm{true}_{i}\, .
\end{equation}
Here, $V^\mathrm{obs}_{i}$  and $V^\mathrm{true}_{i}$ are the $i$th observed visibility and `true' visibility, respectively, and the `true' visibility is the cross-correlation of the signals from antennas $p$ and $q$ at time $t$ and frequency $\nu$ that would have been measured if the direction-independent instrumental gains were unity; $i$ is an index over visibilities which runs over the $n_t n_\nu N_\mathrm{ant}(N_\mathrm{ant}-1)/2$ cross-correlations between the signals from the $N_\mathrm{ant}$ antennas in the array, at the $n_t$ time integrations and $n_\nu$ frequency channels in the data set to be calibrated.

In an array with a redundant antenna layout, such that it measures visibility data derived from the cross-correlation of the voltage responses from sets of antennas with identical baseline separation vectors and beam patterns, it is advantageous to reparametrise the antenna gain parameters as $g_{p} = h_{p}f(A, \phi_{l}, \phi_{m})$ where $h_{p}$ is the relative gain of the antenna and $f(A, \phi_{l}, \phi_{m})$ is a function of degenerate gain amplitude, $A$, and tip-tilt phases, $\phi_{l}$ and $\phi_{m}$, respectively. In this case, we can rewrite the instrumental gain matrix as,
\begin{equation}
\label{Eq:GHDmatrices}
\mathbfss{G} = \mathbfss{H}\mathbfss{D} \ ,
\end{equation}
where the relative gain matrix $\mathbfss{H}$ is a diagonal matrix encoding the antenna, frequency and time dependent relative antenna gains, and, for a single frequency and time, it has elements $H_{ij} = \delta_{ij}h_{p}h_{q}^{*}$, with $p$ and $q$ the antennas associated with the $i$th visibility, and the degenerate gain matrix  $\mathbfss{D}$ is a diagonal matrix encoding the degeneracy function $f(A, \phi_{l}, \phi_{m})$ (see \autoref{Sec:RedundantCalibration} for details).

Defining an interferometric fringe matrix with elements\footnote{To eliminate any ambiguity between the earlier use of $i$ as a counting index and its use here as an imaginary number we advise the reader that $i$ in all instances of the variable combination $2\pi i$ in this paper is the imaginary unit.} $\mathbfss{F}_{\mathrm{fr}, jk}  = \gamma \mathrm{e}^{ -2\pi i (\bm{u} \otimes \bm{l} + \bm{v} \otimes \bm{m} + \bm{w} \otimes \bm{n})_{jk}}$, which encodes the transformation from the image domain to the $uvw$-coordinates sampled by the interferometric array, where $\gamma$ is a normalisation factor (see \autoref{Sec:InterferometricArray}, and paper I for details), and polarised primary beam matrices, $\mathbfss{P}_{\mathrm{nn},I}$, $\mathbfss{P}_{\mathrm{nn},Q}$, $\mathbfss{P}_{\mathrm{nn},U}$ and $\mathbfss{P}_{\mathrm{nn},V}$, which encode the polarisation, frequency and direction-dependent primary beam profiles, we can rewrite $\mathbfit{V}^\mathrm{true}$ as\footnote{The specific form that the component matrices in \autoref{Eq:VtrueExpanded} take for the sky and instrument models considered in this paper will be discussed in Sections \ref{Sec:InstrumentModel} and \ref{Sec:SkyModel}.},
\begin{equation}
\label{Eq:VtrueExpanded}
\mathbfit{V}^\mathrm{true} = \mathbfss{F}_{\mathrm{fr}}[\mathbfss{P}_{\mathrm{nn},I}\mathbfit{T}_{I} + \mathbfss{P}_{\mathrm{nn},Q}\mathbfit{T}_{Q} + \mathbfss{P}_{\mathrm{nn},U}\mathbfit{T}_{U} + \mathbfss{P}_{\mathrm{nn},V}\mathbfit{T}_{V}] \ .
\end{equation}
We can rewrite \autoref{Eq:MEqSingePolShorthandElement} using Equations \ref{Eq:GHDmatrices} and \ref{Eq:VtrueExpanded} as,
\begin{equation}
\label{Eq:MEqSingePolExpanded}
\mathbfit{V}^\mathrm{obs} = \mathbfit{n} + \mathbfss{H}\mathbfss{D}\mathbfss{F}_{\mathrm{fr}}[\mathbfss{P}_{\mathrm{nn},I}\mathbfit{T}_{I} + \mathbfss{P}_{\mathrm{nn},Q}\mathbfit{T}_{Q} + \mathbfss{P}_{\mathrm{nn},U}\mathbfit{T}_{U} + \mathbfss{P}_{\mathrm{nn},V}\mathbfit{T}_{V}] \ .
\end{equation}

The data calibration problem requires solving for gain parameters $h_{p}^\mathrm{m}$ and $f(A, \phi_{l}, \phi_{m})$, given constraints deriving from array redundancy and sky modelling. Approaches to achieving this are summarised in Sections \ref{Sec:RedundantCalibration} -- \ref{Sec:BayesCal}.

\subsection{Redundant calibration}
\label{Sec:RedundantCalibration}

When an array has a redundant antenna layout, such that it measures visibility data derived from the cross-correlation of the voltage responses from sets of antennas with identical baseline separation vectors and beam patterns, the fact that the true visibilities, $V^\mathrm{true}$, that would be measured if all instrument gains were unity are identical on redundant baselines enables one to constrain the relative gain parameters of the array independently of the sky model (e.g. \citealt{1992ExA.....2..203W, 2010MNRAS.408.1029L}). This enables redundant arrays to be fully calibrated via a two-stage process: 
\begin{enumerate}                                                                                               \item \textit{'relative' calibration}, in which baseline redundancy is used to derive redundant gain parameters, $h_{p}^\mathrm{m}$, that are equal to the general direction-independent gain parameter, $g_{p}^\mathrm{m}$,  associated with antenna $p$, to within a frequency, time and polarisation dependent complex degeneracy factor. This complex degeneracy factor is described by a set of degenerate calibration parameters that cannot be solved for using baseline redundancy.
\item \textit{'absolute' calibration}, in which the degenerate calibration parameters associated with relative calibration are solved for with reference to a sky model.
\end{enumerate}

For further details of relative calibration, see e.g. \citealt{2010MNRAS.408.1029L} or paper I. Here, we will assume that this first stage has been completed and, by fixing the gain amplitude of a reference antenna and the gain phases of two reference antennas to an arbitrary choice of values (e.g. \citealt{2018ApJ...863..170L, 2020arXiv200308399D}), accurate solutions have been obtained relative to these reference parameters for the redundant calibration parameters of the antennas of the array, $\hat{h}_{p}^\mathrm{m}$, encoded in a diagonal relative calibration matrix, $\hat{\mathbfss{H}}^\mathrm{m}$.

To fully calibrate the array, one must subsequently solve for the degenerate gain parameters $A$, $\phi_{l}$, $\phi_{m}$ corresponding to the per-frequency, time and polarisation dependent difference between the arbitrary choice of reference parameters used during relative calibration and the true values of these parameters\footnote{The degenerate calibration amplitude $A$, and tip-tilt phases, $\phi_{l}$, $\phi_{m}$, determine the per-frequency, time and polarisation dependent flux-scale and pointing center of the data, respectively.} in the data set under consideration.

\subsection{Imperfect array redundancy}
\label{Sec:ImperfectArrayRedundany}

For a truly redundant array, relative calibration recovers unbiased relative gain solutions. If there are non-redundancies in an array that is assumed to be redundant, relative calibration will introduce additional biases in the gain solutions (e.g. \citealt{2018AJ....156..285J}). However, in \citet{2021MNRAS.506.2066C} it is found that primary beam non-redundancies, of a variety of forms, yield a low level of additional spectral structure in the gain solutions. Additionally, in \citet{2019MNRAS.487..537O} it is shown that spurious spectral structure in recovered relative gain solutions introduced through antenna-position and primary beam non-redundancy can be largely mitigated by relative upweighting of short baselines during calibration. These biases can also be mitigated by treating nominally redundant baselines as correlated rather than identical to within the instrumental noise (e.g. \citealt{2017arXiv170101860S, 2021MNRAS.503.2457B}). 

In contrast, in \citet{2019ApJ...875...70B} it is demonstrated that even in the absence of relative calibration errors, sky-model-incompleteness-sourced biases in the gain parameters solved-for with absolute calibration are sufficiently large to produce calibration-error-induced foreground systematics at a level that can overwhelm the high-redshift 21 cm signal and prevent a detection of of the 21 cm power spectrum on a range of spatial scales. In this work we demonstrate the effectiveness of \textsc{BayesCal} for resolving the sky-model incompleteness problem in the context of calibration of redundant interferometric arrays. The application of \textsc{BayesCal} to resolving the sky-model incompleteness problem in the context of a partially- or non-redundant array or when calibrating with a sky model alone, is discussed in paper I. We defer application of \textsc{BayesCal} in this context to future work.

\subsection{Standard absolute calibration}
\label{Sec:StandardAbscal}

To derive absolute calibration solutions one can define an absolutely calibrated model for the data as,
\begin{equation}
\label{Eq:PureSkyVisModel}
\mathbfit{V}^\mathrm{model} = \hat{\mathbfss{H}}^\mathrm{m}\mathbfss{D}^\mathrm{m}\mathbfit{V}^\mathrm{sim} \ .
\end{equation}
Here, $\mathbfss{D}^\mathrm{m}$ is our model degenerate gain matrix, which is a diagonal matrix that for a single time integration and frequency channel has elements $D_{ij} = \delta_{ij}A\mr{e}^{i(\mathbfit{b}_{i}^{T} \sPhi)}$; $\hat{\mathbfss{H}}^\mathrm{m}$ is the diagonal matrix containing the maximum likelihood redundant gain parameters, $\hat{h}_{p}^\mathrm{m}$, and the elements of the model vector are given by,
\begin{equation}
\label{Eq:MEqSingePolShorthandRedundantModelNonDegenerate} 
V^\mathrm{model}_{\alpha,j} = \hat{h}_{p}^\mathrm{m}\hat{h}_{q}^{\mathrm{m}*}A\mr{e}^{i(\mathbfit{b}_{i}^{T} \sPhi)}V^\mathrm{sim}_{\alpha}\, ,
\end{equation}
where $V^\mathrm{sim}_{\alpha}$ is a simulated model for the true visibility on the $\alpha$th unique baseline,
\begin{equation}
\label{Eq:VsimExpanded}
\mathbfit{V}^\mathrm{sim} = \mathbfss{F}_{\mathrm{fr}}[\mathbfss{P}_{\mathrm{nn},I}\mathbfit{T}_{I}^\mathrm{sim} + \mathbfss{P}_{\mathrm{nn},Q}\mathbfit{T}_{Q}^\mathrm{sim} + \mathbfss{P}_{\mathrm{nn},U}\mathbfit{T}_{U}^\mathrm{sim} + \mathbfss{P}_{\mathrm{nn},V}\mathbfit{T}_{V}^\mathrm{sim}] \ .
\end{equation}
Here $\mathbfit{T}_{j}^\mathrm{sim}$, with $j \in [I,Q,U,V]$, are our simulated sky models for emission in Stokes $I$, $Q$, $U$ and $V$ (see \autoref{Sec:SkyModel} for details).

A Gaussian likelihood for our full model of the data as a function of amplitude and tip-tilt gain phase parameters, $\mathbfit{A}$, $\bm{\mathit{\Phi}}_{l}$ and $\bm{\mathit{\Phi}}_{m}$, encoded in the degeneracy calibration matrix, $\mathbfss{D}^\mathrm{m}$, is thus given by,
\begin{multline}
\label{Eq:AbsCalVisLike}
\mathrm{Pr}(\bm{V}^\mathrm{obs} \;|\; \bm{A}, \bm{\mathit{\Phi}}_{l}, \bm{\mathit{\Phi}}_{m}) = \frac{1}{\pi^{N_{\mathrm{vis}}}\mathrm{det}(\mathbfss{N})} \\
\times\exp\left[-\left(\bm{V}^\mathrm{obs} - \hat{\mathbfss{H}}^\mathrm{m}\mathbfss{D}^\mathrm{m}\bm{V}^\mathrm{sim}\right)^\dagger\mathbfss{N}^{-1}\left(\bm{V}^\mathrm{obs} - \hat{\mathbfss{H}}^\mathrm{m}\mathbfss{D}^\mathrm{m}\bm{V}^\mathrm{sim}\right)\right] \ .
\end{multline} 

In \autoref{Sec:Results}, when using \autoref{Eq:AbsCalVisLike} to solve for the maximum likelihood values of $\mathbfit{A}$, $\bm{\mathit{\Phi}}_{l}$, $\bm{\mathit{\Phi}}_{m}$ for a given set of visibilities, we use a Generalized Simulated Annealing algorithm (\citealt{SimulatedAnnealing}), as implemented in \textsc{scipy}\footnote{https://docs.scipy.org/doc/scipy/reference/generated/\newline scipy.optimize.dual\_annealing.html}.

Since $\mathbfit{A}$, $\bm{\mathit{\Phi}}_{l}$, $\bm{\mathit{\Phi}}_{m}$ are all frequency dependent, their maximum likelihood solutions will still absorb chromatic errors in $\mathbfit{V}^\mathrm{sim}$ relative to $\mathbfit{V}^\mathrm{true}$ as spurious spectral structure. Within this framework, \citet{2019ApJ...875...70B} have shown that model incompleteness can couple a prohibitive level of spectral structure into the gain solutions for 21 cm cosmology applications. We also find this to be the case for the sky and instrument model considered here, when calibrating simulated observed data using redundant calibration (see \autoref{Sec:Results} for details).

\subsection{Absolute calibration with gain priors}
\label{Sec:AbscalWithGainPriors}

In addition to using a sky model to constrain the spectral structure of the antenna gain solutions, as described in \autoref{Sec:StandardAbscal}, structure present in the gains can also be estimated with alternate techniques, such as electric and electromagnetic co-simulations of the receiver system (e.g. \citealt{2021MNRAS.500.1232F}) and reflectometry measurements of the feed-dish system (e.g. \citealt{2018ExA....45..177P}) that are independent of an astrophysical sky model. 

In general, to leverage this a priori knowledge, one can incorporate these constraints as spectral priors on the gain solutions. For absolute calibration of a redundantly calibrated data set, this corresponds to placing priors on the spectral dependence of the redundant gain degeneracy function $f_\mathrm{vis}(A, \Phi_{l}, \Phi_{m})$. In this work, we focus on incorporating information about the power in the amplitude fluctuations of the redundant gain degeneracy function. 

The Fourier domain of the degenerate gain amplitudes provides a natural space to encode priors on the power in these gain fluctuations as a function of frequency since, if power in the degenerate amplitudes is limited to the spectral scales that are Nyquist sampled in the data set being calibrated\footnote{If, additionally, there is power in the degenerate amplitudes on spectral scales larger than the bandwidth of the calibration data set, the degenerate gain parameters can instead be jointly modelled by a set of Fourier modes on Nyquist sampled scales and, for example, a quadratic large spectral scale model that prevents distortion of the expected power spectrum due to spectral leakage (see \citealt{2019MNRAS.484.4152S} for an example of this approach applied to large spectral scale modelling in 21 cm power spectrum estimation).}, the diagonal elements of the covariance matrix of the Fourier modes encode the power in the fluctuations as a function of spatial scale. Although here we do not impose an informative prior on the distribution of the tip-tilt phases, we, nevertheless, follow Paper I and for mathematical convenience apply to them the same reparametrisation. We, therefore, rewrite our degeneracy function parameters $\mathbfit{A}(\nu)$, $\bm{\mathit{\Phi}}_{l}(\nu)$ and $\bm{\mathit{\Phi}}_{m}(\nu)$ using a new set of parameters that describe the amplitudes of their Fourier decompositions, $\mathbfit{A}_\mathrm{F}(\eta)$, $\bm{\mathit{\Phi}}_{l,\mathrm{F}}(\eta)$ and $\bm{\mathit{\Phi}}_{m,\mathrm{F}}(\eta)$, with, 
\begin{align}
\label{Eq:DegeneracyFunctionFourierParameters}
\bm{A} &= \mathbfss{F}^{-1}\bm{A}_\mathrm{F} \\ \nonumber
\bm{\mathit{\Phi}}_{l} &= \mathbfss{F}^{-1}\bm{\mathit{\Phi}}_{l,\mathrm{F}} \\ \nonumber
\bm{\mathit{\Phi}}_{m} &= \mathbfss{F}^{-1}\bm{\mathit{\Phi}}_{m,\mathrm{F}} \ .
\end{align}
Here, $\mathbfss{F}^{-1}$ is a one dimensional inverse Fourier transform matrix mapping from the 'halfcomplex' representation of the $\eta$-domain of the gain parameters to their frequency-domain representations, and $\bm{A}_\mathrm{F}$, $\bm{\mathit{\Phi}}_{l,\mathrm{F}}$ and $\bm{\mathit{\Phi}}_{m,\mathrm{F}}$ are $N_{\nu} \times 1$ real column vectors encoding the 'halfcomplex' representation of Fourier degenerate gain amplitude, tip- and tilt-phase parameters, respectively (see paper I for details).

We assume a uniform prior on the parameters defining the Fourier decomposition of the tip-tilt phases such that $\mathrm{Pr}(\bm{\mathit{\Phi}}_{l,\mathrm{F}}) = \mathrm{Pr}(\bm{\mathit{\Phi}}_{m,\mathrm{F}})= 1$ and impose a Gaussian prior encoding the power in amplitude fluctuations of the redundant gain degeneracy function as,
\begin{multline}
\label{Eq:ProbAgivenSigmaA}
\mathrm{Pr}(\bm{A}_\mathrm{F} \;|\; \bm{\sigma}_{A_\mathrm{F}}^{2}) \; \propto \; \frac{1}{\sqrt{\mathrm{det}(\bm{\sSigma}_{A_\mathrm{F}})}} \\
\times \exp\left[-\frac{1}{2}(\bm{A}_\mathrm{F} -\bar{\bm{A}}_\mathrm{F})^{T}\bm{\sSigma}_{A_\mathrm{F}}^{-1}(\bm{A}_\mathrm{F} -\bar{\bm{A}}_\mathrm{F})\right] \ .
\end{multline}
Here, $\bar{\mathbfit{A}}_\mathrm{F}$ is a vector of amplitudes defining the expectation value of the Fourier decomposition of $\mathbfit{A}$, $\bm{\sigma}_{A_\mathrm{F}}$ is the vector of expected variances of $\mathbfit{A}_\mathrm{F}$ and $\bm{\sSigma}_{A_\mathrm{F}}$ is a diagonal matrix with elements $\sSigma_{A_\mathrm{F},ij} = \delta_{ij}\sigma_{A_\mathrm{F}}^{2}$.

\begin{figure*}
	\centerline{
	\includegraphics[width=0.50\textwidth]{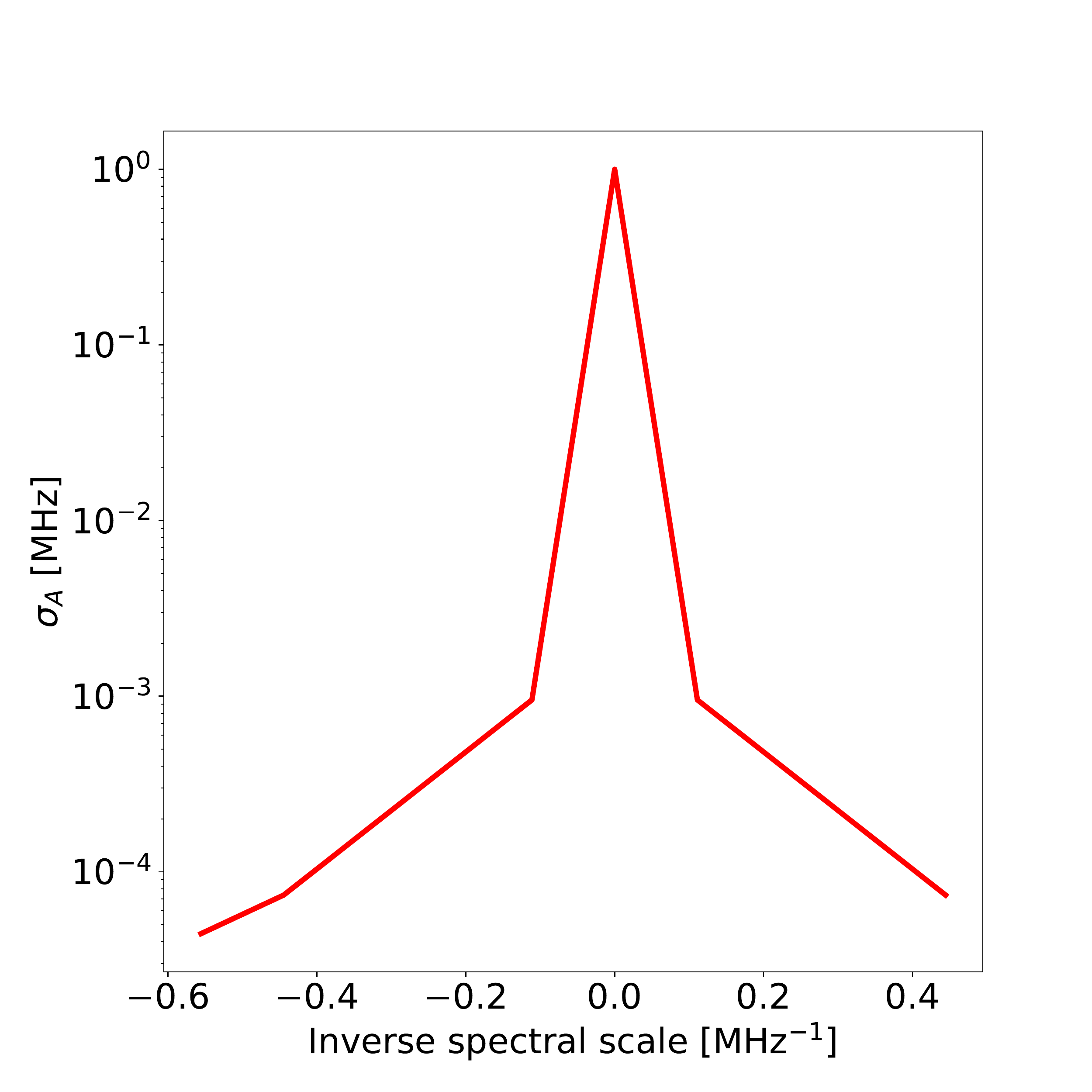}
	\includegraphics[width=0.50\textwidth]{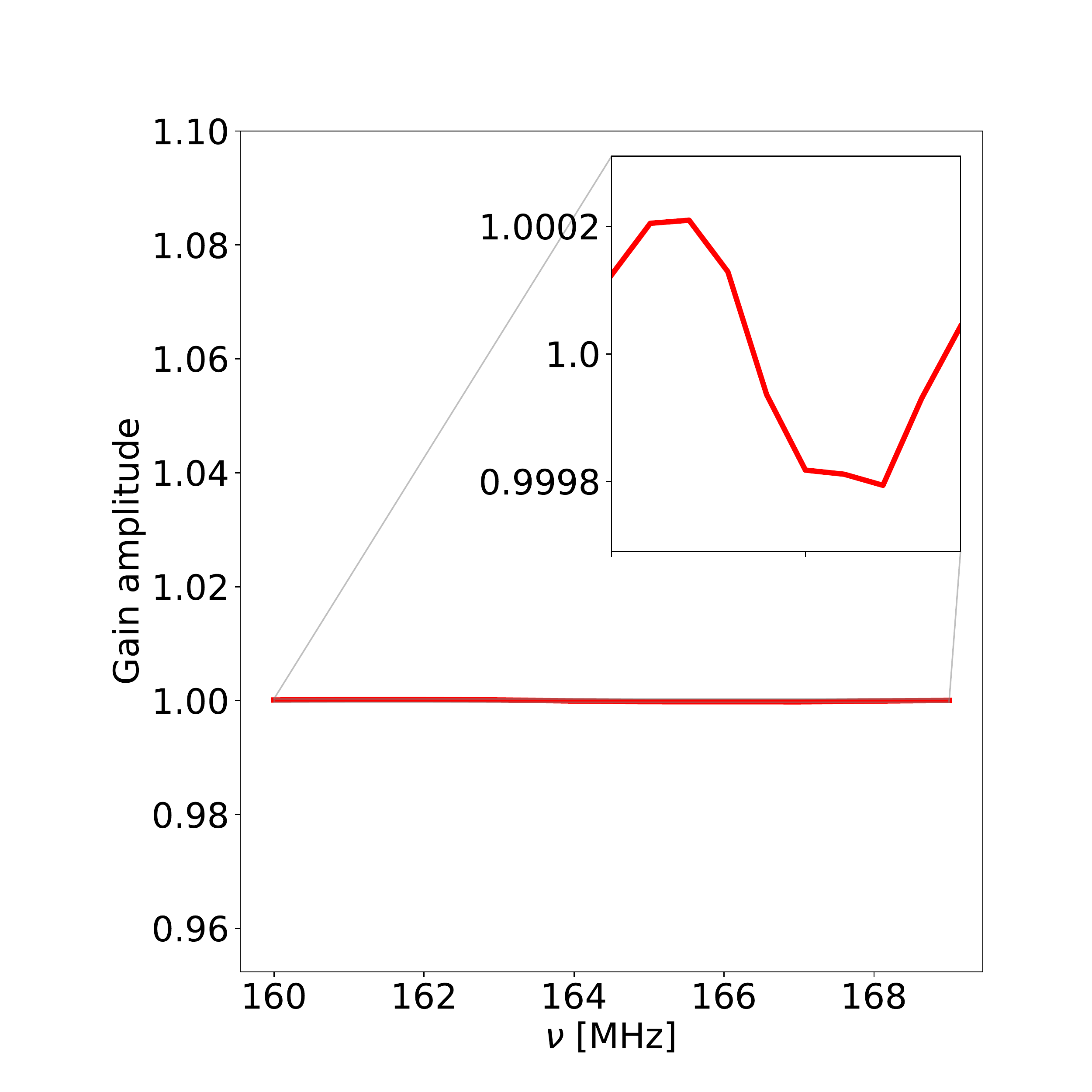}}
\caption{[Left] Prior on the standard deviation of the amplitudes of the basis vectors comprising the Fourier decomposition of the amplitude parameters of the redundant gain degeneracy function, $\sigma_\mathrm{A}$. [Right] Degenerate gain amplitude realisation used in the construction of the simulated observed visibilites calibrated in \autoref{Sec:Results}.} 
\label{Fig:gain_amplitude_prior} 
\end{figure*}

In this work, we assume that the mean amplitude of the gain solution is known a priori, from e.g. electromagnetic co-simulations of the receiver system and reflectometry measurements of the feed-dish system, with a one-sigma fractional uncertainty of $10\%$. We construct our simulated observed data such that it is described by a true redundant gain degeneracy function with a frequency-dependent bandpass amplitude, $\mathbfit{A}^\mathrm{true} = \mathbfss{F}\mathbfit{A}_\mathrm{F}^\mathrm{true}$ (\autoref{Fig:gain_amplitude_prior}, right), and, motivated by simulations of the first generation HERA feed (\citealt{2021MNRAS.500.1232F}), we draw $\mathbfit{A}_\mathrm{F}^\mathrm{true}$ from a Gaussian distribution with an exponentially decreasing spectral-scale dependent standard deviation (\autoref{Fig:gain_amplitude_prior}, left), which we assume is known a priori, from a combination of electric and electromagnetic co-simulations of the receiver system and reflectometry measurements of the feed-dish system, for the purposes of applying the prior defined in \autoref{Eq:ProbAgivenSigmaA}.

Using Bayes' theorem, we can thus derive the posterior probability distribution for the degenerate gain parameters conditioned on the raw data and the power in fluctuations in the degenerate gain amplitude as (see Paper I for details),
\begin{multline}
\label{Eq:ProbCalVNA}
\mathrm{Pr}(\mathbfit{A}_\mathrm{F}, \bm{\mathit{\Phi}}_{l,\mathrm{F}}, \bm{\mathit{\Phi}}_{m,\mathrm{F}} \;|\; \mathbfit{V}^\mathrm{obs}, \bm{\sigma}_{A_\mathrm{F}}^{2}) \; \propto
\; \mathrm{Pr}(\mathbfit{V}^\mathrm{obs} \;|\; \mathbfit{A}_\mathrm{F}, \bm{\mathit{\Phi}}_{l,\mathrm{F}}, \bm{\mathit{\Phi}}_{m,\mathrm{F}}) \;  \\
\times \mathrm{Pr}(\mathbfit{A}_\mathrm{F} \;|\; \bm{\sigma}_{A_\mathrm{F}}^{2}) \; \mathrm{Pr}(\mathbfit{A}_\mathrm{F}) \; \mathrm{Pr}(\bm{\mathit{\Phi}}_{l,\mathrm{F}}) \; \mathrm{Pr}(\bm{\mathit{\Phi}}_{m,\mathrm{F}}) \ ,
\end{multline}
where $\mathrm{Pr}(\mathbfit{V}^\mathrm{obs} \;|\; \mathbfit{A}_\mathrm{F}, \bm{\mathit{\Phi}}_{l,\mathrm{F}}, \bm{\mathit{\Phi}}_{m,\mathrm{F}})$ is given by \autoref{Eq:AbsCalVisLike} reparametrised in terms of the Fourier gain parameters of \autoref{Eq:DegeneracyFunctionFourierParameters}. 
Here, we assume $\bm{\sigma}_{A_\mathrm{F}}^{2}$ is known a priori. If the uncertainties associated with estimates of $\bm{\sigma}_{A_\mathrm{F}}^{2}$ are significant, one can, instead, sample from $\bm{\sigma}_{A_\mathrm{F}}$ jointly with $\mathbfit{A}_\mathrm{F}$, $\bm{\mathit{\Phi}}_{l,\mathrm{F}}$, $\bm{\mathit{\Phi}}_{m,\mathrm{F}}$ and $\bm{\varepsilon}$ and fit for the most probable degenerate gain parameters and power in the degenerate gain amplitude given the data (see Paper I for details).

\subsection{Absolute calibration using \textsc{BayesCal}}
\label{Sec:BayesCal}

To extend the approach to absolute calibration of redundantly calibrated visibilities described in \autoref{Sec:AbscalWithGainPriors} to explicitly account for sky-model incompleteness, we subdivide our sky model into two components, \begin{enumerate}\item a simulated component, $\mathbfit{V}^\mathrm{sim}$, derived using our a priori knowledge of the brightness distribution of known Galactic emission and extragalactic sources in the region of sky under observation and \item a fitted component, $\mathbfit{V}^\mathrm{fit} (\bm{\varepsilon})$, with $\bm{\varepsilon}$ a set of image-domain parameters of the fitted model, that accounts for the contribution to $\mathbfit{V}^\mathrm{true}$ missing in $\mathbfit{V}^\mathrm{sim}$. \end{enumerate}
We write our model visibilites as the sum of these components,
\begin{equation}
\label{Eq:SkyPlusFittedVisModel}
\mathbfit{V}^\mathrm{model} = \hat{\mathbfss{H}}^\mathrm{m}\mathbfss{D}^\mathrm{m}(\mathbfit{V}^\mathrm{sim} + \mathbfit{V}^\mathrm{fit}(\bm{\varepsilon})) \ .
\end{equation}
Following paper I, we write our fitted visibility model over all LSTs of our calibration data set as,
\begin{equation}
\label{Eq:FittedVisibilitiesFullModel}
\mathbfit{V}^\mathrm{fit} = \mathbfss{F}_{\mathrm{fr}}\mathbfss{P}_{\mathrm{nn},I}\mathbfss{S}\mathbfss{C}\bm{\varepsilon}\ .
\end{equation}
Here, 
\begin{itemize}
\item $\mathbfss{F}_{\mathrm{fr}}$ and $\mathbfss{P}_{\mathrm{nn},I}$ are as defined in \autoref{Eq:VtrueExpanded}. 
\item $\mathbfss{S}$ encodes our spectral model for the fitted component of the emission. In \autoref{Sec:Results}, we use \textsc{BayesCal} to calibrate simulated data on sub-30 m HERA baselines\footnote{
While there is no fundamental limitation restricting the use of \textsc{BayesCal} to calibration of visibilities on short baselines, calibrating on short baselines has the benefit of reducing the spatial resolution necessary for $\mathbfit{V}^\mathrm{fit}$ to accurately model the incomplete component of the data and, thus, decreasing the computation time required for the calibration. Additionally, 21 cm cosmology experiments are generally most sensitive on large angular scales; thus, if sky-model incompleteness can be mitigated by \textsc{BayesCal}, barring potential complications such as increased cross-talk between antennas, short baselines constitute a preferred subset if one wishes to use the same data during calibration and data analysis.
} (see \autoref{Sec:InstrumentModel} for details of the instrument model). In this baseline range, Galactic diffuse synchroton emission (GDSE) accounts for a large fraction of the power in the visibilities and, owing to the level of uncertainty associated with models for this emission, it is also the dominant source of model incompleteness. As such, we use a power law model for the spectral structure of the fitted emission in this regime with power law index $\beta_\mathrm{m} = \langle \beta \rangle_\mathrm{GDSE}$, where ${\langle \beta \rangle_\mathrm{GDSE}}$ is the mean spectral index of the a priori known component of the GDSE brightness temperature distribution. In general, the Bayesian evidence can be used to select an optimal spectral model from a set of models for the data or, for this fixed spectral model, to determine the optimal choice of model spectral index. However, in \autoref{Sec:Results}, we will show that this a priori choice of fixed spectral model is already sufficiently robust for high fidelity recovery of calibration solutions with the \textsc{BayesCal} calibration framework, for the observational setup described in \autoref{Sec:InstrumentModel}, so we do not consider such a model selection analysis further here. 
\item $\bm{\varepsilon}$ are our full set of sky-model amplitude parameters for describing the Stokes I brightness temperature of the component of emission in $\mathbfit{V}^\mathrm{true}$ that is unaccounted for in $\mathbfit{V}^\mathrm{sim}$. We define these parameters on a \textsc{HEALPix} grid (\citealt{2005ApJ...622..759G}) at a reference frequency, $\nu_{0}$. 
\item $\mathbfss{C}$ is a matrix that maps between the parameters of our fitted calibration model, $\bm{\varepsilon}$, and $\bm{\varepsilon}_\mathrm{concat}$, where $\bm{\varepsilon}_\mathrm{concat}$ is the vectorisation of the concatenation of model maps over the LSTs in the data set to be calibrated. The $N_\mathrm{pix}$ pixels of the \textsc{HEALPix} grid cover the full sky; however, if the primary beam of a visibility downweights emission at large zenith angles, a subset of the pixels within the observer's horizon may be sufficient to model the visibilities. To take advantage of this and reduce the computational cost of the calibration, when applying \textsc{BayesCal}, we define our set of parameters, $\bm{\varepsilon}$, such that they represent the amplitudes in the $N_\mathrm{pix, s}$ length subset of pixels defined by the union of pixels that fall within a fixed zenith angle, $\theta_\mathrm{cut}$, such that, within this region, the primary beam lies above a given threshold weight, at each central local sidereal time (LST) of the integrations over which the calibration solutions are being jointly estimated. Commonly, one wishes to calibrate interferometric data comprised of a number of shortly spaced successive time integrations, in which case, the region of sky contributing to successive integrations is strongly correlated in time. $\bm{\varepsilon}$ represents the amplitudes of fixed LST-independent regions of the sky; however, $\mathbfss{S}$ and $\mathbfss{F}_{\mathrm{fr}}\mathbfss{P}_{\mathrm{nn},I}$ operate on vectorisations of the concatenation, over the LSTs in the data set to be calibrated, of model maps ($\bm{\varepsilon}_\mathrm{concat}$) and model image cubes ($\mathbfss{S}\bm{\varepsilon}_\mathrm{concat}$), respectively. $\mathbfss{C}$ is defined such that $\bm{\varepsilon}_\mathrm{concat} = \mathbfss{C}\bm{\varepsilon}$ (see paper I for details).
\end{itemize}

\subsubsection{\textsc{BayesCal} likelihood}
\label{Sec:DataLikelihood}

Our likelihood at this stage can be written as (see paper I),  
\begin{multline}
\label{Eq:BayesCalVisLike}
\mathrm{Pr}(\bm{V}^\mathrm{obs} \;|\; \bm{A}, \bm{\mathit{\Phi}}_{l}, \bm{\mathit{\Phi}}_{m}, \bm{\varepsilon}) =\\ \frac{1}{\pi^{N_{\mathrm{vis}}}\mathrm{det}(\mathbfss{N})} \exp\left[-\left(\bm{V}^\mathrm{obs} - \hat{\mathbfss{H}}^\mathrm{m}\mathbfss{D}^\mathrm{m}(\bm{V}^\mathrm{sim} + \bm{V}^\mathrm{fit})\right)^\dagger \right.\\ \left. \mathbfss{N}^{-1} 
 \left(\bm{V}^\mathrm{obs} - \hat{\mathbfss{H}}^\mathrm{m}\mathbfss{D}^\mathrm{m}(\bm{V}^\mathrm{sim} + \bm{V}^\mathrm{fit})\right)\right] \ .
 \end{multline}
Here, $\mathbfss{N}$ is the expected covariance matrix of the raw visibilities and can be estimated by,
\begin{equation}
\label{Eq:CovarianceCalVis}
\mathbfss{N} = (\mathbfss{H}\mathbfss{D})^{\dagger}\mathbfss{N}^{\prime}(\mathbfss{H}\mathbfss{D}) \ ,
\end{equation}
where $\mathbfss{N}^{\prime}$ is the expected covariance matrix of the calibrated data. In this paper, we assume the noise is uncorrelated between baselines and thus $\mathbfss{N}$ has elements,
\begin{equation}
\label{Eq:CalibratedCovarianceElements}
N_{ij}^{\prime} = \left< n_i n_j^{*}\right> = \delta_{ij}\sigma_{j}^{2} \ .
\end{equation}
Here, $\left< .. \right>$ represents the expectation value and $\sigma_{j}$ is the rms value of the noise term for visibility $j$, which, for a pair of identical antennas individually experiencing equal system noise, is given by (e.g. \citealt{1999ASPC..180.....T}),
\begin{equation}
\label{Eq:VisabilityNoise}
\sigma=\dfrac{1}{\eta_{s}\eta_{a}}\dfrac{2k_{B}T_{\mathrm{sys}}}{A\sqrt{2\Delta\nu\tau}} \ ,
\end{equation}
where $k_{B}=1.3806\times10^{-23}~\mathrm{JK^{-1}}$ is Boltzmann's constant, $T_{\mathrm{sys}}$ is the system noise temperature, $A$ is the antenna area, $\Delta\nu$ is the channel width, $\tau$ is the integration time and $\eta_{s}$ and $\eta_{a}$ are the system and antenna efficiencies, respectively. 

In \autoref{Sec:RedundantCalibration}, the gain solutions at each frequency can be solved for independently. The introduction in \textsc{BayesCal} of the fitted frequency dependent visibility model, $\mathbfit{V}^\mathrm{fit}$, as well as the inclusion of spectrally dependent gain priors in both the absolute calibration with gain priors and \textsc{BayesCal} approaches, means that in each of these cases the instrumental gains are correlated in frequency and, correspondingly, the gain parameters of each channel of the data set must be fit for jointly. Thus, for both of these cases, we redefine our model degenerate gain parameter matrix such that for a single time integration, $\mathbfss{D}^\mathrm{m}$ is a block diagonal matrix where each block, $\mathbfss{D}^\mathrm{m}_{r}$, is a diagonal matrix with elements $D_{r, ij} = \delta_{ij}A_{r}\mr{e}^{i\mathbfit{b}_{i}^{T} \sPhi_{r}}$ and the subscript $r$ runs over the $N_\nu$ channels of the data set. We also redefine $\hat{\mathbfss{H}}^\mathrm{m}$ in an analogous manner to now encode the maximum likelihood redundant gain parameters for each channel of the data set.

\subsubsection{Fitted calibration model power priors}
\label{Sec:PowerSpectralPriors}

Paper I details how one can place informative priors on the power in the fitted component of the \textsc{BayesCal} calibration model, parametrised in terms of:
\begin{enumerate} \item a model for the spatial power spectrum of $\bm{\varepsilon}$, or
\item a model for the covariance matrix of $\bm{\varepsilon}$.
\end{enumerate}
These can be used to eliminate the degeneracies between $\bm{\varepsilon}$ and instrumental calibration parameters that are present when $\bm{\varepsilon}$ is unconstrained. A diagonal model for the covariance matrix of $\bm{\varepsilon}$ encoding the expected power in the image domain attributable to emission missing in the simulated sky model from which $\mathbfit{V}^\mathrm{sim}$ derives provides a minimally informative prior that is sufficient for this goal. For the incomplete sky models considered in this paper (see \autoref{Sec:SkyModel}) and on sub-30 m baselines (which are the baseline-lengths included in data sets on which we demonstrate absolute calibration with \textsc{BayesCal} in \autoref{Sec:Results}), the total sky-model error is dominated by the power in unmodelled or mismodelled diffuse emission that derives from the discrepancy between our model for the full-Stokes diffuse emission and the true full-Stokes brightness distribution. In this paper, we assume a uniform uncertainty on the diffuse emission in the field being calibrated and approximate the covariance matrix of $\bm{\varepsilon}$ as diagonal, with elements,
\begin{equation}
\label{Eq:PowerPrior}
\sSigma_{lm, \delta T, ii} = \sigma_{\varepsilon}^{2} \ .
\end{equation}
Here, $\sigma_{\varepsilon}$ is the expected RMS error on the sky model used to construct $\bm{V}^\mathrm{sim}$, which we approximate as $\sigma_{\varepsilon} = f (\sum (T_i - \overline{T})^2 / N_{\varepsilon})^{1/2} $, where $f$ and $(\sum (T_i - \overline{T})^2 / N_{\varepsilon})^{1/2}$ are the fractional uncertainty (see \autoref{Sec:SkyModel}) and RMS of our simulated calibration sky model, respectively. We can correspondingly write our power spectral prior as,
\begin{multline}
\label{Eq:ProbObsPSofEpsilon}
\mathrm{Pr}(\bm{\varepsilon} \;|\; \sigma_{\varepsilon}^{2}) \; \propto \; \frac{1}{\sqrt{\mathrm{det}(\bm{\sSigma}_{lm, \delta T})}}  \exp\left[-\frac{1}{2}~\bm{\varepsilon}^{T}\bm{\sSigma}_{lm, \delta T}^{-1}\bm{\varepsilon}\right] \ .
\end{multline}

Incorporating this prior on the power in the parameters of our fitted visbility model and not assuming any a priori correlation between $\bm{A}$, $\bm{\mathit{\Phi}}_{l}$, $\bm{\mathit{\Phi}}_{m}$, $\bm{\varepsilon}$, we can write the joint probability density of our calibration parameters and our image-space model coefficients conditioned on the raw data and the power in $\bm{\varepsilon}$ as,
\begin{multline}
\label{Eq:ProbCalSkyGivenPS}
\mathrm{Pr}(\mathbfit{A}, \bm{\mathit{\Phi}}_{l}, \bm{\mathit{\Phi}}_{m}, \bm{\varepsilon} \;|\; \mathbfit{V}^\mathrm{obs}, \sigma_{\varepsilon}^{2}) \; \propto
\; \mathrm{Pr}(\mathbfit{V}^\mathrm{obs} \;|\; \mathbfit{A}, \bm{\mathit{\Phi}}_{l}, \bm{\mathit{\Phi}}_{m}, \bm{\varepsilon}) \\ \times \mathrm{Pr}(\bm{\varepsilon} \;|\; \sigma_{\varepsilon}^{2}) \; \mathrm{Pr}(\mathbfit{A}) \; \mathrm{Pr}(\bm{\mathit{\Phi}}_{l}) \; \mathrm{Pr}(\bm{\mathit{\Phi}}_{m}) \ . 
\end{multline}
As discussed in paper I, in a scenario where the uncertainty on $f$ is assumed to be significant, one could jointly sample from a parametrised model for $\sigma_{\varepsilon}^{2}$, in addition to $\bm{\varepsilon}$ and the calibration parameters. Here, we calculate $\sigma_{\varepsilon}^{2}$ as described below \autoref{Eq:PowerPrior}; however, we have confirmed that the maximum a posteriori calibration parameters recovered in \autoref{Sec:Results} are relatively insensitive to this assumption, with errors of up to $20\%$ on the value of $f$ used to define the power spectral prior altering the level of spurious spectral fluctuations in the derived calibration solutions by less than a factor of two.

\subsubsection{Analytic marginalisation over the sky-model parameters}
\label{Sec:AnalyticMarginalisationOverTheSkyModelParameters}

Using Bayes' theorem, we can write the posterior probability distribution of the degenerate gain parameters and sky-model parameters conditioned on the raw data, the power in spectral fluctuations in the degenerate gain amplitude (\autoref{Eq:ProbAgivenSigmaA}) and the power in $\mathbfit{V}^\mathrm{fit}$ (\autoref{Eq:ProbObsPSofEpsilon}) as,
\begin{multline}
\label{Eq:ProbCalSkyGivenPSVNA}
\mathrm{Pr}(\bm{A}_\mathrm{F}, \bm{\mathit{\Phi}}_{l,\mathrm{F}}, \bm{\mathit{\Phi}}_{m,\mathrm{F}}, \bm{\varepsilon} \;|\; \bm{V}^\mathrm{obs}, \sigma_{\varepsilon}^{2}, \bm{\sigma}_{A_\mathrm{F}}^{2}) \; \propto \\
\; \mathrm{Pr}(\bm{V}^\mathrm{obs} \;|\; \bm{A}_\mathrm{F}, \bm{\mathit{\Phi}}_{l,\mathrm{F}}, \bm{\mathit{\Phi}}_{m,\mathrm{F}}, \bm{\varepsilon}) \; \mathrm{Pr}(\bm{\varepsilon} \;|\; \sigma_{\varepsilon}^{2}) \;  \\
\times \mathrm{Pr}(\bm{A}_\mathrm{F} \;|\; \bm{\sigma}_{A_\mathrm{F}}^{2}) \; \mathrm{Pr}(\bm{\mathit{\Phi}}_{l,\mathrm{F}}) \; \mathrm{Pr}(\bm{\mathit{\Phi}}_{m,\mathrm{F}}) \ ,
\end{multline}
In principle, one could sample from the posterior probability distribution given in \autoref{Eq:ProbCalSkyGivenPSVNA}. However, since we are interested specifically in the calibration parameters, 
we can analytically marginalise over the sky-model parameters and directly sample from the far lower dimensional space of the posterior probability distribution for the calibration parameters. This gives our marginal posterior probability distribution, $\mathrm{Pr}(\bm{A}_\mathrm{F}, \bm{\mathit{\Phi}}_{l,\mathrm{F}}, \bm{\mathit{\Phi}}_{m,\mathrm{F}} \;|\; \bm{V}^\mathrm{obs}, \sigma_{\varepsilon}^{2}, \bm{\sigma}_{A_\mathrm{F}}^{2})$. For a detailed derivation of this distribution see paper I. Here, we quote the solution for the marginalised distribution for the case of uniform priors on the degenerate tip-tilt phases,
\begin{multline}
\label{Eq:Margin}
\mathrm{Pr}(\bm{A}_\mathrm{F}, \bm{\mathit{\Phi}}_{l,\mathrm{F}}, \bm{\mathit{\Phi}}_{m,\mathrm{F}} \;|\; \bm{V}^\mathrm{obs}, \sigma_{\varepsilon}^{2}, \bm{\sigma}_{A_\mathrm{F}}^{2}) \propto \\
\frac{\mathrm{det}(\mathbfss{N})^{-1}}{\sqrt{\mathrm{det} (\bm{\upUpsilon}) ~ \mathrm{det}(\bm{\sSigma}_{lm, \delta T}) ~ \mathrm{det}(\bm{\sSigma}_{A_\mathrm{F}})}} \\
\times \exp \biggl[-(\delta\bm{V}^\mathrm{obs})^{\dagger}\mathbfss{N}^{-1}(\delta\bm{V}^\mathrm{obs}) + (\overline{\delta \bm{V}}^\mathrm{obs})^{T} \bm{\upUpsilon}^{-1}(\overline{\delta \bm{V}}^\mathrm{obs}) \\
 - \frac{1}{2}(\bm{A}_\mathrm{F} -\bar{\bm{A}}_\mathrm{F})^{T}\bm{\sSigma}_{A_\mathrm{F}}^{-1}(\bm{A}_\mathrm{F} -\bar{\bm{A}}_\mathrm{F} )\biggr].
\end{multline} 
Here, $\overline{\delta \mathbfit{V}}^\mathrm{obs} = \bm{\upLambda}^{\dagger}\mathbfss{N}^{-1}\delta\mathbfit{V}^\mathrm{obs}$ represents the weighted gridded projection of $\delta\mathbfit{V}^\mathrm{obs}$ on the parameter space of the fitted visibility model,
$\delta\bm{V}^\mathrm{obs} = \bm{V}^\mathrm{obs} - \hat{\mathbfss{H}}^\mathrm{m}\mathbfss{D}^\mathrm{m}\bm{V}^\mathrm{sim}$ is the residual between the visibilities and the fixed simulated component of our calibration model and $\bm{\upUpsilon} \equiv \bm{\upLambda}^{\dagger}\mathbfss{N}^{-1}\bm{\upLambda} + \bm{\sSigma}_{lm, \delta T}^{-1}$ is the covariance matrix of $\overline{\delta \mathbfit{V}}^\mathrm{obs}$, with $\bm{\upLambda} \equiv \hat{\mathbfss{H}}^\mathrm{m}\mathbfss{D}^\mathrm{m}\mathbfss{F}_\mathrm{fr}\mathbfss{P}\mathbfss{S}\mathbfss{C}$ the system matrix mapping from our image domain parameters to the fitted model visibilities.

The dense matrix inversion of $\bm{\upUpsilon}$ and the calculation of the determinants are the most computationally expensive elements in \autoref{Eq:Margin}. However, terms in the exponent without a calibration parameter dependence can be precomputed prior to sampling, the determinant prefactors that are independent of the calibration parameters can be neglected for the purposes of parameter estimation and $\mathrm{det}(\mathbfss{N})$ is diagonal and thus can be calculated inexpensively as the product of the diagonal elements. The remaining diagonal matrix inversions can also be calculated trivially as the inverse of the matrix diagonals.

When calibrating simulated observed data using the \textsc{BayesCal} framework in \autoref{Sec:Results}, we compute $\bm{\upUpsilon}^{-1}$ via Cholesky decomposition and we use \autoref{Eq:Margin} to sample directly from the marginalised posterior for the degenerate gain Fourier parameters, $\mathrm{Pr}(\bm{A}_\mathrm{F}, \bm{\mathit{\Phi}}_{l,\mathrm{F}}, \bm{\mathit{\Phi}}_{m,\mathrm{F}} \;|\; \bm{V}^\mathrm{obs}, \sigma_{\varepsilon}^{2}, \bm{\sigma}_{A_\mathrm{F}}^{2})$, using nested sampling as implemented by the \textsc{PolyChord} algorithm \citep{2015MNRAS.453.4384H, 2015MNRAS.450L..61H}.

\subsection{Temporal model}
\label{Sec:TemporalModel}

Simultaneous calibration of multiple time integrations, such that the calibration solutions are described by a given temporal model, provides a more stringent constraint on the calibration parameters, in general, relative to calibrating each integration independently, and allows one to take advantage of a priori knowledge of the expected temporal evolution of the calibration parameters. When calibrating with \textsc{BayesCal}, it additionally enables one to better constrain the parameters of the fitted sky model.

In \autoref{Sec:Results}, we will use absolute calibration, absolute calibration with gain priors and \textsc{BayesCal} to calibrate five minute data sets comprised of ten snapshot observations (see \autoref{Sec:InstrumentModel}). To achieve this, we define a degenerate gain parameter matrix, $\mathbfss{D}^\mathrm{m}$, that encodes the degenerate gain parameters for all time integrations in the data set being calibrated. In this case, $\mathbfss{D}^\mathrm{m}$ is a block diagonal matrix where each block, $\mathbfss{D}^\mathrm{m}_{s}$, is itself a block diagonal matrix comprised of blocks, $\mathbfss{D}^\mathrm{m}_{r}$, which are diagonal matrices with elements $D_{s,r, ij} = \delta_{ij}A_{s,r}\mr{e}^{i\mathbfit{b}_{i}^{T} \sPhi_{s,r}}$, where the subscripts $r$ and $s$ run over the $N_\nu$ channels and $N_\mathrm{t}$ integrations in the data set, respectively (see paper I for details). In an analogous manner, $\hat{\mathbfss{H}}^\mathrm{m}$ is redefined to comprise $N_\mathrm{t}$ block diagonal matrices with the $s$th block diagonal encoding the maximum likelihood redundant gain parameters for the $s$th integration in the data set; we correspondingly redefine our total visibility data as $\mathbfit{V}^\mathrm{obs} = [(\mathbfit{V}^\mathrm{fit}_{t_{0}})^{T}, (\mathbfit{V}^\mathrm{fit}_{t_{1}})^{T}, \ldots (\mathbfit{V}^\mathrm{fit}_{t_{n}})^{T}]^{T}$, where the visibility data is comprised of $n$ time integrations and $\mathbfit{V}^\mathrm{fit}_{t_{i}}$ is a vector of the raw visibilites corresponding to the $i$th time integration. $\mathbfit{V}^\mathrm{sim}$ and $\mathbfit{V}^\mathrm{fit}$ are redefined in an analogous manner to $\mathbfit{V}^\mathrm{obs}$. Substituting these redefinitions, the likelihoods and posteriors described in Sections \ref{Sec:RedundantCalibration} -- \ref{Sec:BayesCal} can be applied, as written, to simultaneously calibrate multiple-integration data sets.

In this paper, we will assume that the instrument gains are stable over the five minute duration of the data to be calibrated, use time-stationary gains in the simulated observed data and use a time-stationary gain model when calibrating the data. On longer time scales, variations in the environmental conditions of the receiver and signal chain, such as changing ambient temperature, will impart temporal fluctuations in the antenna gains on the temporal scales of those variations. To address this, a model can be introduced for the temporal evolution of the calibration solutions, the parameters of which can be constrained by their own priors, and the optimal temporal model for the data can be determined using Bayesian model selection (see paper I for details).

\section{Instrument Model}
\label{Sec:InstrumentModel}

To explore the effectiveness of the \textsc{BayesCal} framework for interferometric calibration derived in the preceding section, in \autoref{Sec:Results} we demonstrate its applications to the absolute calibration of a simulated observed data set. In \autoref{Sec:SkyModel}, 
we will define the diffuse and point source components of the complete and incomplete sky models used to construct our simulated observed visibility data set, $\mathbfit{V}^\mathrm{obs}$, and calibration visibility model, $\mathbfit{V}^\mathrm{sim}$, respectively. But first, here, we describe the instrument model we use to construct $\mathbfit{V}^\mathrm{obs}$ and $\mathbfit{V}^\mathrm{sim}$, as well as the instrumental forward model component of our fitted sky model, $\mathbfit{V}^\mathrm{fit}$.

\subsection{Interferometric array}
\label{Sec:InterferometricArray}

\begin{figure}
	\centerline{
	\includegraphics[width=0.50\textwidth]{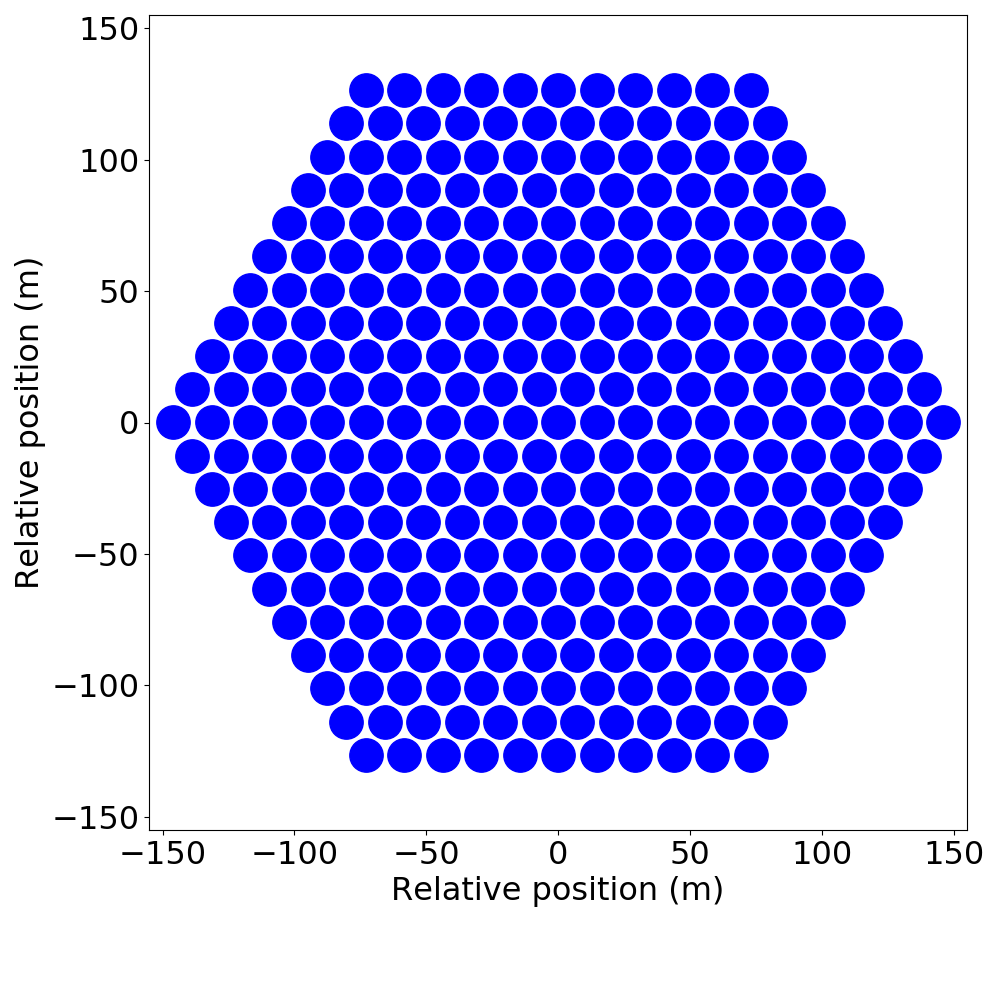}}
\vspace{-1cm}
\caption{Simulated hexagonal close-packed antenna configuration with 331 antennas. Antenna positions plotted relative to $30\fdg721$S $21\fdg411$E at the SKA-South Africa site, Karoo desert.}
\label{Fig:antenna_layout}
\end{figure}

\begin{figure*}
\centerline{
\includegraphics[width=0.50\textwidth]{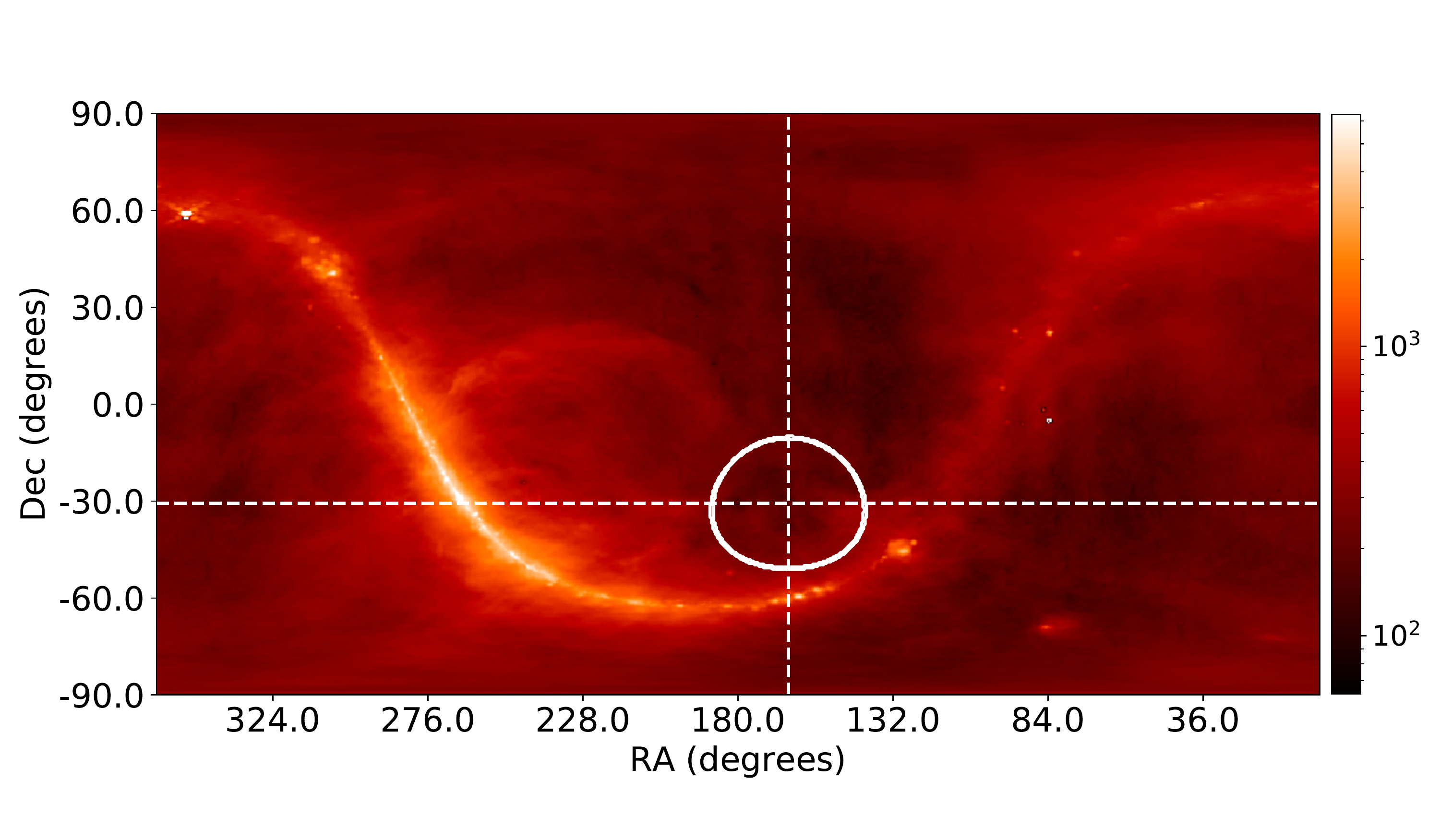}
\includegraphics[width=0.50\textwidth]{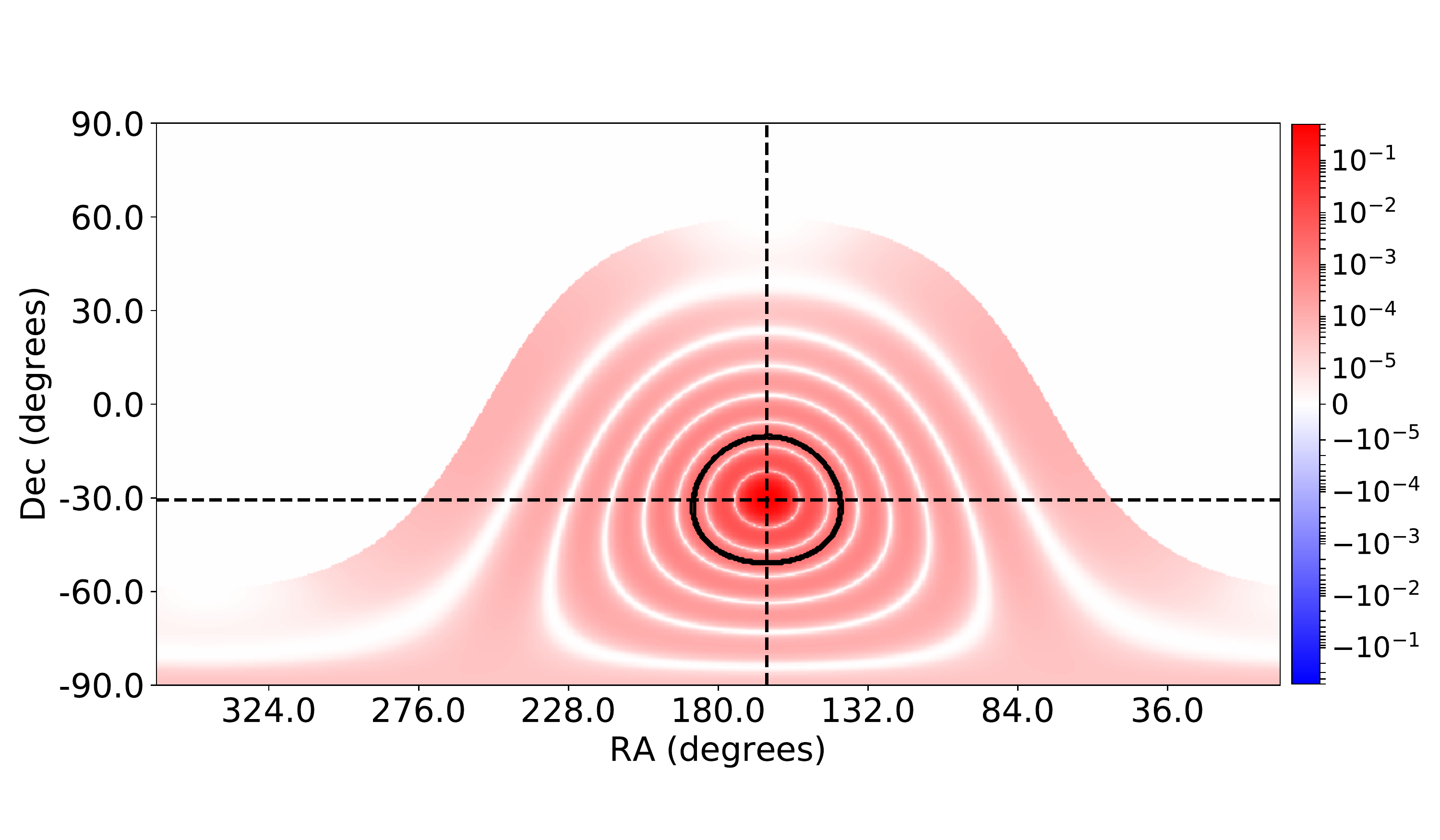}}
\centerline{
\includegraphics[width=0.50\textwidth]{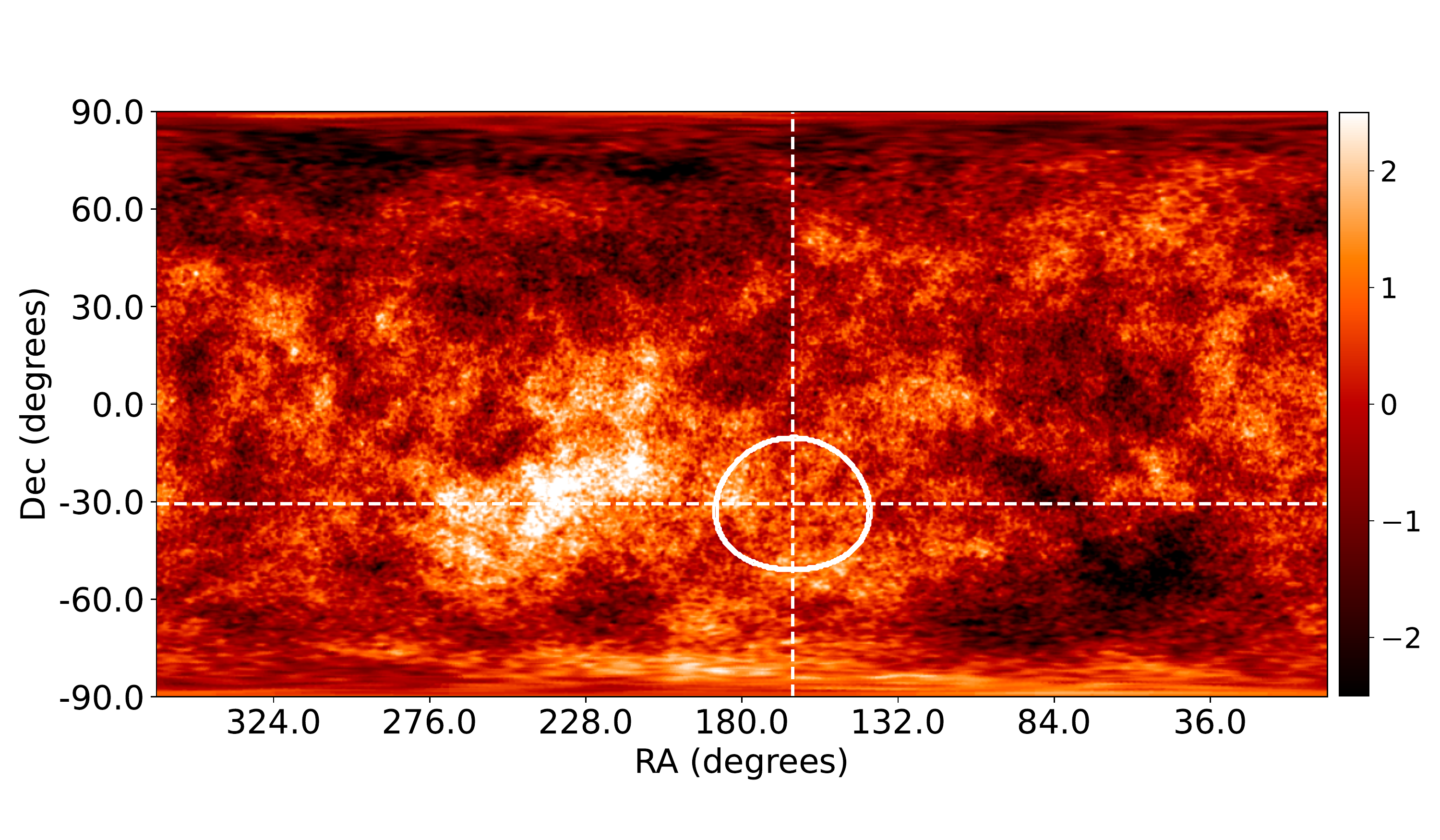}
\includegraphics[width=0.50\textwidth]{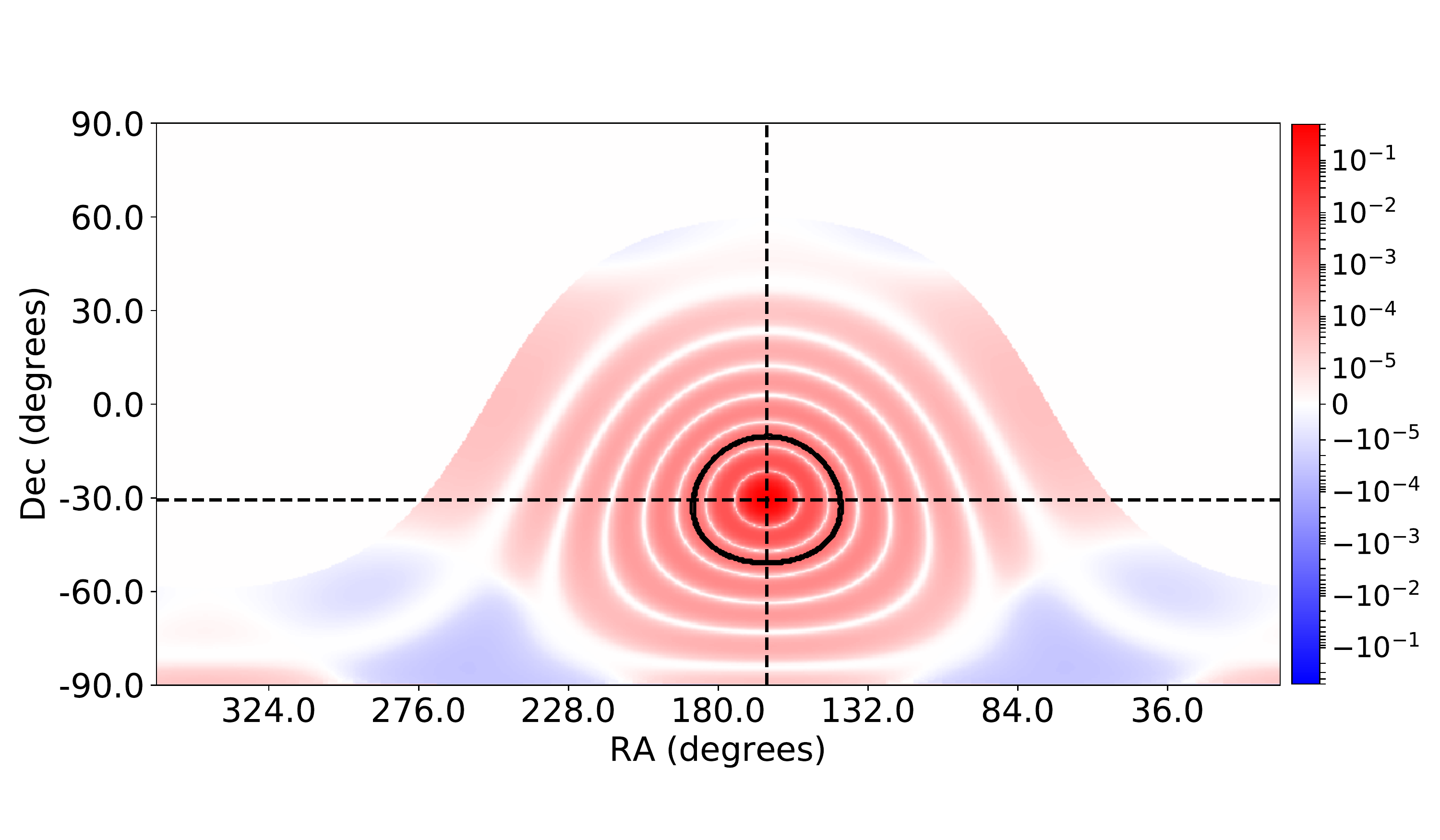}}
\centerline{
\includegraphics[width=0.50\textwidth]{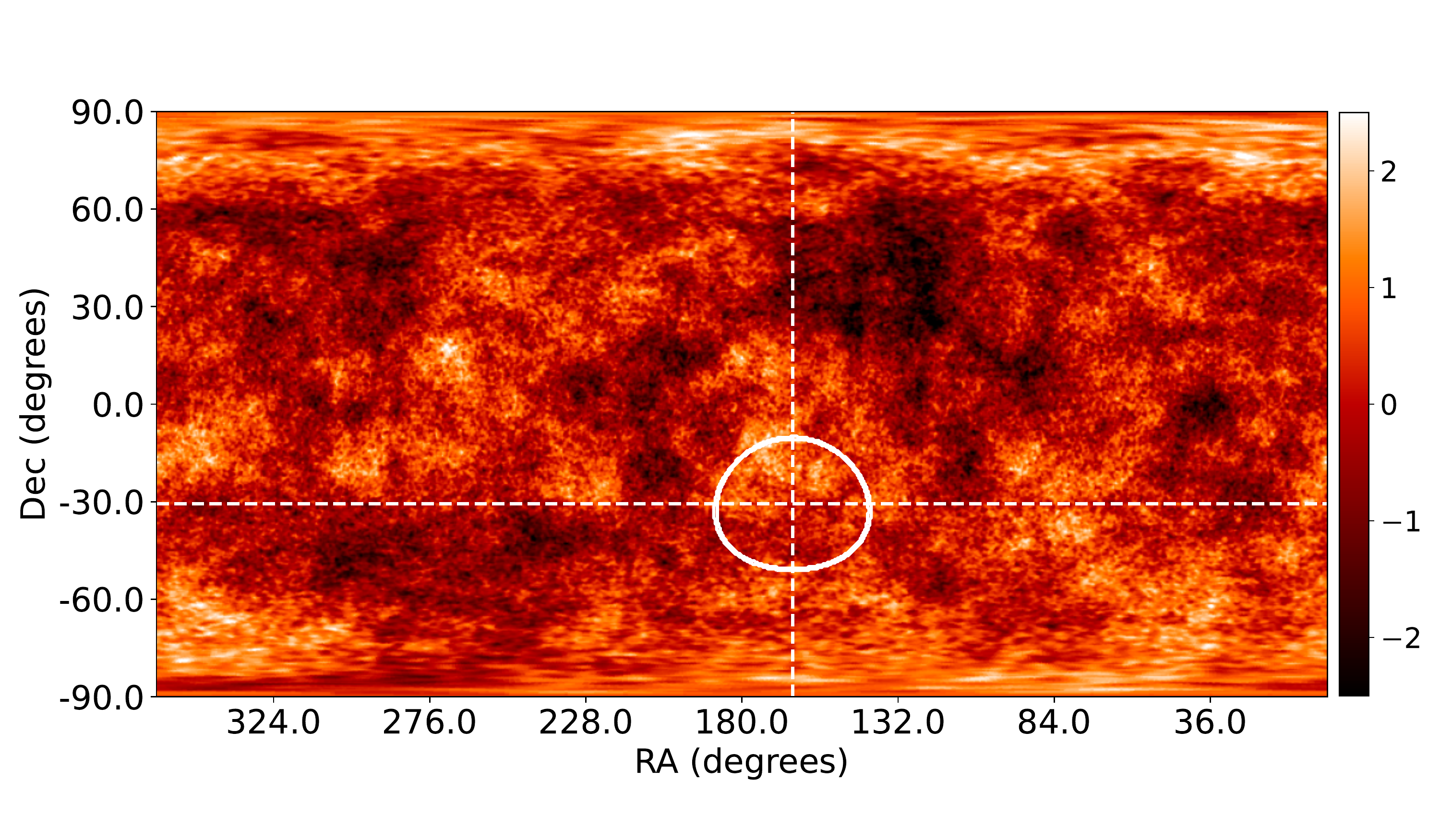}
\includegraphics[width=0.50\textwidth]{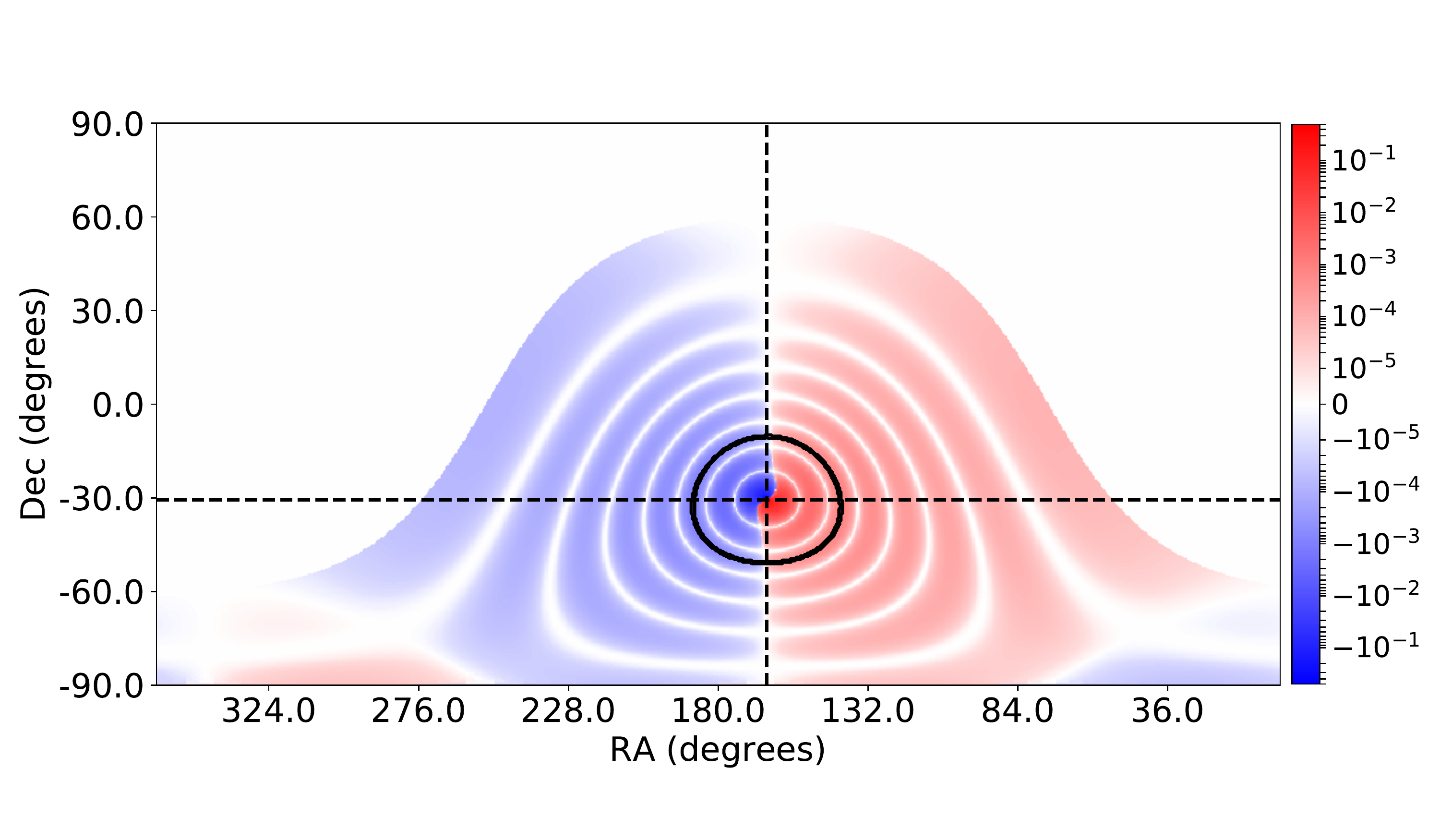}} 
\caption{Diffuse emission brightness temperature in Kelvin (left) in Stokes $I$ (top), $Q$ (middle) and $U$ (bottom) and polarised primary beam patterns (right), encoding the projections of Stokes $I$ (top), $Q$ (middle) and $U$ (bottom) emission onto south-north oriented feeds of the antenna in the array, calculated from \autoref{Eq:VoltageBeamEquatorial} via \autoref{Eq:PolarisedPrimaryBeamMatrix}, at the longitude and latitude of our simulated array and LST of the start of our simulated observed data set. In each subplot, the intersections of the dashed white (left) and black (right) lines is at zenith for the simulated array in the first integration of the simulated observed data and the solid white (left) and black (right) circles show the perimeter of the $40\fdg0$-diameter FoV used in the simulated observations.}
\label{Fig:diffuse_emission_model_and_beam_with_FoV} 
\end{figure*}

For our instrumental model, we use a 331 antenna close-packed hexagonal array configuration\footnote{In the built HERA antenna layout, the hexagonal grid is fractured into three sectors and incorporates outrigger antennas for improved imaging capabilities (\citealt{2016ApJ...826..181D}). The 331 antenna close-packed hexagonal array configuration that we use here is a simplified version of that array layout but preserves the feature relevant to calibration comparison carried out in \autoref{Sec:Results}.} with 14.6 m antenna spacing (see \autoref{Fig:antenna_layout}), situated at the geographic location of HERA (\citealt{2017PASP..129d5001D}). We simulate a five minute observation, commencing at 02:52:00 UTC on the 1st of January 2020 and comprised of ten snapshot observations\footnote{Here, we model the visibilities based on a snapshot at the center of the time step and add noise in the $uv$-domain appropriate for the stated integration time.}. We select this commencement time so that a cold region of the Galaxy, lying away from the Galactic plane (zenith pointing $\mathrm{RA} = 164\fdg42$, $\mathrm{Dec} = -30\fdg61$  at the start of the observation; see \autoref{Fig:diffuse_emission_model_and_beam_with_FoV}, left) is passing through zenith\footnote{This is a preferred choice of field for 21 cm cosmology because it reduces the dynamic range between the foreground emission and 21 cm signal, mitigating the level of foreground systematics that will be introduced to the calibrated data for a given level of spurious spectral structure in the calibration solutions.}. The choice of $n_{t} = 10$ relatively long 30 s integration times maximises the calibration signal to noise, while minimising the computational expense associated with joint estimation of the calibration solutions over multiple integrations\footnote{Nevertheless, we emphasise that the calibration formulation presented here can be applied to an arbitrary number of integrations, each with arbitrarily short integration time, subject to available computational resources.
}.

In general, astrophysical emission from all parts of the sky where the primary beam is non-zero contributes to the measured visibilities. Since the beam is non-zero for zenith angles above the horizon, ideally, one would include all discretised emission elements with zenith angle coordinates $\theta(t) \lesssim 90\fdg0$ when calculating $V(\mathbfit{u}_{pq},\nu, t)$. However, because the computational expense associated with calculating $\bm{\upUpsilon}^{-1}$ in \autoref{Eq:Margin} scales approximately as $N_\mathrm{pix}^{3}$, where $N_\mathrm{pix}$ is the number of sky-model parameters associated with $\mathbfit{V}^\mathrm{fit}$, we limit our model to a restricted FoV centered on the RA and Dec of zenith at the longitude and latitude of our simulated array at each LST of our simulated observations. In the simulations considered in \autoref{Sec:Results}, we use a $40\fdg0$-diameter FoV at each LST, which is approximately double the $\sim20\fdg0$ diameter of the first null of our beam model (see \autoref{Sec:PrimaryBeamModel}). We have confirmed that repeating the analyses for FoVs smaller than the fiducial $40\fdg0$ considered here does not affect the conclusions of this paper. We plan to explore computationally efficient means to extend the FoV of the \textsc{BayesCal} model in future work. 

For this fixed FoV, at a fixed LST, $t$, we define $\mathbfss{F}_{\mathrm{fr}, t}$ as the non-uniform discrete Fourier transform, to the $uvw$-domain coordinates sampled by our array, of our image-space flux-density distribution defined at a discrete set of $lmn$ coordinates that describe either the pixel centers of our image cube, for our diffuse emission simulation, or the coordinates of the catalogued point sources, for our EGS simulation. As such, $\mathbfss{F}_{\mathrm{fr}, t}$ has elements,
\begin{equation}
\label{Eq:PrimaryBeam}
F_{\mathrm{fr},t, ij} = \gamma \exp(-2\pi i (u_{i}l_{j} + v_{i}m_{j} + w_{i}n_{j})) \ ,
\end{equation}
where $\gamma$ is a unit conversion factor which, \begin{enumerate*}\item for our image domain brightness temperature diffuse emission simulation (see \autoref{Sec:SkyBrightnessTemperatureDistribution}) in units of $\mathrm{K}$, takes the form $\gamma=2\times10^{26}\nu^2k_\mathrm{B}\Delta\Omega/c^2$, with $\Delta\Omega$ the image domain pixel area and, \item for our EGS simulations (see \autoref{Sec:PointSourceEmission}) with source flux-densities in $\mathrm{Jy}$ takes a value of $\gamma = 1$ \end{enumerate*}. In both cases, we derive visibilities in units of $\mathrm{Jy}$.

To derive $lmn$ coordinates at a given LST, we follow the approach in \textsc{healvis} (\citealt{2019ascl.soft07002L, 2019MNRAS.487.5840L}). We begin by defining a $40\fdg0$ FoV about the (RA,Dec)-coordinates of the LST-dependent zenith pointing at the longitude and latitude of the simulated observatory. We derive the $lmn$ coordinates as, 
\begin{align}
\label{Eq:lmncoords}
l &= \sin(\phi) \sin(\theta) \\ \nonumber
m &= \cos(\phi) \sin(\theta) \\ \nonumber
n &= \cos(\theta) \ ,
\end{align} 
where $\theta$ and $\phi$ are the zenith angle and azimuth coordinates of the emission sources (point source coordinates or \textsc{HEALPix} pixel centers in the EGS and diffuse emission components of our simulations, respectively) in our selected FoV. For each emission source, we calculate the zenith angle and azimuth coordinates as,
\begin{align}
\label{Eq:zaazcoords}
\theta &= \arccos(\hat{\mathbfit{s}} \cdot \hat{\bm{e}}_\mathrm{u}) \\ \nonumber
\phi &= \arctan(s_\mathrm{e}, s_\mathrm{n}) \ .
\end{align}
Here, $\hat{\mathbfit{s}} = [s_\mathrm{n}, s_\mathrm{e}, s_\mathrm{u}]^{T}$, $\hat{\bm{e}}_\mathrm{n}$, $\hat{\bm{e}}_\mathrm{e}$ and $\hat{\bm{e}}_\mathrm{u}$ are unit vectors pointing from the antenna to the (RA, Dec) coordinates of the emission source and north, east, and up relative to the antenna in the local tangent plane coordinate system, respectively. 

The frequency dependent $uvw$ coordinates of the visibilities sampled by the $i$th baseline in the array are given by $\mathbfit{b}_{i}/\lambda$, where $\mathbfit{b}_{i}$ is the $i$th baseline vector and $\lambda$ is the central wavelength of the radiation measured by a given channel. When constructing our simulated observations to be calibrated in \autoref{Sec:Results}, we consider two cases:
\begin{enumerate}
 \item \textit{The $\mathit{14.5-290~\mathrm{m}}$ ($\mathit{7.73-154.66~\lambda}$ at $\mathit{160~\mathrm{MHz}}$) 'long baseline' data set --} which includes visibilities calculated on the full range of baseline lengths in our simulated array and results in 630 uniquely sampled $uvw$-coordinates per channel.
 \item \textit{The $\mathit{14.5-29~\mathrm{m}}$ ($\mathit{7.73-15.47~\lambda}$ at $\mathit{160~\mathrm{MHz}}$) 'short baseline' data set --} which includes visibilities calculated on the three shortest baseline lengths in our simulated array and results in 9 uniquely sampled $uvw$-coordinates per channel. 
\end{enumerate}

There are several reasons why calibrating the short baseline subset of the visibilities is of interest. For the array layout considered here, these baselines have the highest redundancy and highest power spectrum sensitivities. This makes them a promising subset for 21 cm analyses that seek to constrain the redshifted 21 cm power spectrum on a per-baseline basis (e.g. \citealt{2012ApJ...756..165P}). If one intends to use such a baseline subset for science analysis, the simplest approach to ensuring internal consistency of the calibration and power spectrum analysis is to derive calibration solutions for the data one intends to use from the corresponding raw data\footnote{A data set that is calibrated using only a subset of the data is subject to 'leverage' which manifests itself as excess noise in the component of the data omitted during calibration. See \citet{2016MNRAS.463.4317P, 2017ApJ...838...65P} for a description of leverage in interferometric calibration.} (rather than from a superset or alternate baseline range subset of the data from the array).

Additionally, short baselines have intrinsically lower chromaticity, meaning that, for a fixed calibration model completeness as a function of baseline length, more accurate calibration solutions will be derived by up-weighting the calibration constraints on short baselines (e.g. \citealt{2017MNRAS.470.1849E, 2019MNRAS.487..537O}).

Finally, in Bayesian approaches to recovering the three dimensional power spectrum of the redshifted 21 cm emission from interferometric data by jointly fitting for the 21 cm signal and foregrounds (see e.g. \citealt{2019MNRAS.484.4152S}), the computational efficiency of calculating the power spectrum posterior scales with area of the model $uv$-plane cubed. Thus, here, analysing the high sensitivity short baseline subset of visibilities in a HERA-like array configuration also represents a good compromise between data sensitivity and computation time.

Despite these considerations, we retain the long baseline data set as a reference and additional point of comparison when calibrating simulated observed data with either standard absolute calibration or absolute calibration with gain priors, since the computation time required to calibrate the long baseline data sets with these approaches scales simply with the data volume\footnote{In contrast to using standard absolute calibration or absolute calibration with gain priors, the long baseline data set additionally necessitates a higher resolution fitted image-domain sky model when using \textsc{BayesCal}, increasing the inversion time of $\bm{\upUpsilon}$ in proportion to $N_\mathrm{pix, s}^{3}$ (see \autoref{Sec:BayesCal}).}. 

In both cases, we assume that the antennas lie in a plane at fixed height, such that $w=0$, and we average over the visibilites in redundant baseline groups for the purposes of absolute calibration in \autoref{Sec:Results}.

\subsection{Visibility noise}
\label{Sec:VisibilityNoise}

\begin{table}
\caption{Instrumental and observational parameters.}
\centerline{
\begin{tabular}{l l l}
\toprule
Parameter                & Description                   & Value  \\
\midrule
$B$               & Central frequency range             & $160-169~\mathrm{MHz}$    \\
$N_{\nu}$               & Number of channels             & 10    \\
$\Delta\nu$              & Channel width                 & $1~\mathrm{MHz}$     \\
$\tau$                   & Integration time              & $30~\mathrm{s}$  \\
$\eta_{s}$               & System efficiency             & 1    \\
$\eta_{a}$               & Antenna efficiency            & 1    \\
$A$                      & Antenna area                  & $150~\mathrm{m^2}$\\
\bottomrule
\end{tabular}
}
\label{Tab:NoiseParameters}
\end{table}

The noise level on a visibility resulting from a pair of identical antennas individually experiencing equal system noise is given by \autoref{Eq:VisabilityNoise}. We list the values of the parameters used in our instrument simulation in \autoref{Tab:NoiseParameters}. For these parameter values and assuming a constant system noise temperature of $T_{\mathrm{sys}}=200~\mathrm{K}$ across our $9~\mathrm{MHz}$ bandwidth, \autoref{Eq:VisabilityNoise} yields a visibility noise $\sigma_{V}=0.5~\mathrm{Jy}$ per integration prior to redundant baseline averaging, which we add independently to the real and imaginary components of each of the sampled visibilities.

\subsection{Primary beam model}
\label{Sec:PrimaryBeamModel}

The linearly polarised primary beam response of the baseline between two antennas $p$ and $q$ to polarised sky emission parametrised in terms of the Stokes coherency vector is given by (see paper I for details),
\begin{align}
\label{Eq:PolarisedPrimaryBeamMatrix}
\mathbfss{P}_{pq} = \mathbfss{E}_{pq}\mathbfss{T}^\mathrm{S}
\ .
\end{align}
Here,
\begin{align}
\label{Eq:Epq}
\mathbfss{E}_{pq} = \mathbfss{E}_{p}  \otimes \mathbfss{E}_{q}^{*}
\ ,
\end{align}
is a $4 \times 4$ matrix encoding the polarisation-, frequency- and direction-dependent primary beam response of the baseline to the autocorrelation of the vector electric field associated with the sky brightness distribution; $\mathbfss{E}_{p}$ and $\mathbfss{E}_{q}$ are the voltage beams of antennas $p$ and $q$, respectively; $\otimes$ denotes a Kronecker product and the Stokes transformation matrix,
\begin{equation}
\label{Eq:TS}
\mathbfss{T}^\mathrm{S} =  
\begin{pmatrix}
1 & 1 & 0 & 0\\ 
0 & 0 & 1 & i\\ 
0 & 0 & 1 & i\\ 
1 & -1 & 0 & 0 
\end{pmatrix}
\end{equation}
encodes the mapping between the interferometric coherency vector for a baseline comprised of linearly polarised antennas,
\begin{align}
\label{Eq:CoherencyVector}
\mathbfit{C} &= \begin{pmatrix}
I + Q  \\ \nonumber
U + iV  \\ \nonumber
U - iV  \\ \nonumber
I - Q
\end{pmatrix} \ ,
\end{align}
and the Stokes coherency vector,
\begin{equation}
\label{Eq:StokesCoherencyVector}
\mathbfit{C}^\mathrm{S} 
= \begin{pmatrix}
I  \\ \nonumber
Q  \\ \nonumber
U  \\ \nonumber
V
\end{pmatrix} \ ,
\end{equation}
such that $\mathbfit{C} = \mathbfss{T}^\mathrm{S}\mathbfss{C}^\mathrm{S}$.

In general, $\mathbfss{E}_{p}$ can be derived from electromagnetic simulations of the response of the antenna in an array, or measurements of the antenna response via holographic beam forming can be used to estimate the primary beam response associated with the baselines of an interferometric array.

However, for simplicity, in this work, we consider an analytic, zenith pointing, polarised Airy beam primary beam response of an array comprised of identical linearly polarised antennas with west-east and south-north oriented feeds with no polarisation leakage\footnote{Here, we mean polarisation leakage to refer to voltage excited in nominally south-north and west-east oriented feeds by the east and north components of the vector electric field at the antenna (written in an ENU basis) $E_\mathrm{e}$ and $E_\mathrm{n}$, respectively, as well as any coupling between the voltages along the signal chains between the feeds and the correlator. In contrast, we do account for instrumentally induced polarisation due to the different voltage beams of the antenna (see \autoref{Eq:VoltageBeamEquatorial}). This polarisation manifests, for example, as a non-zero contribution of linearly polarised sky emission to pseudo-stokes $I$ visibilities and, correspondingly, to an image reconstructed from these visibilities.}. In an ($\hat{\mathbfit{e}}_\mathrm{n}$, $\hat{\mathbfit{e}}_\mathrm{e}$) two-component vector basis of an east-north-up (ENU) coordinate system, this corresponds to a polarised voltage pattern of the form,
\begin{equation}
\label{Eq:VoltageBeamENU}
\mathbfss{E}(\theta_{j}, \phi_{j}) = \dfrac{2J_{1}(ka\sin(\theta_{j}))}{ka\sin(\theta_{j})} 
\begin{pmatrix}
1 & 0 \\ 
0 & 1
\end{pmatrix}
\ .
\end{equation}
Here, $J_{1}$ is the Bessel function of the first kind of order one, $\theta_{j}$ and $\phi_{j}$ are the zenith angle and azimuth coordinates of the discretised emission element $j$ (i.e. the sources in our EGS simulation and \textsc{HEALPix} pixels in our diffuse emission simulations), $k=2\pi/\lambda$ is the wavenumber, with $\lambda$ the wavelength of the emission, and $a$ the aperture radius. In the simulations in this paper, we use an aperture radius $a = 7.25~\mathrm{m}$, such that \autoref{Eq:VoltageBeamENU} is a reasonable first-order approximation to the voltage beam of a 14.5 m diameter fixed dish antenna used by HERA.  

While a vector basis tied to the orientations of the antenna feeds provides the simplest vector space for defining the polarised voltage beam, a sky-centric basis tied to the celestial sphere provides a more natural vector space in which to define the polarised sky models from which $\mathbfit{V}^\mathrm{true}$, $\mathbfit{V}^\mathrm{sim}$ and $\mathbfit{V}^\mathrm{fit}$ are derived. In this work, we use an equatorial basis for this purpose. To evaluate \autoref{Eq:MEqSingePolExpanded} one must transition the beam and the sky to a mutual vector basis. To this end, we re-write \autoref{Eq:VoltageBeamENU} with respect to an equatorial basis as, 
\begin{equation}
\label{Eq:VoltageBeamEquatorial}
\mathbfss{E}(\theta_{j}, \phi_{j}) = \dfrac{2J_{1}(ka\sin(\theta_{j}))}{ka\sin(\theta_{j})}  \mathbfss{T}_\mathrm{ENU-equatorial}
\ ,
\end{equation}
where $\mathbfss{T}_\mathrm{ENU-equatorial}$ is the transition matrix between 
the ($\hat{\mathbfit{e}}_\mathrm{n}$, $\hat{\mathbfit{e}}_\mathrm{e}$) and ($\hat{\mathbfit{e}}_\mathrm{Dec}$, $\hat{\mathbfit{e}}_\mathrm{RA}$) vector bases,
\begin{equation}
\label{Eq:TransitionMatrixZaazEquatorialBasisVectors}
\mathbfss{T}_\mathrm{ENU-equatorial} = 
\begin{pmatrix}
\hat{\mathbfit{e}}_\mathrm{n} \cdot \hat{\mathbfit{e}}_\mathrm{Dec}  & \hat{\mathbfit{e}}_\mathrm{n} \cdot \hat{\mathbfit{e}}_\mathrm{RA} \\
\hat{\mathbfit{e}}_\mathrm{e} \cdot \hat{\mathbfit{e}}_\mathrm{Dec}  & \hat{\mathbfit{e}}_\mathrm{e} \cdot \hat{\mathbfit{e}}_\mathrm{RA} 
\end{pmatrix}
\ .
\end{equation}
The analytic form that \autoref{Eq:TransitionMatrixZaazEquatorialBasisVectors} takes, as a function of antenna location, hour and parallactic angle, is derived in \ref{Sec:TENUequatorial}.

Using \autoref{Eq:PolarisedPrimaryBeamMatrix}, our polarised primary beam matrix takes the form,
\begin{equation}
\label{Eq:PolarisedPrimaryBeamEquatorial}
\mathbfss{P}(\theta_{j}, \phi_{j}) = \left[\dfrac{2J_{1}(ka\sin(\theta_{j}))}{ka\sin(\theta_{j})}\right]^{2}  \bar{\mathbfss{T}}
\ .
\end{equation}
Here, $\bar{\mathbfss{T}} = (\mathbfss{T}_\mathrm{ENU-equatorial} \otimes \mathbfss{T}_\mathrm{ENU-equatorial}^{*})\mathbfss{T}^\mathrm{S}$, with $\mathbfss{T}^\mathrm{S}$ the Stokes transformation matrix given in \autoref{Eq:TS}, and $\mathbfss{P}(\theta_{j}, \phi_{j})$ is a $4 \times 4$ matrix with elements,
\begin{align}
\label{Eq:PolarisedPrimaryBeamMatrixElements}
\mathbfss{P} =  
\begin{pmatrix}
P_{\mathrm{nn},I} & P_{\mathrm{nn},Q} & P_{\mathrm{nn},U} & P_{\mathrm{nn},V}\\ 
P_{\mathrm{ne},I} & P_{\mathrm{ne},Q} & P_{\mathrm{ne},U} & P_{\mathrm{ne},V}\\ 
P_{\mathrm{en},I} & P_{\mathrm{en},Q} & P_{\mathrm{en},U} & P_{\mathrm{en},V}\\ 
P_{\mathrm{ee},I} & P_{\mathrm{ee},Q} & P_{\mathrm{ee},U} & P_{\mathrm{ee},V} 
\end{pmatrix}
\ .
\end{align}

\autoref{Fig:diffuse_emission_model_and_beam_with_FoV}, right, shows the projections on the sky of the $\mathbfss{P}_{\mathrm{nn},I}$, $\mathbfss{P}_{\mathrm{nn},Q}$ and $\mathbfss{P}_{\mathrm{nn},U}$ elements of the polarised primary beam matrix, relevant to the $V^\mathrm{nn}$ visibilities calibrated in \autoref{Sec:Results}, calculated from \autoref{Eq:PolarisedPrimaryBeamEquatorial}, at the longitude and latitude of our simulated array and LST of the start of our simulated observed data set.

\subsection{Degenerate gain parameters}
\label{Sec:DegenerateGainParameters}

In addition to the array layout, instrument noise and primary beam models described in Sections \ref{Sec:InterferometricArray} -- \ref{Sec:PrimaryBeamModel}, another component of the instrument model that must be defined is the antenna gains. As described in \autoref{Sec:RedundantCalibration}, for a redundant array, the instrument gains can be divided into redundant gains, solved for with relative calibration, and a smaller set of degenerate gain parameters that must be solved for with reference to a sky model in order to absolutely calibrate the visibilities. When calibrating the simulated observed data sets in \autoref{Sec:Results}, we will assume that the maximum likelihood redundant gain parameters have been determined, leaving only the degenerate gain parameters to be solved for. Correspondingly, when performing absolute calibration, we set the relative gain solutions in the calibration model equal to the expectation value of the solutions that would be obtained for these parameters with relative calibration (matching their values in the simulated data) and solve for the remaining gain parameters corresponding to the degenerate gain degrees of freedom that cannot be solved for with baseline-redundancy-based constraints.

For accurate relative calibration, as assumed here, the absolute calibration solutions are independent of the specific choice of relative calibration parameters. Therefore, for simplicity we set them to unity in both the simulated data and the relative calibration component of the calibration model. The degenerate gain amplitude parameters used to construct the simulated observed visibilites that are calibrated in \autoref{Sec:Results} are described in \autoref{Sec:AbscalWithGainPriors} and shown in \autoref{Fig:gain_amplitude_prior}, right, and we define the true degenerate tip-tilt phase parameters used to construct the simulated observed visibilites to be zero in all channels of the data set. Accurate calibration in \autoref{Sec:Results} corresponds to accurate recovery of these degenerate gain parameters by the calibration algorithm.

\section{Sky Model}
\label{Sec:SkyModel}

In this section, we define the diffuse and point source components of the full sky model used to construct $\mathbfit{V}^\mathrm{obs}$ and the incomplete versions of that sky model (assuming uncertainty on the diffuse emission model and a flux density limited catalogue comprising a subset of the full point source model) used to construct $\mathbfit{V}^\mathrm{sim}$. We then apply the instrument models defined in \autoref{Sec:InstrumentModel} to those sky models to generate $\mathbfit{V}^\mathrm{obs}$ and $\mathbfit{V}^\mathrm{sim}$.

\subsection{Sky brightness temperature distribution}
\label{Sec:SkyBrightnessTemperatureDistribution}

At the radio frequencies relevant to 21 cm cosmology observations ($\sim10^2$--$10^3~\mathrm{MHz}$), the time and frequency dependent polarised sky brightness temperature distribution observed by the interferometer, $\mathbfit{C}$, is dominated by Stokes $I$, which, in turn, is comprised of  three major emission components: \begin{enumerate*}\item GDSE radiated through the acceleration of high energy cosmic-ray electrons in the magnetic field of the Galaxy, \item synchrotron emission from extragalactic sources (EGS), and \item free-free emission from within the Galaxy.
\end{enumerate*}

Of these components, the power in GDSE and free--free emission as a function of angular scale on the sky is reasonably approximated by inverse power laws. Since the angular scale probed by a baseline is inversely proportional to its length, GDSE and free-free emission contribute most significantly to the visibilities on short baselines. Depending on the field observed, GDSE can be the dominant astrophyical contributor to the measured visibilities on baseline lengths up to several tens of meters at $\sim100~\mathrm{MHz}$ (see e.g. \citealt{2016MNRAS.462.3069S}). The contribution from free-free emission is subdominant to GDSE on all angular scales in the relevant frequency range and is expected to account for approximately $1\%$ of the sky temperature at $150~\mathrm{MHz}$ (e.g. \citealt{1999A&A...345..380S}). The EGS two-dimensional spatial power spectrum is approximately flat, subject to perturbations due to source clustering, and is the dominant contributor to the measured visibilities on baseline lengths $\gtrsim10\mathrm{s}$ of meters, at $\sim10^2~\mathrm{MHz}$, in regions excluding the brightest point sources on the sky.

Total intensity, Stokes $I$, emission makes up the dominant fraction of the sky temperature from these emission sources; however, polarised emission arises generically in processes involving magnetised plasmas, and the linear  polarisation fraction of diffuse emission has been measured at levels between $\sim1-5\%$ at $154~\mathrm{MHz}$ (\citealt{2016ApJ...830...38L}) and $\sim15-20\%$ at $408~\mathrm{MHz}$ (\citealt{1965AuJPh..18..635M}).   

To construct $\mathbfit{V}^\mathrm{obs}$, we separate the emission components described above into two categories based on whether we parametrise our simulation of their contribution to $\mathbfit{V}^\mathrm{obs}$ in terms of a brightness distribution on a regular model grid or in terms of a set of source flux-densities defined at coordinates given by source surveys. In the first category, we include diffuse emission comprised of GDSE, diffuse free-free emission, and a diffuse sea of unresolved point sources below the completeness limit of cataloged sources in the region of sky under observation; in the second, we include emission from catalogued point sources.

\subsection{Diffuse emission}
\label{Sec:DiffuseEmission}

\subsubsection{Stokes $I$} 
\label{Sec:StokeI}

We define our diffuse emission sky on a \textsc{HEALPix} grid (\citealt{2005ApJ...622..759G}). For our assumed a priori known component of the  Stokes $I$ diffuse emission, we use the 2016 version of the Global Sky Model (GSM; \citealt{2017MNRAS.464.3486Z}), parametrised in terms of sky brightness temperature, $T$, as implemented in the \textsc{PyGSM} package (\citealt{2016ascl.soft03013D}). Here, $T$ is related to the spectral brightness distribution via the Rayleigh--Jeans law, 
\begin{equation}
\label{Eq:SynchEmissivity}
T(\nu) = \dfrac{c^2 B}{2\nu^2k_\mathrm{B}} \ ,
\end{equation}
with $c$ and $k_\mathrm{B}$ the speed of light and Boltzmann constant, respectively.

When constructing our simulations, we use a spatial resolution parametrised by the \textsc{HEALPix} $N_\mathrm{Nside}$ parameter, which is greater than or equal to double the spatial scale probed by the longest baselines in the data set to be calibrated\footnote{This requires $N_\mathrm{Nside} \ge 32$ for the $14.5-29~\mathrm{m}$ data set and $N_\mathrm{Nside} \ge 256$ for the $14.5-290~\mathrm{m}$ data set.}, and consider a 10 channel data set with $1~\mathrm{MHz}$ channel widths and central frequencies spanning the range $160 \le \nu \le 169~\mathrm{MHz}$. This results in an image cube, $T_\mathrm{GSM}$, comprised of $N_\nu$ full-sky GSM maps evaluated at the central frequencies of our simulated channels, each containing $N_\mathrm{pix} = 12N_\mathrm{Nside}^{2}$ pixels.

In order to construct a complete Stokes $I$ simulation, in addition to estimating the Stokes $I$ total intensity from the GSM model, one must account for the expected discrepancy between the GSM model for Stokes $I$ emission and the true brightness distribution of Stokes $I$ emission on the sky associated with the non-zero uncertainty level on the GSM model. To account for this, we construct our Stokes $I$ sky simulation as the sum of $T_\mathrm{GSM}$ and an additional image cube, $T_{\delta I}$, where $T_{\delta I}$ is a model for the difference between the GSM model for Stokes $I$ emission and the true brightness distribution of Stokes $I$ emission that will be present in observational data. To construct $T_{\delta I}$, we assume that its intensity and power law index can be modelled by Gaussian random fields (GRFs). We begin by deriving a sky brightness temperature realisation for the central frequency of the cube, $T_{\delta I}(\nu_\mathrm{c})$, using the \textsc{synfast} method of the \textsc{healpy}\footnote{https://healpy.readthedocs.io/en/latest/index.html} package (\citealt{2019JOSS....4.1298Z}), as a GRF realisation with an angular power spectrum proportional to that of $T_\mathrm{GSM}(\nu_\mathrm{c})$, as estimated with \textsc{healpy} \textsc{anafast}. We scale the absolute power in $T_{\delta I}(\nu_\mathrm{c})$ such that RMS($T_{\delta I}(\nu_\mathrm{c})$) = $f*$RMS($T_\mathrm{GSM}(\nu_\mathrm{c})$). Here, $f$ is the estimate of the average fractional error in $T_{\mathrm{GSM}}$, at frequency  $\nu_\mathrm{c}$, associated with using the imperfect total intensity GSM map as a model of the Stokes $I$ brightness temperature distribution. Given the uncertainty in the true value of $f$, we consider two values:
\begin{enumerate} 
\item $f=0.5$, which we consider to be a conservative upper limit on the fractional error in approximating the true Stokes $I$ brightness temperature distribution of diffuse emission with the imperfect Stokes $I$ GSM sky model in a relatively cold region of sky, lying out of the Galactic plane, in the 160--169 MHz spectral range of our simulated spectral band. We use this value when constructing the $T_{\delta I}(\nu_\mathrm{c})$ contribution to the simulated true visibilities, for our high and moderate model incompleteness calibration scenarios (see \autoref{Sec:Results} for more details). 
\item $f=0.25$, which is a more modest estimate\footnote{Work on an improved version of the GSM that will provide an estimate of the uncertainty on the model at a given interpolation frequency indicates that uncertainty on the improved GSM map is $\sim10\%$ for emission in the frequency range considered here (Adrian Liu, private communication). Assuming comparable errors on the \citet{2017MNRAS.464.3486Z} GSM map, the uncertainties on the Stokes $I$ diffuse model considered here are conservative.} of the fractional error in approximating the true Stokes $I$ intensity distribution with the imperfect Stokes $I$ GSM sky model. We use this value when constructing the $T_{\delta I}(\nu_\mathrm{c})$ contribution to our simulated true visibilities in our low model incompleteness calibration scenario. \end{enumerate}
 
An interferometer is insensitive to the frequency dependent sky monopole, therefore interferometric data calibration is correspondingly insensitive to the  mean of $T_{\delta I}(\nu_\mathrm{c})$, except insofar as it influences the sky noise. Here, we set the mean of the GRF to zero. The foregrounds are detectable at a high signal-to-noise level in the instrumental array and observing setups that we will consider in this paper, and, as such, we expect the impact of this approximation to be small. However, in an observing regime where this was not the case, the approach taken here can be adapted to account for the impact of the mean sky temperature of the a priori unknown component of the emission on the noise level in the data (as well as other sources of uncertainty on the noise level) by jointly estimating the noise in the data with the other parameters of the analysis (see \citealt{2019MNRAS.484.4152S} and \citealt{2019MNRAS.488.2904S}) for applications of this approach in the context of power spectral estimation using interferometric data).  

We model the spectral structure of $T_{\delta I}$ with a spatially dependent Gaussian spectral index distribution, $\beta(l,m,n)$, with mean and standard deviation consistent with measurements of the spectral index distribution of Galactic emission between 90 and 190 MHz with EDGES (\citealt{2017MNRAS.464.4995M}): $\langle \beta \rangle_\mathrm{GDSE} = -2.6$ and $\sigma_{\beta_\mathrm{GDSE}} = 0.02$, respectively. Incorporating this spectral dependence, we construct our full brightness temperature image cube for this emission component as,
\begin{equation}
\label{Eq:StokesIerror}
T_{\delta I}(l,m,n,\nu) = T_{\delta I}(l,m,n,\nu_\mathrm{c})\left(\frac{\nu}{\nu_\mathrm{c}}\right)^{\beta(l,m,n)} \ .
\end{equation}

\subsubsection{Stokes $Q$, $U$ and $V$} 
\label{Sec:StokeQUV}

We follow an equivalent procedure to that described in \autoref{Sec:StokeI} to construct our assumed a priori known components and complete Stokes $Q$ and $U$ diffuse emission skies. However, in place of the GSM model, we construct simulated sky brightness temperature maps for both our assumed a priori known and incomplete components of the Stokes $Q$ and $U$ diffuse emission. Specifically, we assume that the intensity of Stokes $Q$ and $U$ diffuse emission can be spatially modelled as a Gaussian random field (GRF). We construct GRF realisations of the emission intensity fields, $T_{Q}(l,m,n,\nu_\mathrm{c})$ and $T_{U}(l,m,n,\nu_\mathrm{c})$, respectively, with equal angular power spectra; in both cases, they are proportional to the angular power spectrum of our Stokes I model, scaled such that the linear polarisation fraction  of the known component of the emission, $I_\mathrm{L}/I$ (with $I_\mathrm{L}^{2} = Q^{2} + U^{2}$), is approximately $3\%$, consistent with measurements in the frequency range of our simulations by \citet{2016ApJ...830...38L}. The central channels of the Stokes $I$, $Q$, and $U$ maps, $T_\mathrm{GSM}(l,m,n,\nu_\mathrm{c})$, $T_{Q}(l,m,n,\nu_\mathrm{c})$, $T_{U}(l,m,n,\nu_\mathrm{c})$, respectively, used for constructing the full array simulated visibilities are shown in \autoref{Fig:diffuse_emission_model_and_beam_with_FoV}, left. We construct full Stokes $Q$ and $U$ image cubes, $T_{Q}(l,m,n,\nu)$ and $T_{U}(l,m,n,\nu)$, respectively, following an analogous spectral extrapolation to that described by \autoref{Eq:StokesIerror}, now applied to $T_{Q}(l,m,n,\nu_\mathrm{c})$ and $T_{U}(l,m,n,\nu_\mathrm{c})$. Finally, we assume negligible astrophysical emission in Stokes $V$ ($V/I \ll 1$).

To construct $T_{\delta Q}$ and  $T_{\delta U}$ we use the same approach as that described for deriving $T_{\delta I}$ in \autoref{Sec:StokeI} and, for simplicity, assume the same fractional uncertainties, in a given calibration model completeness scenario, for all three Stokes parameters.

\subsubsection{Complete simulated observed data and incomplete simulated calibration model} 
\label{Sec:CompleteVtrueAndIncompleteVsim}

From our diffuse emission cubes, we construct simulated true visibilities corresponding to a single time integration as,
\begin{multline}
\label{Eq:MEqSingePolShorthandExpandedGSMPlusDeltaGSM}
\mathbfit{V}^\mathrm{true}_{\mathrm{diffuse},t_{i}} = \mathbfss{F}_{\mathrm{fr}, t_{i}}[\mathbfss{P}_{\mathrm{nn},I, t_i}(\mathbfit{T}_{\mathrm{GSM},t_{i}}+\mathbfit{T}_{\delta I,t_{i}})\\ + \mathbfss{P}_{\mathrm{nn},Q, t_i}(\mathbfit{T}_{Q,t_{i}}+\mathbfit{T}_{\delta Q,t_{i}}) + \mathbfss{P}_{\mathrm{nn},U, t_i}(\mathbfit{T}_{U,t_{i}}+\mathbfit{T}_{\delta U,t_{i}})] \ ,
\end{multline}
where, as described in \autoref{Sec:InterferometricArray}, we limit our simulation to emission within a $40\fdg0$ field encompassing the highest beam weighted fraction of the sky emission and centered on the RA and Dec of zenith at the longitude and latitude of our simulated array, at each LST of our simulated observations. $\mathbfit{T}_{\mathrm{GSM},t_{i}}$, $\mathbfit{T}_{\delta I,t_{i}}$, $\mathbfit{T}_{Q,t_{i}}$, $\mathbfit{T}_{\delta Q,t_{i}}$, $\mathbfit{T}_{U,t_{i}}$ and $\mathbfit{T}_{\delta U,t_{i}}$ are vectorisations of these subsets of our polarised image cubes, each of length $N_\mathrm{pix, s} \times N_\mathrm{\nu}$, with $N_\mathrm{pix, s}$ the number of pixels at the resolution of the simulation, per frequency channel, falling within our $40\fdg0$ subset fields.

Similarly, we construct simulated model visibilities, corresponding to a single time integration, but now exclusively from our assumed a priori known diffuse emission cubes, as,
\begin{multline}
\label{Eq:MEqSingePolShorthandExpandedGSM}
\mathbfit{V}^\mathrm{sim}_{\mathrm{diffuse},t_{i}} = \mathbfss{F}_{\mathrm{fr}, t_{i}}[\mathbfss{P}_{\mathrm{nn},I, t_i}\mathbfit{T}_{\mathrm{GSM},t_{i}} + \mathbfss{P}_{\mathrm{nn},Q, t_i}\mathbfit{T}_{Q,t_{i}} + \mathbfss{P}_{\mathrm{nn},U, t_i}\mathbfit{T}_{U,t_{i}}] \ .
\end{multline}

To derive versions of $\mathbfit{V}^\mathrm{true}_\mathrm{diffuse}$ and $\mathbfit{V}^\mathrm{sim}_\mathrm{diffuse}$ incorporating all time integrations of our observation, we simply substitute instances of the $\mathbfit{T}_{\mathrm{GSM},t_{i}}$ and $\mathbfit{T}_{\delta \mathrm{diff},t_{i}}$ image vectors for single time integrations with their concatenation over all time integrations and substitute instances of $\mathbfss{F}_{\mathrm{fr}, t_{i}}$ and $\mathbfss{P}_{t_{i}}$ with an analogous pair of block diagonal matrices defined such that each block is given by $\mathbfss{F}_{\mathrm{fr}, t_{i}}$ or $\mathbfss{P}_{t_{i}}$ at a single time integration. When implementing these matrices, as well as matrices associated with the instrument model and gain parameters, we use the \textsc{SCIPY} sparse matrix linear algebra package\footnote{https://docs.scipy.org/doc/scipy/reference/sparse.html} for all diagonal and block diagonal matrices, to minimise their memory footprints\footnote{For the baseline ranges and instrument model considered here, use of sparse matrices reduces the memory footprint of the calibration by 3--6 orders of magnitude.}.

\subsection{Point source emission}
\label{Sec:PointSourceEmission}

We use the sources catalogued in the Murchison Widefield Array GLEAM survey\footnote{https://vizier.u-strasbg.fr/viz-bin/VizieR-3?-source=VIII/100/gleamegc} (\citealt{2015PASA...32...25W, 2017MNRAS.464.1146H}) as the basis of our point source emission simulation.

When constructing our point source emission simulation, we use GLEAM integrated fluxes and their associated uncertainties over the 95--103, 147--154 and 185--193 MHz spectral ranges, which we denote here as $S_{\mathrm{int},99}$, $S_{\mathrm{int},151}$, and $S_{\mathrm{int},189}$ and $\sigma_{S_{99}}$, $\sigma_{S_{151}}$, and $\sigma_{S_{189}}$, respectively. We discard any sources that do not have positive, greater than two sigma detections\footnote{For the data calibrated in \autoref{Sec:Results}, this amounts to discarding $\sim35\%$ of the sources.} in $S_{\mathrm{int},99}$, $S_{\mathrm{int},151}$ and $S_{\mathrm{int},189}$.

We assume that the spectra of the catalogued sources can be approximated as a power law of the form: $S \propto \nu^{-\alpha}$, and we calculate the 99--189 MHz source spectral indices as,
\begin{equation} 
\label{Eq:GLEAMSI}
\alpha_{99-189} = -\log(S_{\mathrm{int},189}/S_{\mathrm{int},99})/\log(189/99) \ .
\end{equation}

We subsequently define the integrated flux densities of the point sources in our EGS simulation, over the 1 MHz spectral channels in the 160--169 MHz spectral range of our simulated spectral band, as,
\begin{equation}
\label{Eq:EGSspectra}
S(\nu) = S_{\mathrm{int},151}\left(\dfrac{\nu}{151~\mathrm{MHz}}\right)^{-\alpha_{99-189}} \ .
\end{equation}

When we construct $\mathbfit{V}^\mathrm{true}_{\mathrm{EGS},t_{i}}$, we include sources that meet the criteria described above and that fall within a $40\fdg0$ field (see \autoref{Sec:InterferometricArray}) encompassing the highest beam weighted fraction of the sky emission and centered on the RA and Dec of zenith at the longitude and latitude of our simulated array at time $t_{i}$,
\begin{equation}
\label{Eq:MEqSingePolShorthandExpandedEGS}
\mathbfit{V}^\mathrm{true}_{\mathrm{EGS},t_{i}} = \mathbfss{F}_{\mathrm{fr}, t_{i}}\mathbfss{P}_{\mathrm{nn},I, t_i}\mathbfit{S}_{\mathrm{EGS}, t_{i}} \ .
\end{equation}
Here, $\mathbfit{S}_{\mathrm{EGS}, t_{i}}$ is the vectorisation of the flux densities of all sources meeting these criteria.

In a practical calibration scenario, to maximise the completeness of the calibration model, one would incorporate all catalogued sources in $\mathbfit{V}^\mathrm{sim}_{\mathrm{EGS},t_{i}}$. In this case, the model EGS visibilities would be incomplete due to the missing contribution from sources below the completeness limit of the catalogue. For the simulations considered here, to avoid needing to use a hypothetical catalogue of sources below the completeness limit of GLEAM to construct $\mathbfit{V}^\mathrm{true}_{\mathrm{EGS},t_{i}}$ and then neglecting these additional sources when constructing an incomplete $\mathbfit{V}^\mathrm{sim}_{\mathrm{EGS},t_{i}}$, we instead take the approach used in e.g. \citet{2016MNRAS.461.3135B} and use GLEAM sources to construct $\mathbfit{V}^\mathrm{true}_{\mathrm{EGS},t_{i}}$ and a subset of GLEAM sources to construct $\mathbfit{V}^\mathrm{sim}_{\mathrm{EGS},t_{i}}$.

When constructing $\mathbfit{V}^\mathrm{sim}_{\mathrm{EGS},t_{i}}$ from a subset of the sources included in $\mathbfit{V}^\mathrm{true}_{\mathrm{EGS},t_{i}}$, in addition to the cuts associated with $\mathbfit{V}^\mathrm{true}_{\mathrm{EGS},t_{i}}$, we impose a flux density cut of the form: $S_{\mathrm{int},151} > S_\mathrm{min}$.
We consider two cases for this minimum flux density cut,
\begin{enumerate}
 \item $S_\mathrm{min} = 1~\mathrm{Jy}$, which we associate with our high foregrounds incompleteness model, and
 \item $S_\mathrm{min} = 100~\mathrm{mJy}$, which we associate with our moderate and low foregrounds incompleteness models.
\end{enumerate}
In both cases, we construct our EGS simulated visibilities as,
\begin{equation}
\label{Eq:MEqSingePolShorthandExpandedEGSSubset}
\mathbfit{V}^\mathrm{sim}_{\mathrm{EGS},t_{i}} = \mathbfss{F}_{\mathrm{fr}, t_{i}}\mathbfss{P}_{\mathrm{nn},I, t_i}\mathbfit{S}_{\mathrm{EGS, s},t_{i}} \ ,
\end{equation}
where $\mathbfit{S}_{\mathrm{EGS, s},t_{i}}$ is the vectorisation of the flux densities of the subset of sources meeting the selection criteria for case (i) or (ii), depending on the model incompleteness of the calibration scenario considered. We assume that the polarised point sources contribution to the visibilities is small relative to unpolarised sources (e.g. \citealt{2020PASA...37...29R}) and, thus, do not include equivalent polarised source terms in \autoref{Eq:MEqSingePolShorthandExpandedEGSSubset}.

We generalise \autoref{Eq:MEqSingePolShorthandExpandedEGSSubset} to multiple time integrations in an analogous manner to the multi-integration diffuse model in \autoref{Sec:DiffuseEmission}. To derive versions of $\mathbfit{V}^\mathrm{true}_\mathrm{EGS}$ and $\mathbfit{V}^\mathrm{sim}_\mathrm{EGS}$ incorporating all time integrations of our observation, we simply substitute instances of the $\mathbfit{S}_{\mathrm{EGS},t_{i}}$ and $\mathbfit{S}_{\mathrm{EGS, s},t_{i}}$ source flux density vectors for single time integrations with their concatenation over all time integrations and substitute point source model instances of $\mathbfss{F}_{\mathrm{fr}, t_{i}}$ and $\mathbfss{P}_{\mathrm{nn},I, t_i}$ with an analogous pair of block diagonal matrices defined such that each block is given by $\mathbfss{F}_{\mathrm{fr}, t_{i}}$ or $\mathbfss{P}_{\mathrm{nn},I, t_i}$ at the $i$th single time integration.

Over the observing duration of the data sets calibrated in \autoref{Sec:Results}, for the two point source model completeness cases described above, $\mathbfit{V}^\mathrm{sim}_\mathrm{EGS}$ includes approximately $940$ and $11,000$, respectively, of the $14,300$ sources that fall within the union of the $40\fdg0$ fields centered on zenith at the 10 LSTs of our 5 minute data set. This corresponds to the inclusion of 41\% and 95\% of the total flux in $\mathbfit{V}^\mathrm{true}_\mathrm{EGS}$, respectively.

\section{Results}
\label{Sec:Results}

In order to test the efficacy of the \textsc{BayesCal} calibration methodology, in this section, we demonstrate its application to simulated observed data, constructed as described in Sections \ref{Sec:InstrumentModel} and \ref{Sec:SkyModel}, and compare the inferred gains to those recovered using standard absolute gain calibration of redundantly calibrated visibilities (\autoref{Sec:StandardAbscal}) and absolute gain calibration supplemented with a prior on the level of power in spectral fluctuations in the degenerate gain amplitude (\autoref{Sec:AbscalWithGainPriors}). 

For a detailed description of the mathematical formalism underpinning the three approaches, we refer the reader to paper I, or, alternatively, for a summary, to \autoref{Sec:BayesCal} of this paper. However, in short, absolute gain calibration supplemented with a prior on the level of power in spectral fluctuations in the degenerate gain amplitude can be viewed as an extension of standard absolute calibration of redundantly calibrated visibilities that incorporates physically motivated informative priors on the spectral structure of the degenerate gain amplitude solutions, sourced from external constraints such as electric and electromagnetic co-simulations of the receiver system and reflectometry measurements of the feed-dish system that are independent of an astrophysical sky model. In both standard absolute calibration or absolute gain calibration with gain priors the calibration visibility model is impacted by incompleteness and uncertainties associated with our a priori knowledge of the brightness distribution of the sky. In contrast, \textsc{BayesCal}, which in turn can be viewed as an extension to the absolute gain calibration with gain priors framework, explicitly incorporates a fitted visibility model, $\mathbfit{V}^\mathrm{fit}$, for the component of the true observed visibilites that is unaccounted for in the simulated calibration visibility model, $\mathbfit{V}^\mathrm{sim}$, used in the first two calibration approaches. In the limit that the spectral and instrument forward model used in $\mathbfit{V}^\mathrm{sim}$ and $\mathbfit{V}^\mathrm{fit}$ is accurate, the joint use of the two models allows one to construct a complete calibration visibility model. 

For all three calibration methodologies, we demonstrate the effect on the recovered calibration solutions of three levels of incompleteness of the a priori known component of the calibration model, $\mathbfit{V}^\mathrm{sim}$, corresponding to different levels of uncertainty on the distribution of low frequency radio emission in the field to be calibrated: high, medium and low. The uncertainties of the diffuse emission models and minimum flux densities of the point sources included in these models are summarised in \autoref{Tab:CompletenessLevels}. The number of sky-model parameters, $N_\mathrm{pix, s}$, associated with $\mathbfit{V}^\mathrm{fit}$ scales quadratically with the angular resolution of the model, which, in turn, is proportional to the maximum baseline length of the data set to be calibrated. The computational complexity associated with inverting $\bm{\upUpsilon}$ scales as $N_\mathrm{pix, s}^{3}$. For this reason, for each level of simulated calibration sky-model incompleteness, we apply relative calibration plus \textsc{BayesCal} (\autoref{Sec:BayesCal}) exclusively to the $14.5-29~\mathrm{m}$ data set. In contrast, the computational complexity associated with relative plus absolute calibration, both with (\autoref{Sec:AbscalWithGainPriors}) and without (\autoref{Sec:StandardAbscal}) placing a prior on the gain solutions, scales linearly with data volume. Thus, we derive calibration solutions for both the $14.5-29~\mathrm{m}$ and $14.5-290~\mathrm{m}$ data sets with these methodologies. For all calibration types, we assume the calibration solutions are stable in time over the five minute duration of the data set (see \autoref{Sec:TemporalModel}).

For all data calibration in this section, when using either absolute calibration or \textsc{BayesCal}, we assume the interferometric array is redundant and that the maximum likelihood redundant gain parameters have been determined, leaving only the degenerate gain parameters to be solved for (see \autoref{Sec:DegenerateGainParameters}).

\begin{table*}
\caption{Simulated calibration model, $\mathbfit{V}^\mathrm{sim}$, completeness levels.}
\centerline{
\begin{tabular}{l l l}
\toprule
Model name        & Fractional error on diffuse                  & $S_\mathrm{min}$ (Jy) of sources included \\
                  & emission in $\mathbfit{V}^\mathrm{sim}$                   &  in $\mathbfit{V}^\mathrm{sim}$  \\
\midrule
low incompleteness      & 0.25    & 0.1    \\
moderate incompleteness & 0.5     & 0.1    \\
high incompleteness     & 0.5     & 1      \\
\bottomrule
\end{tabular}
}
\label{Tab:CompletenessLevels}
\end{table*}

\subsection{Comparison between \textsc{BayesCal} and absolute calibration of redundantly calibrated visibilities}
\label{Sec:ComparisonBetweenBayesCalAndAbscal}

\begin{figure*}
	\centerline{
	\includegraphics[width=0.33\textwidth]{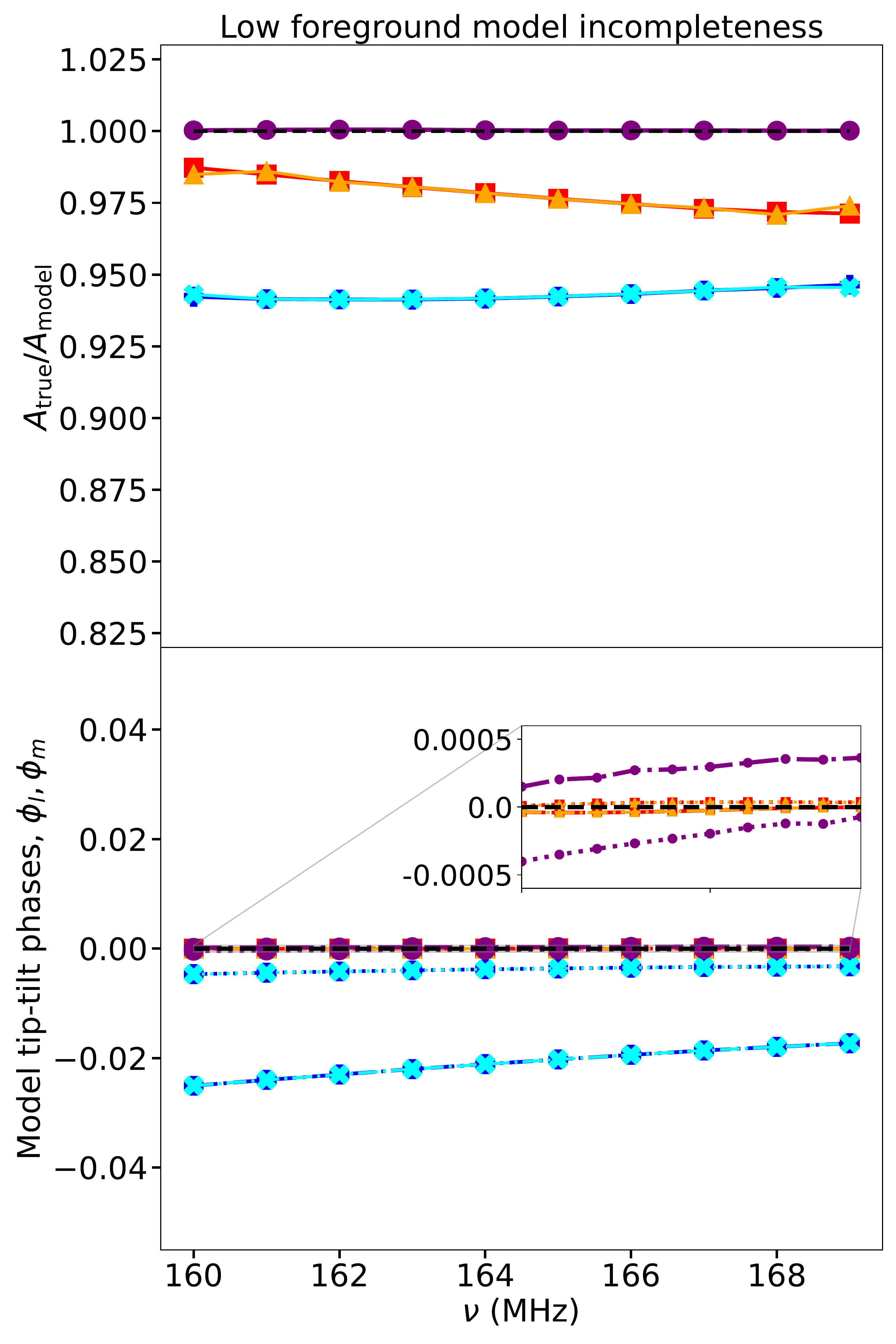}
	\includegraphics[width=0.33\textwidth]{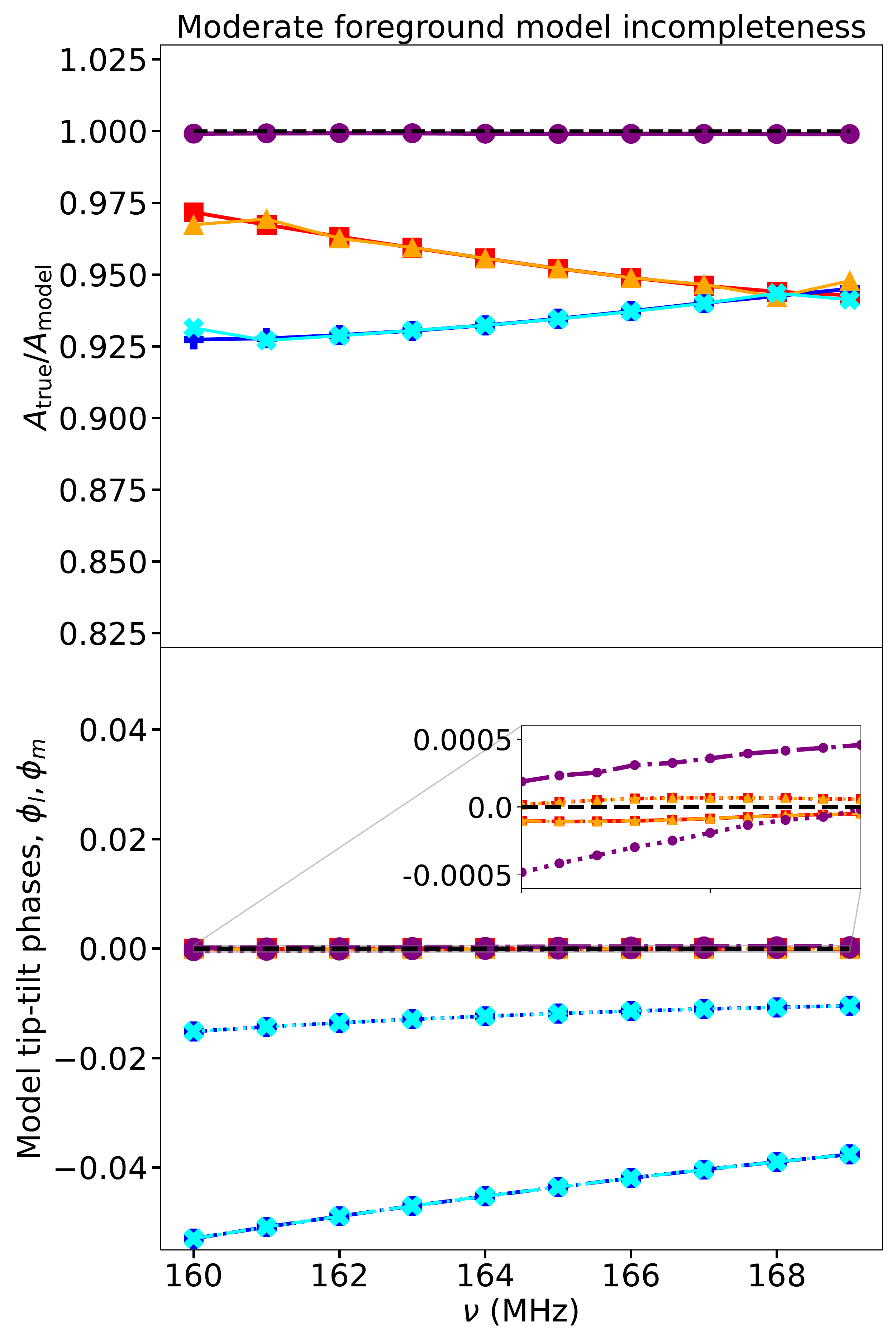}
	\includegraphics[width=0.33\textwidth]{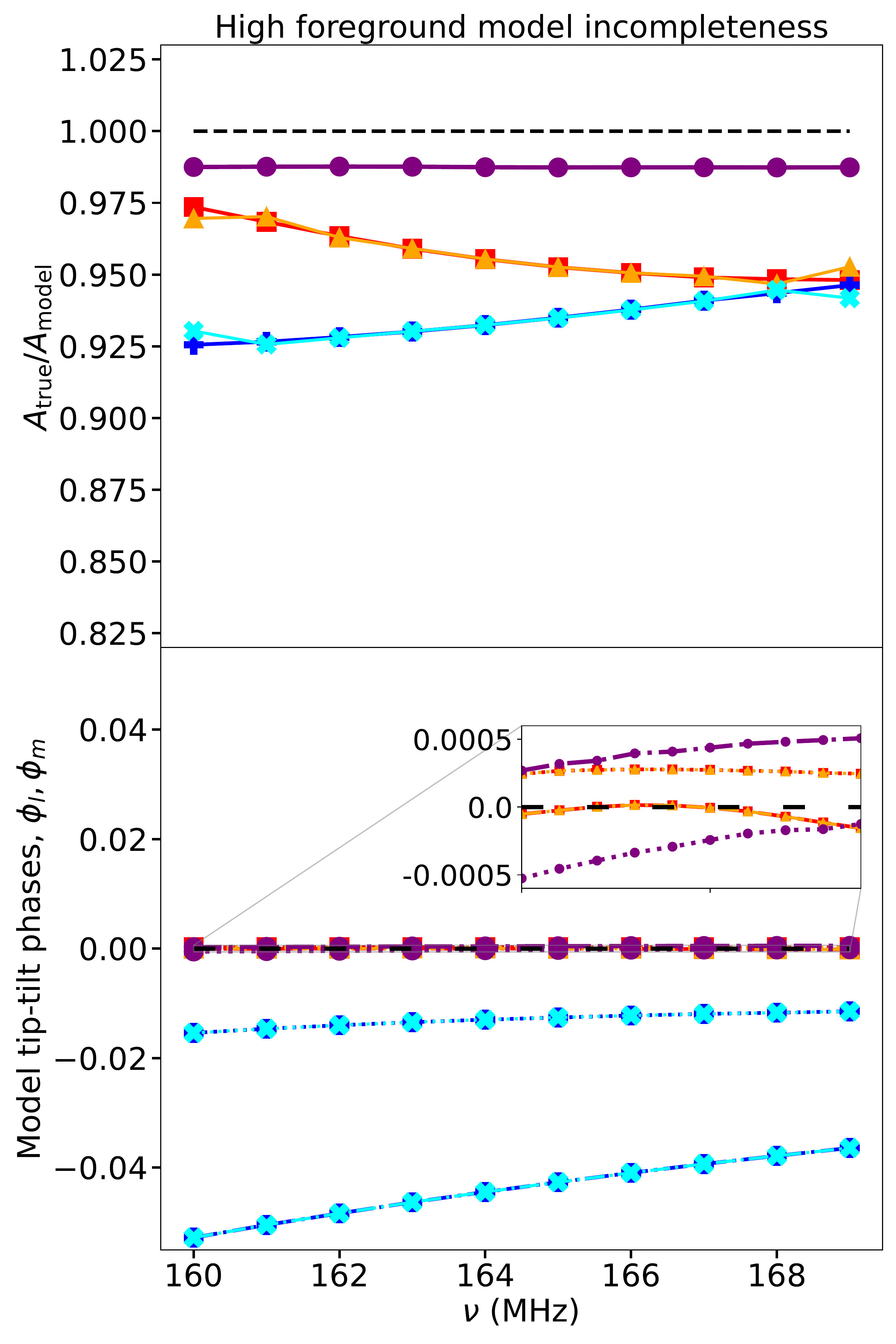}}
	\centerline{
        \includegraphics[width=0.95\textwidth]{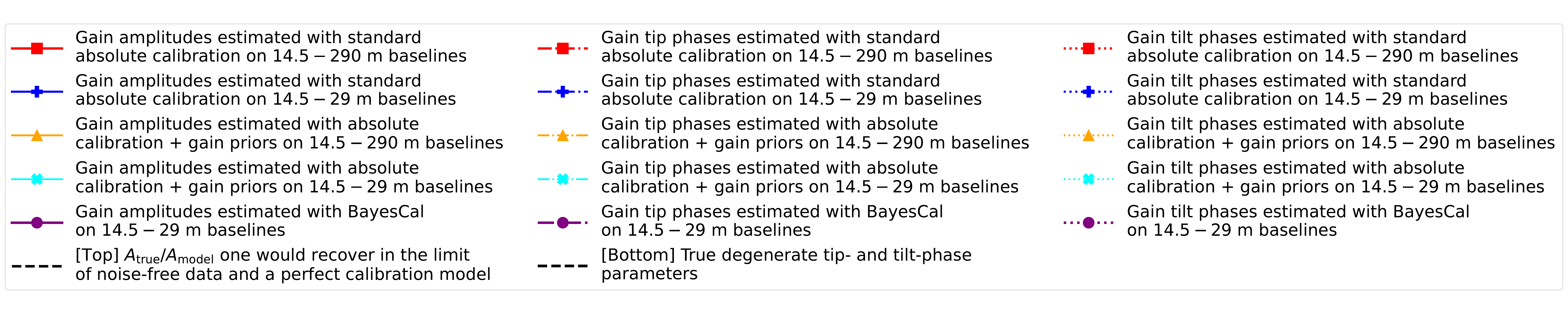}} 
\caption{Calibration solutions derived in three regimes of incompleteness of the simulated component of the calibration model (see \autoref{Tab:CompletenessLevels}): low (left), moderate (middle) and high incompleteness (right). [Top] Ratio of the input degenerate calibration amplitudes and i) the mean \textsc{BayesCal} estimates (see main text for details) of the degenerate calibration amplitudes (purple lines), ii) the absolute calibration solutions for the degenerate calibration amplitudes when calibrating data on 14.5--29 m baselines or 14.5--290 m baselines (blue and red lines, respectively), or iii) the absolute calibration with gain priors solutions for the degenerate calibration amplitudes when calibrating data on 14.5--29 m baselines or 14.5--290 m baselines (cyan and orange lines, respectively). [Bottom] The \textsc{BayesCal} solutions for the degenerate tip-tilt phases (dot-dashed and dotted purple lines, respectively) and the maximum a posteriori and maximum likelihood tip-tilt phases recovered with absolute calibration of visibilities, with and without imposition of informative gain priors, on 14.5--29 m baselines (cyan and blue dot-dashed and dotted lines, respectively) and 14.5--290 m baselines (orange and red dot-dashed and dotted red lines, respectively). The zoomed inset axes display the low-level structure in the \textsc{BayesCal} and maximum a posteriori and maximum likelihood tip-tilt phases recovered with absolute calibration of visibilities, with and without imposition of informative gain priors, on 14.5--290 m baselines. We note that the maximum a posteriori and maximum likelihood parameters recovered with absolute calibration of visibilities, with and without imposition of informative gain priors, have only minor differences in their large spectral scale power and biases and thus red and orange lines and cyan and blue lines, respectively, overlay each other to a significant degree in both the top and bottom sub-plots. The dashed black lines in the top sub-plots show the true unitary value of the ratio of the true and fitted degenerate gain amplitudes, which would be recovered by hypothetical data calibration if the data was noise-free and the calibration model was perfect. In the bottom sub-plots, the dashed black lines display the true null value of the input degenerate tip-tilt phase solutions in the simulated observed data. 
}
\label{Fig:CalibrationSolutionsComparison}
\end{figure*}

\autoref{Fig:CalibrationSolutionsComparison} shows recovered calibration solutions in the low, medium and high simulated visibility model incompleteness regimes, from left to right. In each case, five sets of calibration solutions are displayed: \textsc{BayesCal} solutions for the $14.5-29~\mathrm{m}$ data set (purple lines), absolute calibration solutions for the $14.5-29~\mathrm{m}$ and $14.5-290~\mathrm{m}$ data sets without gain priors (blue and red lines, respectively) and absolute calibration solutions for the $14.5-29~\mathrm{m}$ and $14.5-290~\mathrm{m}$ data sets with gain priors (cyan and orange lines, respectively).

The top panel of each of the sub-figures, displays:
\begin{enumerate}
\item the ratio of the input degenerate calibration amplitudes and the mean of the functional posterior probability distribution of the degenerate gain amplitudes, $\left\langle \mathrm{Pr}(\mathbfit{A}_\mathrm{true}/\mathbfit{A} \; \vert \; \mathbfit{A}_\mathrm{F}, \mathbfit{V}^\mathrm{obs}) \right\rangle$, recovered when using \textsc{BayesCal} to fit the sum of a simulated model for the known flux contributions from diffuse emission and point sources and a statistical model for the a priori unknown component of sky emission, jointly with instrumental gain parameters, to visibilities with 14.5--29 m baselines (solid purple lines), 
\item the ratio of the input degenerate calibration amplitudes and the maximum a posteriori degenerate gain amplitudes, $\mathbfit{A}_\mathrm{true}/\hat{\mathbfit{A}}_{1}$, when fitting an incomplete simulation of the known flux contributions from diffuse and point sources, to 14.5--29 m baselines or 14.5--290 m baselines, while imposing priors on the amplitude of spectral fluctuations in the gain amplitude solutions (solid cyan and orange lines, respectively), and
\item the ratio of the input degenerate calibration amplitudes and the maximum likelihood degenerate gain amplitudes, $\mathbfit{A}_\mathrm{true}/\hat{\mathbfit{A}}_{2}$, when only fitting an incomplete simulation of the known flux contributions from diffuse and point sources to 14.5--29 m baselines (solid blue lines) or 14.5--290 m baselines (solid red lines).                                                                                                                \end{enumerate}
Going forward, we use $\hat{\mathbfit{A}}$ as a stand-in variable for both the maximum a posteriori and maximum likelihood degenerate gain amplitudes, but, in each case, whether one or both variables are being referred to and, if only one, which one is being referred to will be described explicitly or will be clear from context. The dashed black lines represent the unitary value of this ratio that would be recovered by hypothetical data calibration in the limit that the signal-to-noise ratio on the data tends to infinity and the calibration model is perfect.

The bottom panel of each of the sub-figures, displays:
\begin{enumerate}
\item the mean of the functional posterior probability distribution of the degenerate gain tip-tilt phases recovered with \textsc{BayesCal}: $\left\langle \mathrm{Pr}(\bm{\mathit{\Phi}}_{l} \; \vert \; \bm{\mathit{\Phi}}_{l,\mathrm{F}}, \mathbfit{V}^\mathrm{obs}) \right\rangle$ (dot-dashed purple lines) and $\left\langle \mathrm{Pr}(\bm{\mathit{\Phi}}_{m} \; \vert \; \bm{\mathit{\Phi}}_{m,\mathrm{F}}, \mathbfit{V}^\mathrm{obs}) \right\rangle$ (dotted purple lines), respectively,
\item the maximum a posteriori degenerate tip and tilt phases, $\hat{\bm{\mathit{\Phi}}}_{l, 1}$ (dot-dashed cyan and orange lines on 14.5--29 m and 14.5--290 m baselines, respectively) and $\hat{\bm{\mathit{\Phi}}}_{m, 1}$ (dotted cyan and orange lines on 14.5--29 m and 14.5--290 m baselines, respectively), recovered with absolute calibration while imposing priors on the amplitude of spectral fluctuations in the gain amplitude solutions,
\item the maximum likelihood degenerate tip and tilt phases, $\hat{\bm{\mathit{\Phi}}}_{l, 2}$ (dot-dashed blue and red lines on 14.5--29 m and 14.5--290 m baselines, respectively) and $\hat{\bm{\mathit{\Phi}}}_{m, 2}$ (dotted blue and red lines on 14.5--29 m and 14.5--290 m baselines, respectively), recovered with absolute calibration alone.
\end{enumerate}
As with the degenerate gain amplitudes, going forward, we use $\hat{\bm{\mathit{\Phi}}}_{l}$ and $\hat{\bm{\mathit{\Phi}}}_{m}$ as stand-in variables for both the maximum a posteriori and maximum likelihood degenerate tip-tilt phases, but, in each case, whether one or both variables are being referred to and, if only one, which one is being referred to will be described explicitly or will be clear from context. The black dashed line shows the null input value of the tip-tilt phases in the simulated observed data.

When comparing the recovered calibration solutions shown in \autoref{Fig:CalibrationSolutionsComparison}, it is useful to separate errors relative to the true calibration parameters associated with the simulated observed data set into two classes: \begin{enumerate*} \item the mean bias in the recovered parameter estimates and \item the level of spurious spectral fluctuations imparted to the data by imperfect estimation of the gain parameters \end{enumerate*}. In the first case, the average bias in the redundant gain amplitude parameters acts as a multiplicative scale factor to the data. As such, it is comparatively benign in 21 cm cosmology analysis, because it does not complicate the spectral separation of the foregrounds and the 21 cm signal in the data\footnote{The average bias in the redundant gain amplitude parameters is less likely to entirely preclude recovery of the 21 cm signal relative to spurious spectral fluctuations imparted to the data by imperfect estimation of the redundant gain amplitude parameters. Nevertheless, we note that the largest average bias in the redundant gain amplitude parameters is $\sim 6\%$, which, following calibration, corresponds to misestimation of the amplitude of the power spectrum at a $\sim12\%$ level. Thus, the average bias in the redundant gain amplitude can also make a contribution to the power spectrum error budget at a level that is significant for precision cosmology applications.}. In contrast, in the second case, spurious spectral fluctuations imparted to the data by imperfect estimation of the redundant gain amplitude parameters introduce foreground systematics that will bias estimation of the 21 cm signal on the spectral scales of those fluctuations; thus, mitigating spurious spectral fluctuations in the calibrated data is of paramount importance.

\subsubsection{Bias in gain parameter estimates}
\label{Sec:BiasInGainParameterEstimates}

\begin{table*}
\caption{Average absolute bias in gain parameter estimates, expressed as fractional errors, $1 - \left\langle \abs{x_{i}/ x_{i, \mathrm{ideal}}} \right\rangle$, for the ratio of the true and recovered degenerate gain amplitude parameters and absolute errors, $\left\langle \abs{x_{i} - x_{i, \mathrm{ideal}}} \right\rangle$, for the tip and tilt gain phases. In both cases, $x_{i}$ corresponds to the $i$th parameter in the parameter column and $x_{i, \mathrm{ideal}}$ corresponds to the value of $x_{i}$ that would be obtained in the limit that one calibrated noiseless data with a perfect calibration model.}
\centerline{
\begin{tabular}{l l l l l }
\toprule
Parameter        &   Calibration method     & Baseline length range (m)   & Calibration model incompleteness & Average absolute bias \\
\midrule
$\mathbfit{A}_\mathrm{true}/\hat{\mathbfit{A}}$ & Absolute calibration &  14.5--290 & low & $2.19\%$   \\
 & & & moderate & $4.49\%$   \\
 & & & high     & $4.32\%$  \\
 & &  14.5--29 & low & $5.70\%$   \\
 & & & moderate & $6.54\%$   \\
 & & & high     & $6.54\%$  \\
$\left\langle \mathrm{Pr}(\mathbfit{A}_\mathrm{true}/\mathbfit{A} \; \vert \; \mathbfit{A}_\mathrm{F}, \mathbfit{V}^\mathrm{obs}) \right\rangle$ & \textsc{BayesCal} &  14.5--29 & low & $0.03\%$   \\
 & & & moderate & $0.09\%$   \\
 & & & high     & $1.25\%$  \\
\midrule
$\hat{\bm{\mathit{\Phi}}}_{l}$ & Absolute calibration &  14.5--290 & low & $2.4 \times 10^{-5}~\mathrm{rad}$   \\
 & & & moderate & $8.3 \times 10^{-5}~\mathrm{rad}$   \\
 & & & high &     $4.8 \times 10^{-5}~\mathrm{rad}$  \\
 & &  14.5--29 & low & $2.1 \times 10^{-2}~\mathrm{rad}$   \\
 & & & moderate & $4.5 \times 10^{-2}~\mathrm{rad}$   \\
 & & & high     & $4.4 \times 10^{-2}~\mathrm{rad}$  \\
$\left\langle \mathrm{Pr}(\bm{\mathit{\Phi}}_{l} \; \vert \; \bm{\mathit{\Phi}}_{l,\mathrm{F}}, \mathbfit{V}^\mathrm{obs}) \right\rangle$ & \textsc{BayesCal} &  14.5--29 & low & $2.8 \times 10^{-4}~\mathrm{rad}$   \\
 & & & moderate & $3.4 \times 10^{-4}~\mathrm{rad}$   \\
 & & & high     & $4.1 \times 10^{-4}~\mathrm{rad}$  \\
\midrule
$\hat{\bm{\mathit{\Phi}}}_{m}$ & Absolute calibration &  14.5--290 & low & $3.0 \times 10^{-5}~\mathrm{rad}$   \\
 & & & moderate & $5.5 \times 10^{-5}~\mathrm{rad}$   \\
 & & & high     & $2.6 \times 10^{-4}~\mathrm{rad}$  \\
 & &  14.5--29 & low & $3.8 \times 10^{-3}~\mathrm{rad}$   \\
 & & & moderate & $1.2 \times 10^{-2}~\mathrm{rad}$   \\
 & & & high     & $1.3 \times 10^{-2}~\mathrm{rad}$  \\
$\left\langle \mathrm{Pr}(\bm{\mathit{\Phi}}_{m} \; \vert \; \bm{\mathit{\Phi}}_{m,\mathrm{F}}, \mathbfit{V}^\mathrm{obs}) \right\rangle$ & \textsc{BayesCal} &  14.5--29 & low & $2.2 \times 10^{-4}~\mathrm{rad}$   \\
 & & & moderate & $2.3 \times 10^{-4}~\mathrm{rad}$   \\
 & & & high     & $2.9 \times 10^{-4}~\mathrm{rad}$  \\
\bottomrule
\end{tabular}
}
\label{Tab:BiasInRecoveredGains}
\end{table*}

The mean biases in the ratio of the recovered and true degenerate gain amplitude solutions and those in the recovered estimates of the tip-tilt phases, relative to the null tip-tilt phases of the input simulated observed data, are summarised in \autoref{Tab:BiasInRecoveredGains}, for each level of calibration model incompleteness, baseline length range, and for both \textsc{BayesCal} and absolute calibration. The average absolute biases when using absolute calibration both with and without imposing an informative gain amplitude prior are comparable in all cases; thus, in this subsection references to absolute calibration should be read as encompassing both cases.

The largest average absolute biases in the degenerate tip-tilt phase parameters are found in estimates recovered with standard absolute calibration of the $14.5-29~\mathrm{m}$ data set, in which case the average absolute tip-tilt phase offset ranges from a few times $10^{-3}~\mathrm{rad}$, at best, for the degenerate tilt phase, paired with a larger $\sim 2\times10^{-2}~\mathrm{rad}$ offset in average absolute tip phase, when the low incompleteness simulated calibration visibility models is used in the calibration, to greater than $10^{-2}~\mathrm{rad}$ in both tip and tilt phase when either moderate or high incompleteness simulated calibration visibility models are used. Relative to the degenerate tip-tilt phase solutions recovered for the short-baseline data set using absolute calibration, both with and without an informative prior on the degenerate gain amplitudes, the high resolution information available in the $14.5-290~\mathrm{m}$ data set results in a significant reduction in biases in the average absolute degenerate tip-tilt phase offsets recovered with the same calibration approaches.

In general, a less complete calibration model will result in larger tip-tilt phase biases\footnote{On average, a less complete calibration model will result in larger gain errors. However, owing to an interferometric array's incomplete sampling of the $uv$-plane and constructive and destructive interference of sky emission in the visibilities, in addition to the level of calibration model completeness, the gain error is dependent on \begin{enumerate*}\item the spatial distribution of the observed emission and calibration sky model and \item the array layout and null space of the instrument model.\end{enumerate*} As a result of these additional dependencies, gain error is not guaranteed to be a monotonic function of calibration model completeness.}. In fitting the calibration model to the data, one expects bias in the recovered degenerate gain solutions to be a function of signal-to-noise weighted calibration model completeness, which is itself a function of baseline length. The visibility phase bias ($\bm{b} \cdot (\sPhi-\sPhi_\mathrm{true})$) imparted to a model visibility by a given degenerate tip-tilt phase vector bias, ($\sPhi-\sPhi_\mathrm{true}$), is linearly dependent on the baseline length on which the visibility is measured. The greater sensitivity of long baselines to degenerate tip-tilt phase bias thus results in reduced tip-tilt phase bias, for a given calibration model completeness, when the data set includes longer baselines. Additionally, for the data sets calibrated here, the increased completeness of the model on long baselines due to the relatively more complete point source model, when compared with the diffuse model, and the larger extent to which the diffuse emission is resolved out on long baselines of the $14.5-290~\mathrm{m}$ data set further act to reduce the average absolute biases in the degenerate tip-tilt phase parameters, relative to those recovered from the $14.5-29~\mathrm{m}$ data set.

The level of average absolute bias in the degenerate tip-tilt phase offsets recovered  when calibrating the $14.5-29~\mathrm{m}$ data set with \textsc{BayesCal} is 1--2 orders of magnitude lower than in those recovered with absolute calibration, with or without the imposition of an informative degenerate gain amplitude prior, when performed on the same data set in all three calibration model incompleteness regimes, and is within an order of magnitude of the level of average absolute bias in the degenerate tip-tilt phase offsets achieved with absolute calibration on $14.5-290~\mathrm{m}$. This improvement with \textsc{BayesCal} results from the greater completeness of the  combined $\mathbfit{V}^\mathrm{sim} + \mathbfit{V}^\mathrm{fit}$ \textsc{BayesCal} calibration model, with sufficiently high completeness achieved to compensate for the loss of resolution associated with calibrating a short-baseline data set.

The level of the detrimental effect one can expect biases in the tip-tilt phases to have on 21 cm power spectrum estimation is dependent on the extent to which visibility phase information is used in the power spectrum estimator. Existing proposals for the analysis of calibrated data sets for 21 cm cosmology applications can be classified into two distinct families, based on whether they estimate the power spectrum of the measured or reconstructed sky (\citealt{2019MNRAS.483.2207M}). When the power spectrum of the measured sky is estimated, for example, via the delay transform of the visibilites along the frequency axis (e.g. \citealt{2012ApJ...756..165P}), the phase of the visibilities is discarded and, thus, the recovered signal estimates are robust to biases in the degenerate tip-tilt phase parameters. In contrast, in approaches that forward model or reconstruct the signal (e.g. \citealt{2019MNRAS.484.4152S} and \citealt{2019PASA...36...26B}, respectively), bias, and especially spurious spectral fluctuations, in the degenerate tip-tilt phase solutions have the potential to introduce foreground systematics into the recovered 21 cm signal estimates by shifting apparent source positions as a function of frequency and, in so doing, imparting non-smooth spectral structure along any given line of sight. 

In the context of recovered sky power spectrum estimators, if one were to use absolute calibration of the short-baseline data sets, such as those considered here, for 21 cm cosmology applications, the biases in the degenerate tip-tilt phase parameters are as large as several degrees, with gradients of order a degree over the $9~\mathrm{MHz}$ bandwidth of the data set. Such large offsets and gradients will bias the reconstructed sky model and have the potential to introduce additional spectral structure in recovered sky power spectrum estimators that will bias their power spectrum estimates if not accounted for. In contrast, the greatly reduced degenerate tip-tilt phase biases found with \textsc{BayesCal} correspond to shifts in source positions of order an arcminute when the high and moderate incompleteness simulated calibration visibility models are used in the calibration and sub-arcminute shifts in source positions for the low incompleteness simulated calibration visibility model. In both cases, in the frequency range relevant to 21 cosmology during CD and the EoR ($\nu \lesssim 200~\mathrm{MHz}$), these shifts are 1--2 orders of magnitude smaller than the angular scales probed by the 10s of metre baselines on which interferometric experiments are generally attempting to extract the 21 cm signal. The amplitude of fluctuations in the phase parameters is comparable or smaller than this level.  Thus, it is unclear how challenging such fluctuations will be for analyses in this category. Resolving this question is likely to require investigation on a per-analysis basis. 

The maximum bias in $\mathbfit{A}_\mathrm{true}/\hat{\mathbfit{A}}$ occurs when calibrating the $14.5-29~\mathrm{m}$ data set using absolute calibration, with a bias of $\sim 6.5\%$ when calibrating data with the high or moderate incompleteness calibration models. This drops to $\sim 5.7\%$ with the low incompleteness simulated visibility model. The completeness of the $14.5-290~\mathrm{m}$ visibility model varies with baseline length and is dominated by the incompleteness of the diffuse and point source visibility model on short and long baselines, where the respective emission types make more significant contributions to the measured visibilities. The bias in values of $\mathbfit{A}_\mathrm{true}/\hat{\mathbfit{A}}$ when calibrating the $14.5-290~\mathrm{m}$ data set using absolute calibration is $\sim 4\%$ in the high and moderate incompleteness regime and drops to $\sim 2\%$ in the low incompleteness regime. The lower bias in the values of $\mathbfit{A}_\mathrm{true}/\hat{\mathbfit{A}}$ when calibrating the $14.5-290~\mathrm{m}$ data set relative to the $14.5-29~\mathrm{m}$ data set can be understood in terms of the higher completeness of the point source relative to diffuse components of the simulated calibration sky models. As a result, the bias in $\mathbfit{A}_\mathrm{true}/\hat{\mathbfit{A}}$ for the $14.5-290~\mathrm{m}$ data set will be a weighted average of the larger and smaller biases associated with the different baseline lengths of the data set. 

When calibrating either the $14.5-29~\mathrm{m}$ or $14.5-290~\mathrm{m}$ data sets with absolute calibration using the high or moderate incompleteness simulated calibration sky models, both when calibrating with or without an informative prior on the degenerate gain amplitudes, the average absolute bias in $\mathbfit{A}_\mathrm{true}/\hat{\mathbfit{A}}$ is comparable. The only difference between the high or moderate incompleteness simulated calibration sky models is the reduction in the minimum flux-density (from $1$ to $0.1~\mathrm{Jy}$ at $151~\mathrm{MHz}$) of GLEAM sources included in $\mathbfit{V}^\mathrm{sim}$. Thus, the similarity between the average absolute biases in $\mathbfit{A}_\mathrm{true}/\hat{\mathbfit{A}}$ in these two cases implies the limited importance, with respect to reducing bias, of the increased point source completeness resulting from the additional sources included in the model when there is comparatively large incompleteness in the diffuse emission component of the sky model used for $\mathbfit{V}^\mathrm{sim}$.

The $\sim 1.25\%$ bias in $\left\langle \mathrm{Pr}(\mathbfit{A}_\mathrm{true}/\mathbfit{A} \; \vert \; \mathbfit{A}_\mathrm{F}, \mathbfit{V}^\mathrm{obs}) \right\rangle$ when calibrating the $14.5-29~\mathrm{m}$ data set with \textsc{BayesCal} using the high incompleteness simulated calibration sky model and the sub-$0.1\%$ biases when calibrating the $14.5-29~\mathrm{m}$ data set with \textsc{BayesCal} using the moderate and low incompleteness simulated calibration sky models are lower than the corresponding biases found when calibrating the same data set with absolute calibration by factors of $\sim 5$, $73$ and $190$ in the high, moderate and low completeness regimes, respectively. Relative to the biases associated with the average absolute values of the degenerate gain amplitudes recovered using absolute calibration of the $14.5-290~\mathrm{m}$ data set, the  biases in $\left\langle \mathrm{Pr}(\mathbfit{A}_\mathrm{true}/\mathbfit{A} \; \vert \; \mathbfit{A}_\mathrm{F}, \mathbfit{V}^\mathrm{obs}) \right\rangle$ when calibrating the $14.5-29~\mathrm{m}$ data set with \textsc{BayesCal} are also lower by factors of $\sim 4$, $50$ and $73$, in the high, moderate and low completeness regimes, respectively. 

Nevertheless, while important for accurate data normalisation and error propagation, low level bias in the redundant gain amplitude parameters is still comparatively benign in 21 cm cosmology applications, since it does not complicate spectral separation of the foregrounds and the 21 cm signal in the data. In contrast, spurious spectral fluctuations imparted to the data by imperfect estimation of the redundant gain amplitude parameters must be suppressed to a high degree in order to recover the 21 cm signal in both measured and reconstructed sky-based approaches to 21 cm signal estimation.

\subsubsection{Spurious spectral structure}
\label{Sec:SpuriousSpectralStructure}

\begin{figure*}
	\centerline{
	\includegraphics[width=0.5\textwidth]{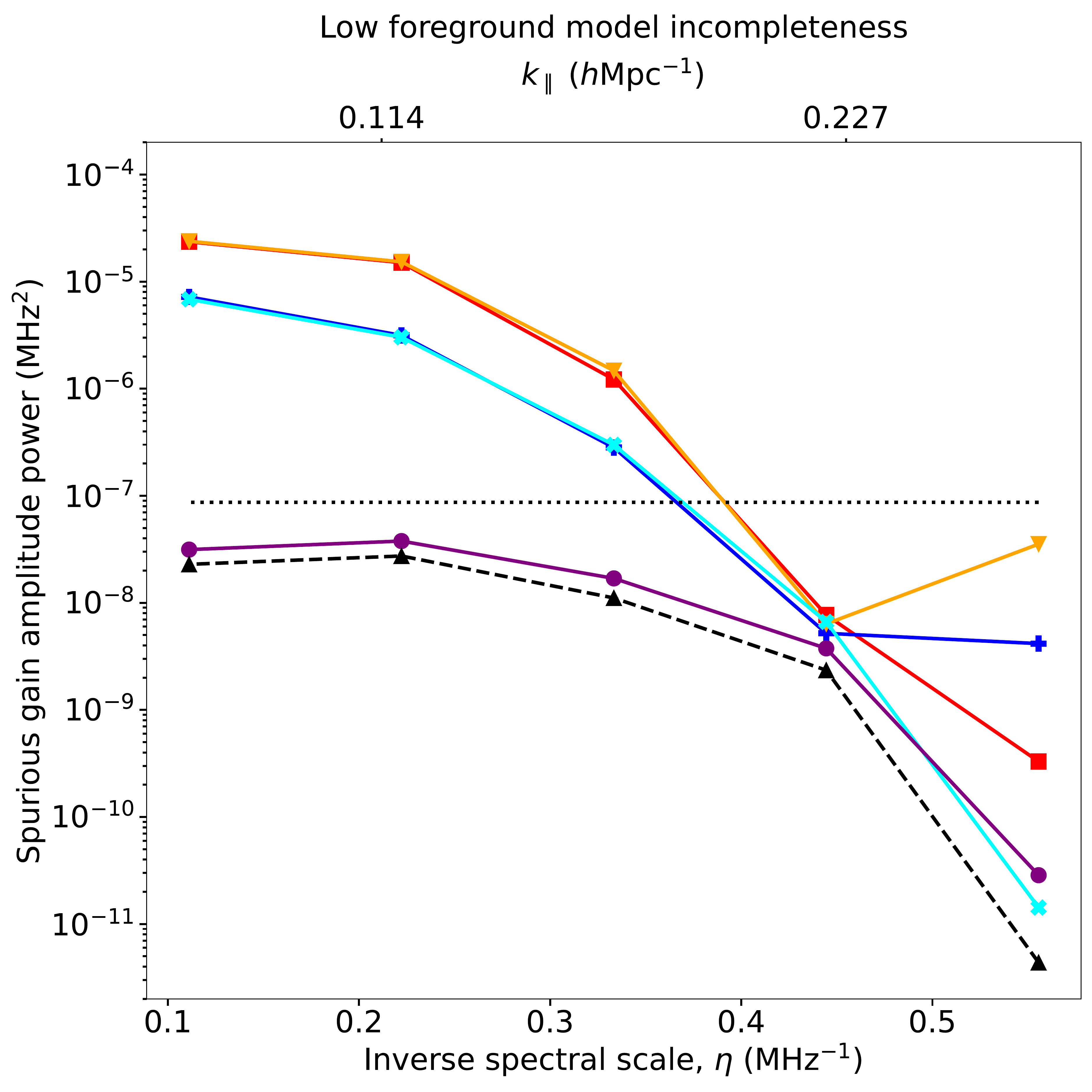}
	\includegraphics[width=0.5\textwidth]{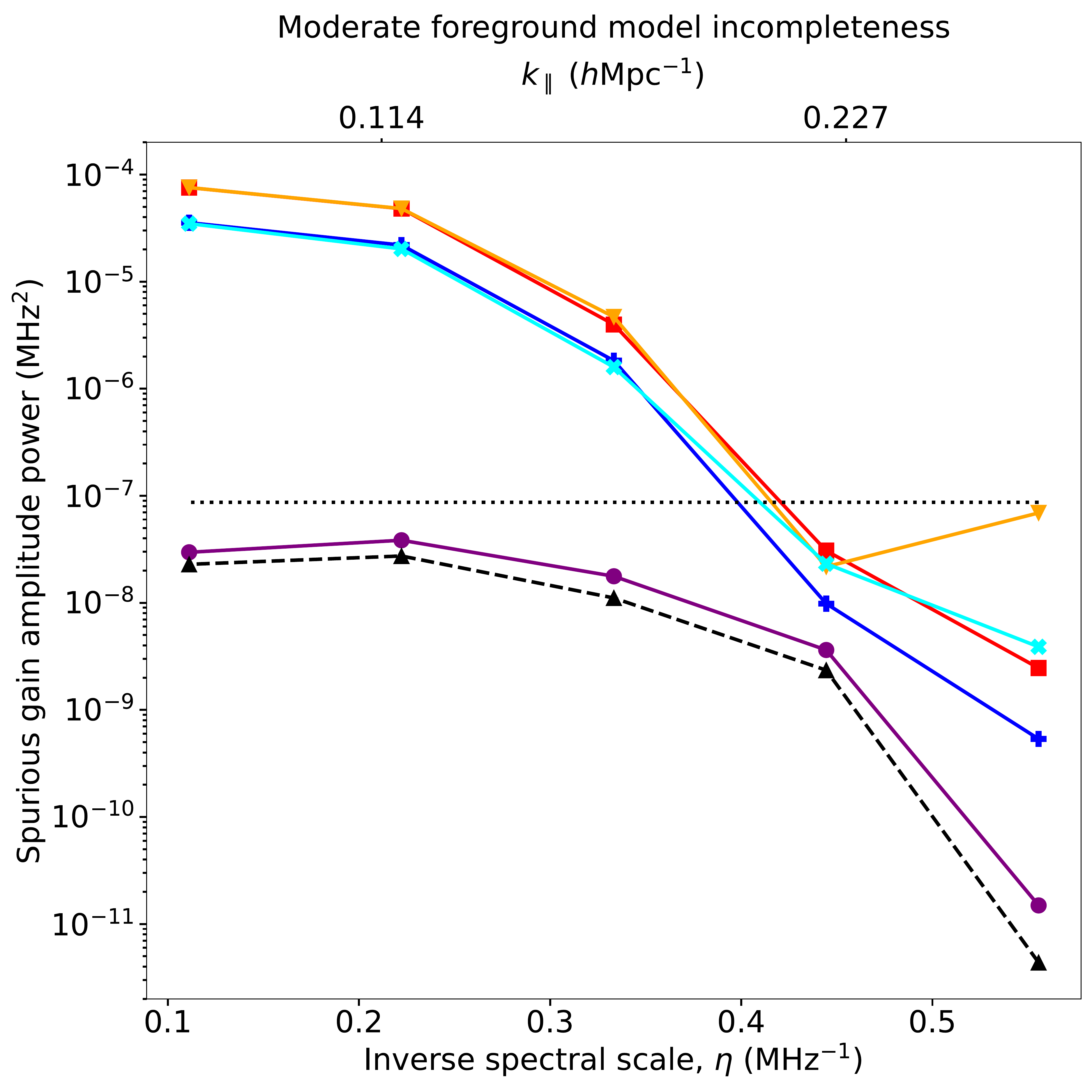}}
	\centerline{
	\includegraphics[width=0.5\textwidth]{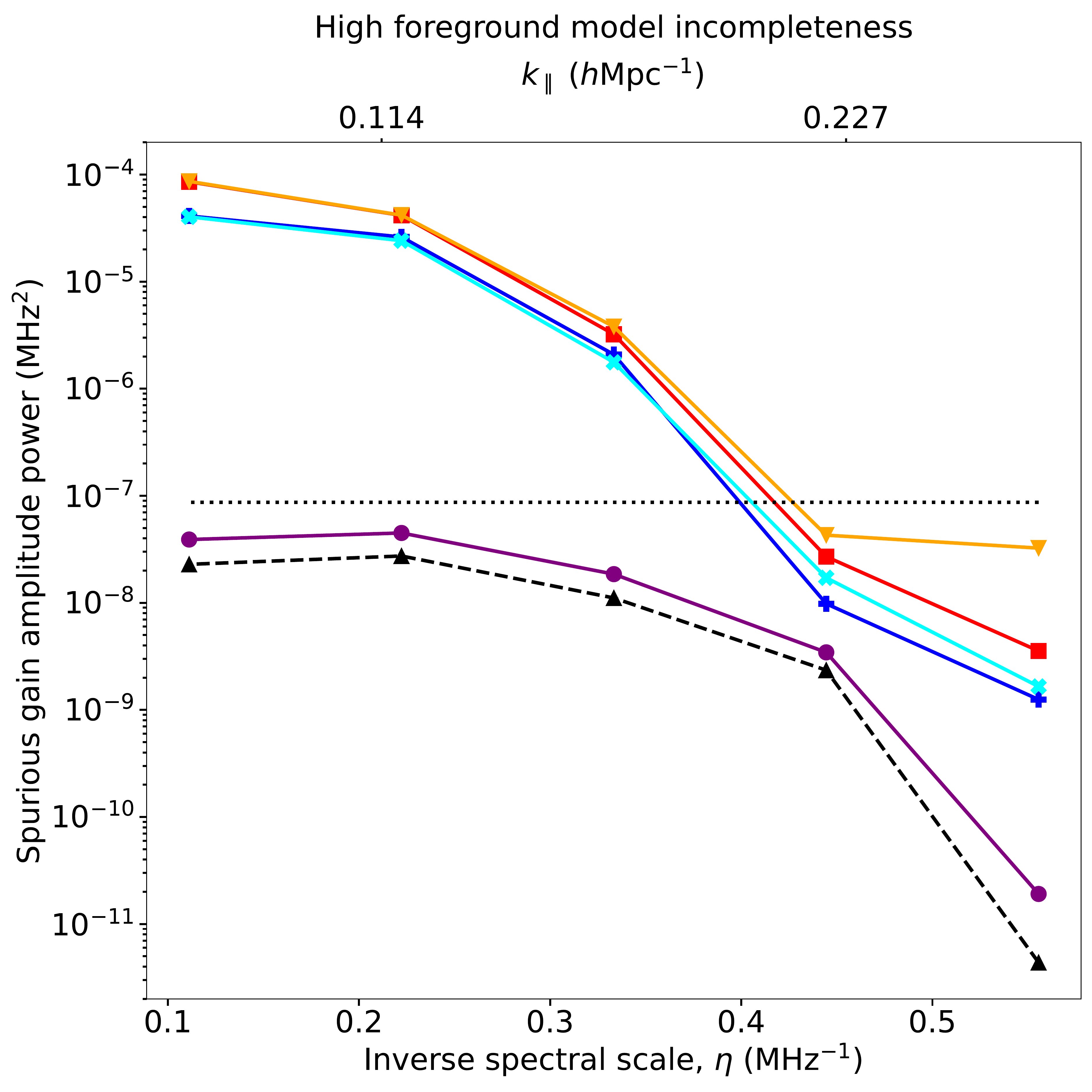}
        \includegraphics[width=0.5\textwidth]{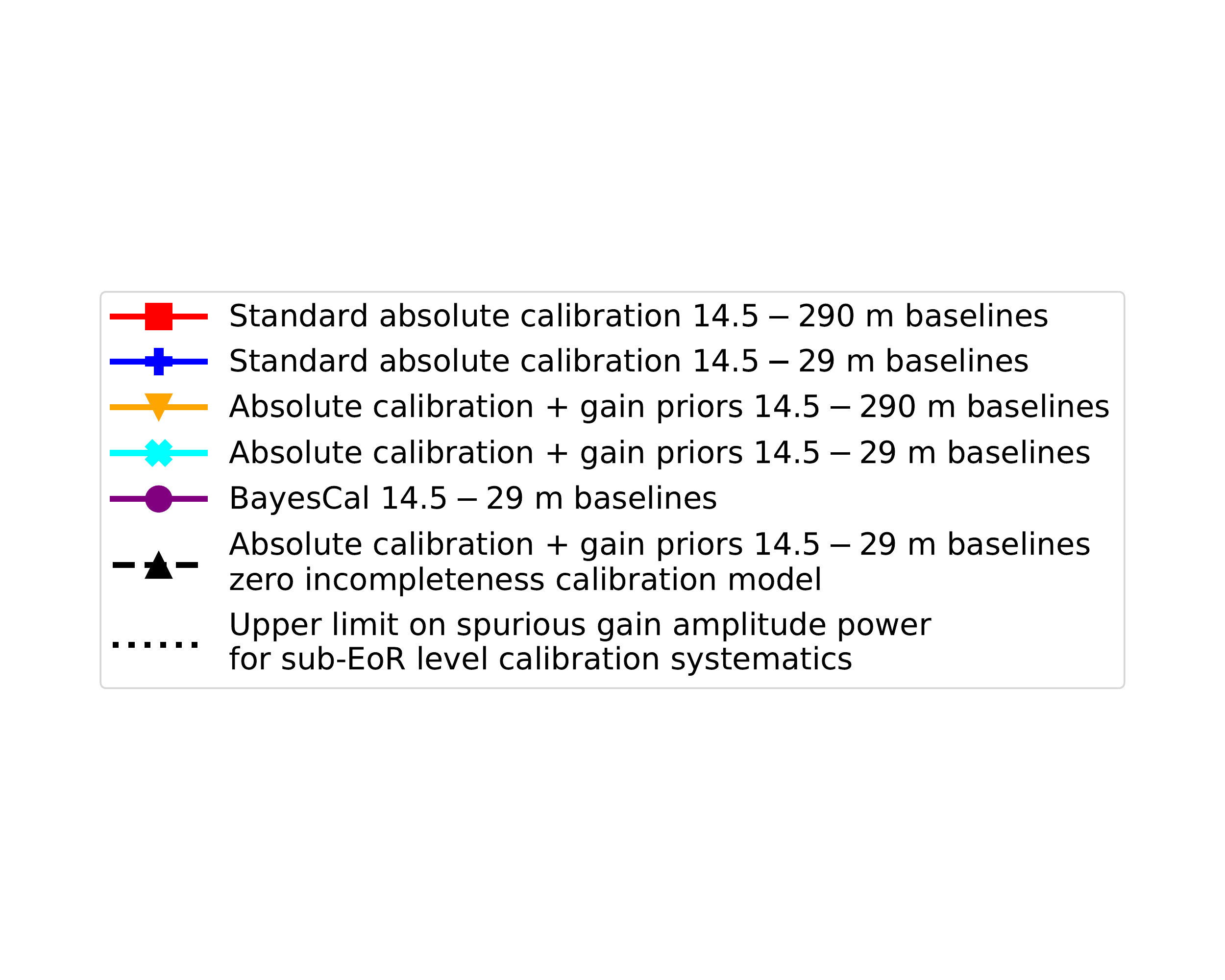}}
\caption{Power spectra, $P_{A}(\eta)$, of $R = A_\mathrm{true}/A_\mathrm{model}$ characterising the level of discrepant spectral structure imparted to the data as a result of imperfect calibration, derived in three regimes of incompleteness of the simulated component of the calibration model (see \autoref{Tab:CompletenessLevels}): low (top left), moderate (top right) and high incompleteness (bottom left). In all cases, $P_{A}(\eta)$ is calculated via \autoref{Eq:GainAmplitudePS}. The degenerate gain amplitude is real and therefore has a symmetric power spectrum; here, we plot half of the power spectrum for positive $\eta$. The line colours, denoting the calibration approach used and the baseline lengths it is applied to are labelled in the legend (see main text for details). The dotted black lines show the level at which spurious spectral fluctuations imparted to the data by imperfect calibration will introduce foreground systematics equal in amplitude, at a given spectral scale, to a white noise proxy for the 21 cm signal with an RMS amplitude of 10 mK. High fidelity calibration solutions with spurious gain amplitude power below the level of the dotted black line are necessary for unbiased recovery of the power spectrum of the 21 cm signal proxy from interferometric data. The dashed black lines show $P_{A}(\eta)$ for the maximum a posteriori degenerate gain amplitude solutions that would hypothetically be recovered using absolute calibration with gain priors in the limit of a perfect zero-imcompleteness calibration model, in which the non-zero spurious gain amplitude power derives purely from the noise in the data. This also corresponds to the \textsc{BayesCal} solutions that would be recovered in the limit of a perfect fitted calibration model component of the \textsc{BayesCal} data model for the contribution to the data from incomplete and imperfectly modelled emission on the sky.
}
\label{Fig:CalibrationSolutionsPSComparison}
\end{figure*}

When the ratio of the mean gain solution to the amplitude of spurious spectral fluctuations in the gains, on a given spectral scale, is equal to the dynamic range between the foregrounds and the 21 cm signal, the foreground systematic introduced will be at the level of the 21 cm signal on that scale. Thus, in order to recover unbiased estimates of the 21 cm signal, at a given spectral scale, one requires the ratio between the mean degenerate gain amplitude and the amplitude of spurious spectral fluctuations present in the data following calibration to be much smaller than the dynamic range between the foregrounds and the 21 cm signal at that scale. For the field calibrated here, and assuming a white noise-like 21 cm signal with an RMS amplitude\footnote{We use an amplitude of 10 mK for our reference white noise-like 21 cm signal to facilitate simple estimation of the corresponding ratio for alternate choices of EoR model. More stringent spectral calibration requirements will be needed for fainter EoR signals, with less power than this on a given spectral scale, and vice versa for brighter EoR signals.} of 10 mK, this corresponds to a requirement that the amplitude of spurious spectral fluctuations in the ratio of the estimated and true degenerate gain amplitudes be smaller than 1 part in $\sim 10^{3.5}$. For an instrument with a mean gain amplitude of unity and a 1 MHz channel width, as considered here, this equates to a requirement that the power in these fluctuations be much smaller than $\sim 10^{-7}~\mathrm{MHz}^2$.
 
To determine whether this threshold is met, we define $R \equiv A_\mathrm{true}/A_\mathrm{model}$, where $A_\mathrm{model}$ is the mean of the functional posterior probability distribution, maximum a posteriori, or maximum likelihood parameter estimates of the degenerate gain amplitudes when discussing the calibration solutions derived with \textsc{BayesCal} and absolute calibration with and without imposing gain priors, respectively. The level of spurious spectral fluctuations that will be imparted to the calibrated data due to imperfect calibration, in the range of cases considered, can be assessed by calculating the power spectrum, $P_{A}(\eta)$, of fluctuations in the ratio of the recovered and true redundant gain amplitude parameters,
\begin{equation}
\label{Eq:GainAmplitudePS}
\left\langle \widetilde{\delta RW}(\eta)\widetilde{\delta RW}^{*}(\eta^\prime) \right\rangle(\Delta\nu)^{2} \equiv \delta_\mathrm{K}(\eta-\eta^\prime)P_{A}(\eta) \ .
\end{equation}
Here, $\eta$ measures inverse spectral scale\footnote{When plotting the power spectra, we also display the values of the inverse radial distance parameter corresponding to the plotted $\eta$-modes relevant to redshifted 21-cm emission in the frequency range of the calibrated data:  $k_{\parallel} \approx \dfrac{2\pi\eta H_\mathrm{0}f_{21}E(z)}{c(1+z)^2}$. Here, $f_{21} \simeq 1420$ MHz is the rest frame frequency of the 21-cm line, $c$ is the speed of light in a vacuum, $H_\mathrm{0} = h100~\mathrm{km~s^{-1}~Mpc^{-1}}$ is the present day value of the Hubble constant, where we take $h=0.6766$, and $E(z) = \sqrt{\Omega_{\rm M}(1+z)^{3} + \Omega_{\sLambda}}$ is the dimensionless Hubble parameter, with $\Omega_{\rm M} = 0.3111$ and $\Omega_{\sLambda} = 0.6889$ the present day matter and dark energy density in units of the critical density, respectively. We use the best fitting values from the \citet{2020A&A...641A...6P} TT,TE,EE+lowE+lensing+BAO analysis for $h$, $\Omega_{\rm M}$ and $\Omega_{\sLambda}$.}. $\widetilde{\delta RW}(\eta)$ is the discrete Fourier transform of $\delta RW$ with $\delta R = (R-\overline{R})$, where $\overline{R}$ is the mean of $R$, and $W$ is a Blackmann--Harris window function applied to down-weight spectral leakage of power from large spectral scale fluctuations in $R$. $\Delta\nu$ is the channel width in MHz and $\delta_\mathrm{K}$ is the Kronecker delta function.

In \autoref{Fig:CalibrationSolutionsPSComparison}, we show $P_{A}(\eta)$, estimated using \autoref{Eq:GainAmplitudePS}, for our five sets of calibration solutions: \textsc{BayesCal} solutions for the $14.5-29~\mathrm{m}$ data set (purple lines), absolute calibration solutions with gain priors for the $14.5-29~\mathrm{m}$ and $14.5-290~\mathrm{m}$ data sets (cyan lines and orange lines, respectively) and absolute calibration solutions without gain priors for the $14.5-29~\mathrm{m}$ and $14.5-290~\mathrm{m}$ data sets (blue lines and red lines, respectively), derived in the low, medium and high simulated visibility model incompleteness regimes (top left, top right and bottom left, respectively). The dotted black lines show the level at which spurious spectral fluctuations imparted to the data by imperfect calibration will introduce foreground systematics equal in amplitude, at a given spectral scale, to a white noise-like 21 cm signal with an RMS amplitude of 10 mK. The dashed black lines shows $P_{A}(\eta)$ for the maximum a posteriori degenerate gain amplitude solutions that would hypothetically be recovered using absolute calibration with gain priors\footnote{Since in the limit of zero incompleteness or error in the a priori known component of the calibration model the corresponding  zero-power prior on the fitted component of the \textsc{BayesCal} calibration model suppresses to zero its contribution to the total calibration model, in this limit, \textsc{BayesCal} simplifies to the absolute calibration with gain priors algorithm described in \autoref{Sec:AbscalWithGainPriors}. Thus, equivalent solutions would be derived in this limit using absolute calibration with gain priors or \textsc{BayesCal}. However, for brevity and to avoid confusion when discussing these solutions with those recovered in the realistic high, moderate and low calibration model incompleteness scenarios, we use the former to refer to the solutions in this zero-error and zero-incompleteness limit going forward.} in the limit of a perfect zero-incompleteness calibration model, in which the non-zero spurious gain amplitude power derives purely from the noise in the data. This also corresponds to the \textsc{BayesCal} solutions that would be recovered in the limit that the $\mathbfit{V}^\mathrm{fit}$ component of the \textsc{BayesCal} data model perfectly modelled the contribution to the data from incomplete and imperfectly modelled emission on the sky missing in the standard a priori known component of the data model, $\mathbfit{V}^\mathrm{sim}$, in the high, moderate and low calibration model incompleteness scenarios.

Both standard absolute calibration and absolute calibration with informative priors on the gain solutions result in spurious spectral fluctuations in $R$ at a level which will result in foreground systematics exceeding the amplitude of our fiducial 21 cm signal on 3, 4.5 and 9 MHz spectral scales (corresponding to $k_{\parallel} \simeq 0.17, 0.11$ and $0.06~h\mathrm{Mpc}^{-1}$, respectively, for redshifted 21 cm emission in the frequency range of our simulated observed data)\footnote{A 1 MHz spectral scale corresponds to a $k_\parallel \approx 0.5~h\mathrm{Mpc}^{-1}$ redshifted 21 cm inverse spatial scale at the center of the 160--169 MHz spectral band of the data sets being calibrated.} in the high, moderate and low calibration model incompleteness simulated visibility models when calibrating data sets comprised of $14.5-29~\mathrm{m}$ baselines or data sets comprised of $14.5-290~\mathrm{m}$ baselines. Calibration of the data with either of these approaches precludes separation of the 21 signal from the foregrounds, in the calibrated data set, on these spectral scales, by spectral means. 

In contrast, significantly higher fidelity solutions are recovered when using \textsc{BayesCal}. Specifically, we find that spurious spectral fluctuations in $R$ when using \textsc{BayesCal} are sufficiently low for foreground systematics that would result from applying spectral scale-based signal separation analysis techniques used in 21 cosmology to be below the level of the fiducial 21 cm signal on all spectral scales accessible in the data set and for all levels of simulated visibility model completeness considered. Comparable performance within a factor of a few is achieved in all three regimes of incompleteness of the a priori known component of the calibration sky model. In each case, the \textsc{BayesCal} calibration solutions reduce spurious redundant gain amplitude fluctuations to a few parts in $10^{4}$ of the amplitude of simulated redundant gain parameters on $9~\mathrm{MHz}$ spectral scales, with continuing reduction down to a few parts in $10^{6}$ on $1.8~\mathrm{MHz}$ spectral scales. 
 
Comparing the \textsc{BayesCal} calibration solutions to those that would hypothetically be recovered using absolute calibration with gain priors in the limit of a perfect zero-imcompleteness calibration model in which the non-zero spurious gain amplitude power derives purely from the noise in the data, it can be seen that \textsc{BayesCal} enables recovery of calibration solutions similar in fidelity to those recovered in this perfect calibration model scenario. In all cases, the \textsc{BayesCal} solutions mitigate spurious gain amplitude power to within a factor of two of the complete zero-error calibration model solutions on all but the smallest ($1.8~\mathrm{MHz}$) Nyquist sampled spectral scale in the calibration data. On the $1.8~\mathrm{MHz}$ spectral scale, the difference in fidelity of the \textsc{BayesCal} and perfect calibration model solutions increases to a factor of ten in power in spurious spectral structure in the calibration solutions; nevertheless, the absolute fidelity of the \textsc{BayesCal} solutions on this spectral scale is sufficiently high for foreground systematics introduced by imperfections in the calibration solutions on this spectral scale to be negligible relative to the power in the fiducial EoR model on the same scale.

In the context of 21 cm cosmology applications, these results show that, in the scenarios modelled, only \textsc{BayesCal} enables recovery of the 21 cm at $k_\parallel \lesssim 0.15~h\mathrm{Mpc}^{-1}$.  For all the other techniques in all completeness regimes, that scale is contaminated. More generally, when calibrating using \textsc{BayesCal}, the level of spurious spectral structure in the calibration solutions implies that the power in foreground systematics that would result from 21 cm signal separation of the foregrounds by spectral means in the presence of spurious spectral fluctuations in $R$ are a factor of a few below the fiducial 21 cm signal level on 9 MHz spectral scales ($k_{\parallel} \simeq 0.06~h\mathrm{Mpc}^{-1}$), and drop to close to 4 orders of magnitude below the power in the fiducial 21 cm signal by $1.8~\mathrm{MHz}$ scales ($k_{\parallel} \simeq 0.28~h\mathrm{Mpc}^{-1}$).

\subsubsection{The impact of gain priors}
\label{Sec:ImpactOfGainPriors}

The generic effect of imposing informative gain priors when using either absolute calibration or \textsc{BayesCal} (i.e. joint estimation of absolute calibration solutions and a sky model) is the up-weighting of gain solutions consistent with one's prior knowledge of the spectral structure in the instrumental gains. However, when an incomplete calibration model is used, imposing informative gain priors will, in general, worsen the fit of the calibration model to the visibility data (since the fit of the model to the data alone is optimised in the absence of informative priors). Comparing the level of spurious spectral structure in degenerate gain amplitude parameters recovered using absolute calibration with and without informative priors on the gain solutions demonstrates how application of gain priors without explicitly addressing the incompleteness of the simulated calibration sky model is not guaranteed to be advantageous on all spectral scales. Rather, it leads to a redistribution of spurious spectral structure with, in some cases, minor improvements on large spectral scales (where the maximum likelihood solutions are least consistent with the prior and true gain solutions and, thus, where the impact of the prior is largest) coming at the cost of less accurate solutions on smaller spectral scales.

Ultimately, the impact of applying gain priors when using an incomplete sky model is a complex interplay between the incompleteness of the specific flux distribution of calibration model of the sky observed in the data being calibrated and the instrumental chromaticity of the baselines being calibrated. In contrast, imposing gain priors while jointly fitting for gain parameters and a model for the contribution to the visibilities from the component of the flux distribution absent in one's simulated calibration sky model, as done with \textsc{BayesCal}, mitigates this problem and optimises for calibration models in which both the recovered gains reflect their priors and the product of the gain parameters and the sum of the fitted and simulated components of the calibration model accurately describes the data.

\subsection{\textsc{BayesCal} functional posterior probability distributions}
\label{Sec:BCCPD} 

\begin{figure*}
	\centerline{
	\includegraphics[width=0.33\textwidth]{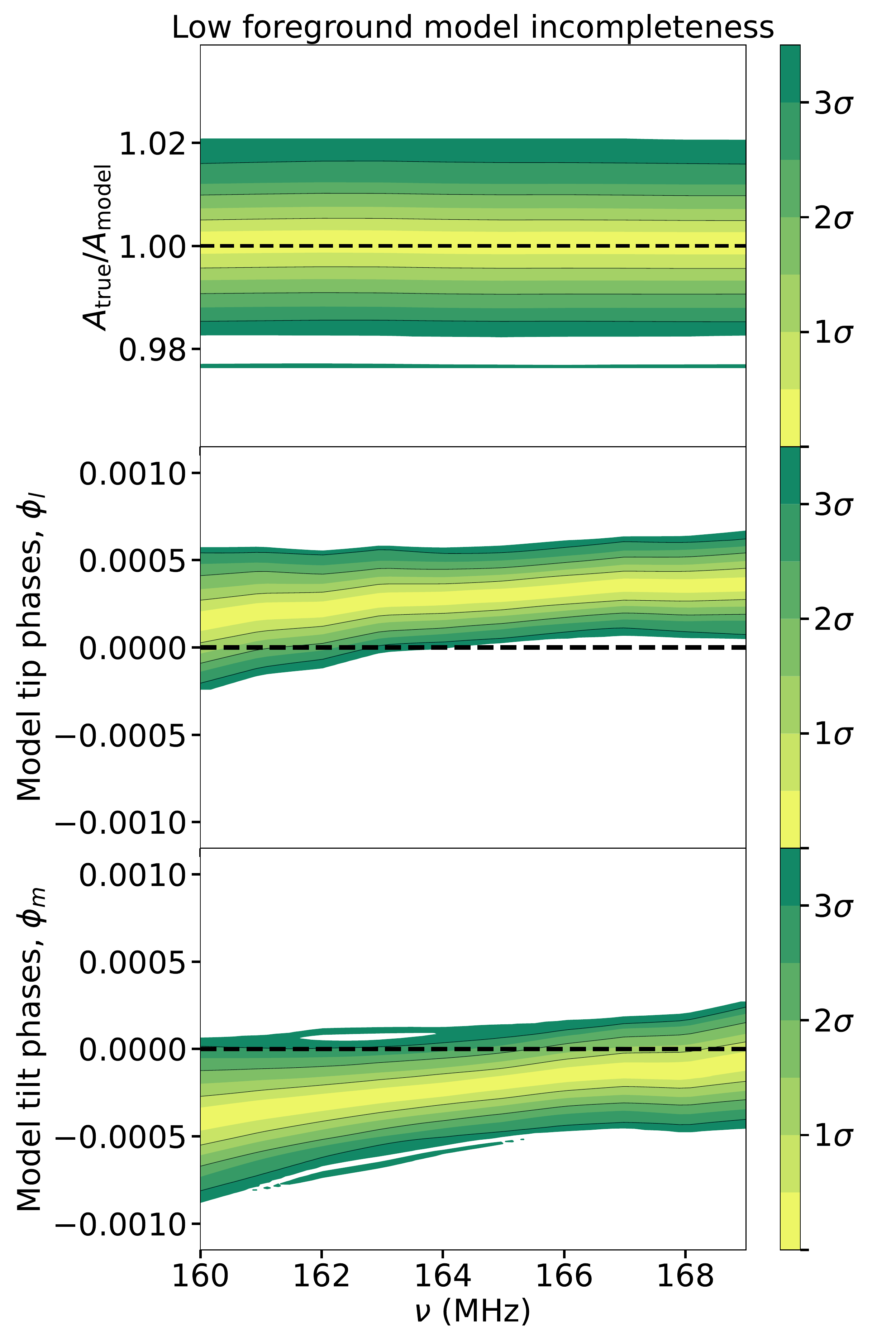}
	\includegraphics[width=0.33\textwidth]{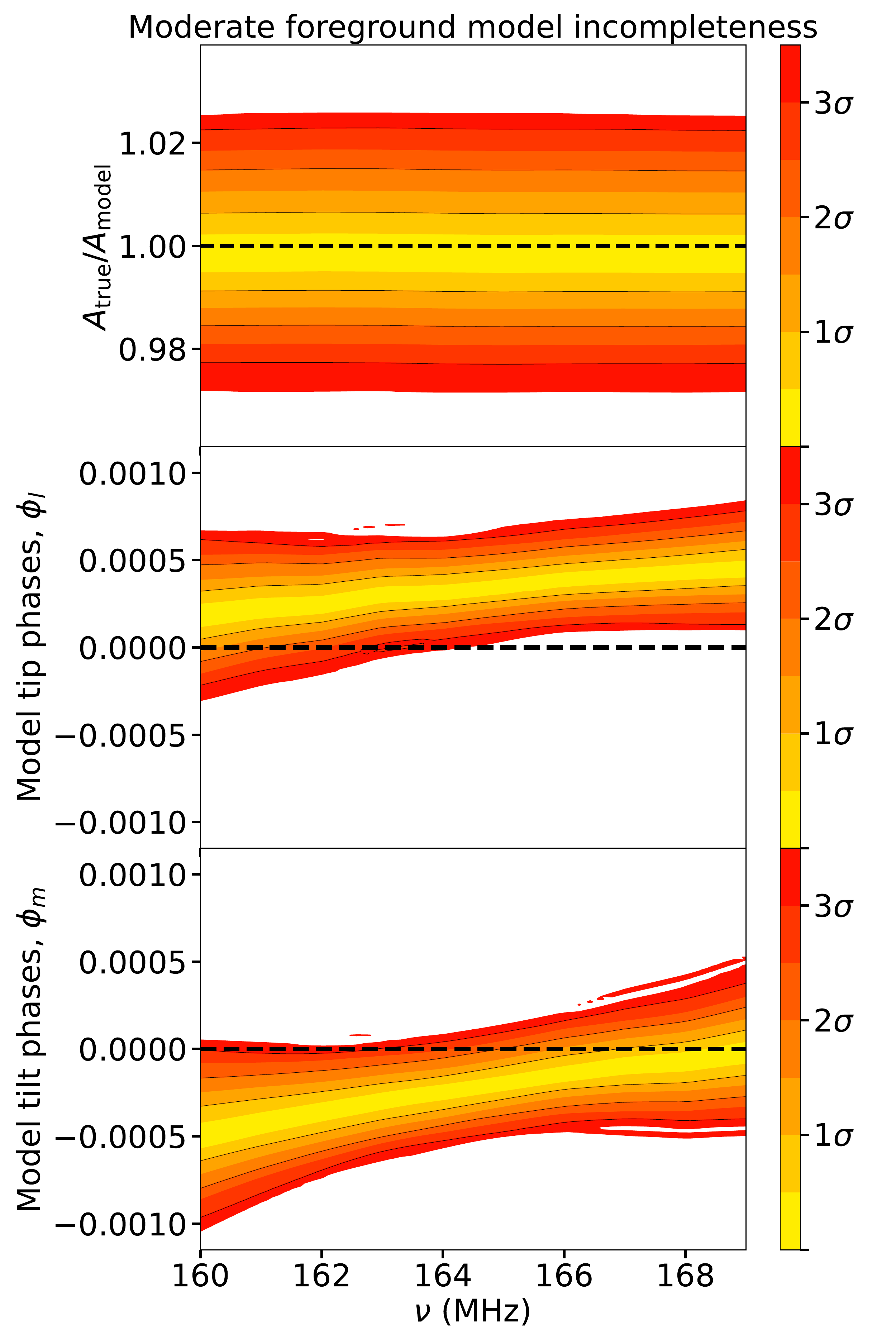}
	\includegraphics[width=0.33\textwidth]{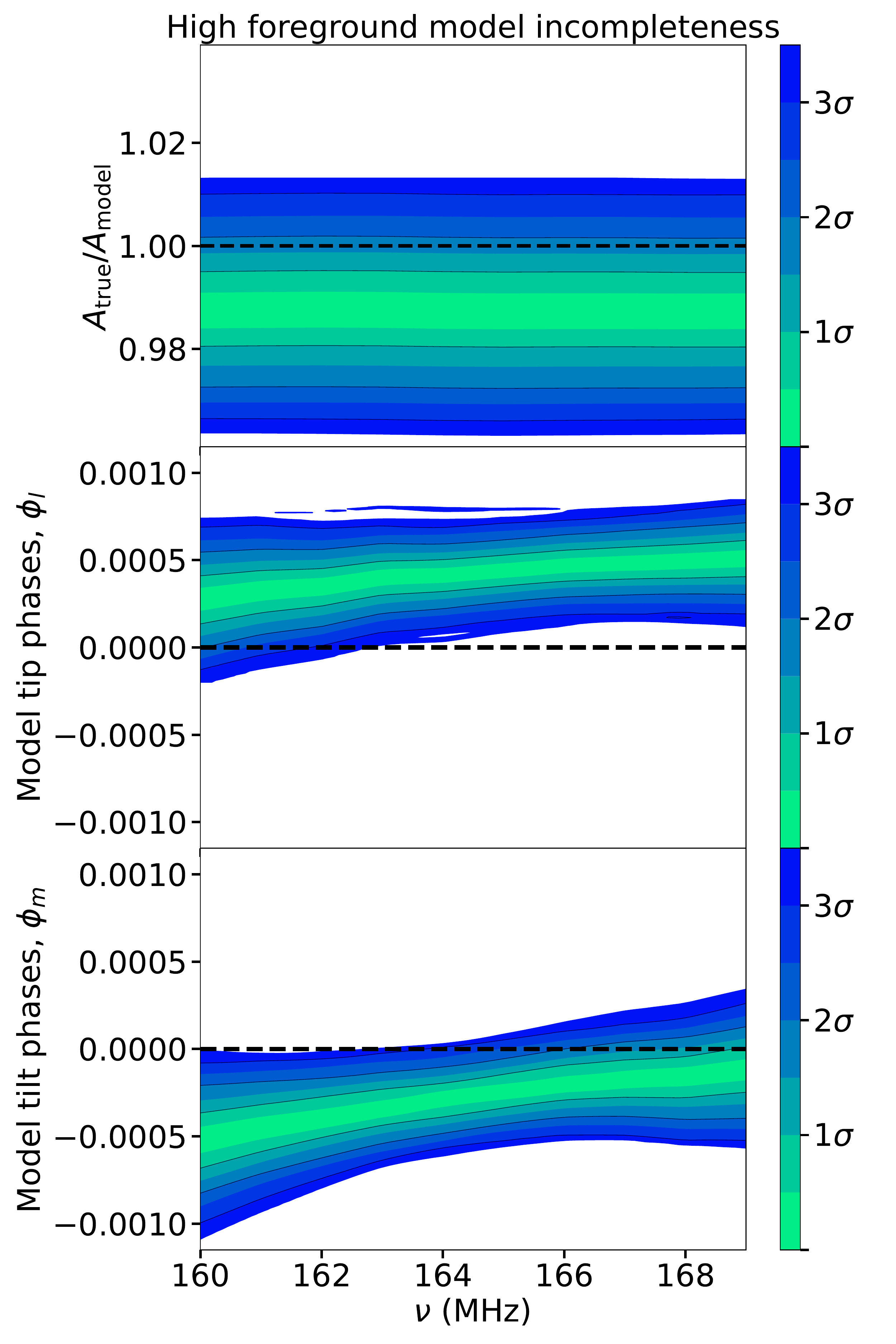}}    
\caption{Contour plots of the functional posterior probability distributions for the ratio of the input degenerate calibration amplitude and the model degenerate calibration amplitude ($\mathrm{Pr}(\mathbfit{A}/\mathbfit{A}_\mathrm{true} \; \vert \; \mathbfit{A}_\mathrm{F}, \mathbfit{V}^\mathrm{obs})$; top), tip-phase ($\mathrm{Pr}(\bm{\mathit{\Phi}}_{l} \; \vert \; \bm{\mathit{\Phi}}_{l,\mathrm{F}})$, center) and tilt-phase ($\mathrm{Pr}(\bm{\mathit{\Phi}}_{l} \; \vert \; \bm{\mathit{\Phi}}_{l,\mathrm{F}})$, bottom), when fitting the sum of a simulated model for the known flux contributions from diffuse and point sources and a statistical model for the unknown component of sky emission, jointly with the degenerate gain solutions when the simulated sky model has low, moderate and high incompleteness from left to right, respectively. The dashed black lines show the input values of the parameters. 
}
\label{Fig:CalibrationSolutionsConditionalPosteriors}
\end{figure*} 

In \autoref{Sec:ComparisonBetweenBayesCalAndAbscal}, we showed the frequency dependent means of the functional posterior probability distributions of the parameters of the \textsc{BayesCal} solutions, in order to simplify comparison with the maximum a posteriori and maximum likelihood solutions recovered with absolute calibration of visibilities, with and without imposition of informative gain priors. However, there is significantly greater information content in the functional posterior probability distributions of the parameters than in the mean of the distribution alone. 

Estimation of the functional posterior probability distributions allows one to go beyond a single estimate of the calibrated visibility data, and, to instead derive the probability distribution of the data conditional on the calibration solutions. This provides a direct means of robustly propagating the statistical uncertainties in the calibration solutions through to the estimates of the 21 cm signal, which, in turn, is important if one wishes to avoid underestimation of the uncertainties on 21 cm estimates and mitigate the risk of either a false detection of the signal or of inflating the significance of a signal detection.

In \autoref{Fig:CalibrationSolutionsConditionalPosteriors}, we show contour plots\footnote{We derive the contour plots of the functional posterior probability distributions from the posteriors of the Fourier mode parametrisations of the degenerate gain parameters using the \textsc{fgivenx} software package (\citealt{2018JOSS....3..849H}).} of the functional posterior probability distributions for the ratio of the input degenerate calibration amplitude and the model degenerate calibration amplitude ($\mathrm{Pr}(\mathbfit{A}/\mathbfit{A}_\mathrm{true} \; \vert \; \mathbfit{A}_\mathrm{F}, \mathbfit{V}^\mathrm{obs})$, top), tip-phase ($\mathrm{Pr}(\bm{\mathit{\Phi}}_{l} \; \vert \; \bm{\mathit{\Phi}}_{l,\mathrm{F}}, \mathbfit{V}^\mathrm{obs})$; center) and tilt-phase ($\mathrm{Pr}(\bm{\mathit{\Phi}}_{m} \; \vert \; \bm{\mathit{\Phi}}_{m,\mathrm{F}}, \mathbfit{V}^\mathrm{obs})$; bottom) recovered when using the low, moderate and high simulated visibility models in the left, center and right panels, respectively.

For all levels of sky-model incompleteness considered, we find that estimates of the degenerate gain amplitude and tilt phase solutions recovered with \textsc{BayesCal} calibration are consistent with the true calibration solutions to within the 3-$\sigma$ limits of their conditional probability distributions. The tip phase solutions recovered with \textsc{BayesCal} are consistent to within the 3-$\sigma$ limits of their conditional probability distributions in the lower half of the band and marginally inconsistent in the upper half of the band. This reflects the fact that $\mathbfit{V}^\mathrm{fit}$ provides a good but not perfect model of the contribution to the observed visibilities from emission that is missing or imperfectly modelled in the a priori known component of the calibration model. However, as noted in \autoref{Sec:BiasInGainParameterEstimates}, the absolute bias in the degenerate tip phase solutions is small relative to the angular scales on which interferometers designed to detect the power spectrum of redshifted 21 cm emission are maximised. Thus, we expect the small absolute level of the bias to limit its impact on 21 cm signal recovery in reconstructed sky based frameworks. Furthermore, it has no impact on measured sky-based power spectrum estimation based frameworks, such as delay-spectrum power spectrum estimation, that do not use the phase information in the data.

We also note that in addition to the means of the conditional probability distributions for the ratio of the input degenerate calibration amplitude and the recovered degenerate calibration amplitude being spectrally smooth, as shown in \autoref{Fig:CalibrationSolutionsPSComparison}, the iso-probability-contours of the distributions are also spectrally smooth, and it is the variance in the mean amplitudes of the samples, rather than their shape, which dominates the calibration uncertainties. This derives from the individual parameter samples from the posterior probability distributions of the Fourier mode parametrisations of the degenerate gain amplitudes corresponding to individual model gain solutions with spectral structure that is consistent with the true instrumental gains in the data.

This is favourable for 21 cm cosmology applications, because, when the \textsc{BayesCal} calibration uncertainties are propagated through to the signal estimates, not only will the mean calibrated data solutions have high spectral fidelity, but the individual samples from the conditional probability distribution of the calibrated data set will as well. This will result in reduced uncertainties on estimates of the 21 cm signal recovered from the data, relative to a situation where the mean of the distribution is flat but other iso-probability-contours have spectral structure.

Nevertheless, we note that the optimality, with respect to bias minimisation, of propagating the samples from the posteriors for the degenerate gain parameters through to the target statistic one wishes to estimate from the data depends on the convexity of the function that takes one from the raw data to that statistic. Jensen's inequality states that for a convex function (such as the power spectrum of the 21 cm emission),$f(x)$ on $\mathbb{R}_{X}$, where $X$ is a discrete random variable, assuming $E[f(X)]$ and $f(E[X])$ are finite, we have,
\begin{equation}
\label{Eq:JensenInequality}
E[f(X)] \ge f(E[X]) \ .
\end{equation}
This implies that bias in the mean of the power spectra of samples from the probability distribution of the data conditional on the calibration solutions will be equal to or exceed the bias in the power spectrum of the data derived from the means of the posteriors on the degenerate gain parameters. This derives from random errors in samples from the posteriors on the degenerate gain adding up constructively in the former case and  averaging down in the latter. Thus, averaging the functional posteriors prior to forming the power spectrum\footnote{In practice, this can be done by averaging the samples from the posteriors on the degenerate gain parameters, as in \autoref{Sec:ComparisonBetweenBayesCalAndAbscal}, or the conditional posteriors at any subsequent stage of data processing before non-linear operations equivalent to convex functions on the data are applied.} will minimise imperfect-calibration-derived spurious spectral structure in the final power spectral estimates.

\section{Summary \& Conclusions}
\label{Sec:SummaryAndConclusions}

\textsc{BayesCal}, a new Bayesian framework for interferometric calibration designed to recover high fidelity gain solutions by mitigating sky-model incompleteness, was introduced in paper I of this series. The \textsc{BayesCal} calibration model is comprised of \begin{enumerate*}\item a simulated visibility model, $\mathbfit{V}^\mathrm{sim}$, which describes the expected contribution to the observed visibilities associated with a priori known sky emission (this model component is standard in sky-referenced calibration) and \item a fitted visibility model, $\mathbfit{V}^\mathrm{fit}$, that models the contribution to the observed visibilities from emission not included in $\mathbfit{V}^\mathrm{sim}$ due to the incompleteness and uncertainties associated with our a priori knowledge of the brightness distribution of the sky.\end{enumerate*} When calibrating data with \textsc{BayesCal}, one jointly fits for instrumental gains, constrained by available priors on their spectral structure, and the most probable sky-model parameter values for $\mathbfit{V}^\mathrm{fit}$, constrained by a prior on their total power.

\textsc{BayesCal} provides a powerful new tool for high fidelity interferometric calibration that is particularly well suited for application to analyses for which either the accuracy or spectral fidelity of the instrumental calibration is important to the science goal. Estimation of a faint 21 cm signal in the presence of bright spectrally smooth foregrounds is an example of such an analysis.

Here, we have demonstrated the application of \textsc{BayesCal} in the context of absolute calibration of data from redundant interferometric arrays, in which relative gains between antennas can be solved for by using sky-model-independent constraints based on the internal consistency of the redundant groups of baselines. To assess the efficacy of the \textsc{BayesCal} calibration methodology, we have applied it to simulated observations of full-Stokes sky emission in the 160--169 MHz spectral band, with a HERA-like hexagonal close-packed redundant array, for three levels of completeness of the a priori known component of the calibration model (corresponding to three levels of uncertainty and completeness of our model for the distribution of diffuse low frequency radio emission and point sources in the data to be calibrated). 

We have shown that \textsc{BayesCal} enables recovery of significantly higher fidelity  calibration solutions than standard absolute calibration or absolute calibration with physical gain priors. Relative to these approaches, we find the largest improvements on the largest spectral scales. On small and intermediate spectral scales, power in spurious gain amplitude fluctuations is comparable to or lower that these alternate approaches. On larger spectral scales (3, 4.5 and 9 MHz spectral scales, corresponding to $k_{\parallel} \simeq 0.17, 0.11$ and $0.06~h\mathrm{Mpc}^{-1}$, respectively, for redshifted 21 cm emission in the frequency range of our simulated observed data calibrated here) power in spurious gain amplitude fluctuations is suppressed by an additional 1--4 orders of magnitude, relative to those approaches. 

In the context of 21 cm cosmology applications, our results show that, in the scenarios modelled, only \textsc{BayesCal} yields sufficiently high fidelity calibration solutions for recovery of the 21 cm signal at $k_\parallel \lesssim 0.15~h\mathrm{Mpc}^{-1}$.  For all the other techniques in all completeness regimes, that scale is contaminated. The overall fidelity of the calibration solutions recovered with \textsc{BayesCal} is such that foreground systematics imparted by imperfections in the calibration solutions are below a white noise proxy for the 21 cm signal with an RMS amplitude of 10 mK, on all spectral scales in the calibrated data set, ranging from a factor of a few below the fiducial 21 cm signal level on 9 MHz spectral scales ($k_{\parallel} \simeq 0.06~h\mathrm{Mpc}^{-1}$) to close to 4 orders of magnitude below the power in the fiducial 21 cm signal on $1.8~\mathrm{MHz}$ scales ($k_{\parallel} \simeq 0.28~h\mathrm{Mpc}^{-1}$).

\section*{Acknowledgements}

PHS and JCP acknowledge support from NSF Award \#1907777 and Brown University's Richard B. Salomon Faculty Research Award Fund. PHS and JLS acknowledge funding from the Canada 150 Research Chairs Program. PHS acknowledges fellowship funding from the McGill Space Institute. This research was conducted using computational resources and services at the Center for Computation and Visualization, Brown University. PHS thanks James Aguirre, Zach Martinot, Adrian Liu and Irina Stefan for valuable discussions and Irina Stefan, Adrian Liu, Chris Carilli and Josh Dillon for helpful comments on a draft of this manuscript.

\section*{Data Availability}

The data from this study will be shared on reasonable request to the corresponding author.



\appendix

\section{The transition matrix between ENU and equatorial vector bases}
\label{Sec:TENUequatorial}

Careful choice of vector bases enables significant simplification of the component matrices of the RIME. A vector basis tied to the orientations of the antenna feeds provides the simplest vector space for defining the polarised voltage beam and a sky-centric basis tied to the celestial sphere provides a more natural vector space in which to define one's polarised sky model. However, to evaluate \autoref{Eq:MEqSingePol}, one must transition the beam and the sky to a mutual vector basis.

\subsection{Two dimensional vector field basis transformations}
\label{Sec:TwoDimensionalVectorFieldBasisTransformations}

Suppose $A = \{ \bm{\hat{\alpha}}_{1}, \bm{\hat{\alpha}}_{2}\}$ and $B = \{\bm{\hat{\beta}}_{1}, \bm{\hat{\beta}}_{2}\}$ are two ordered bases for a 2-dimensional vector space V and $\mathbfit{v}_{A} = (v_{\alpha_{1}}, v_{\alpha_{2}})$ and $\mathbfit{v}_{B} = (v_{\beta_{1}}, v_{\beta_{2}})$ are the coordinate 2-tuples of a vector $\mathbfit{v}$ in V with respect to basis A and B, respectively. The components of the coordinate 2-tuples of $\mathbfit{v}$ are given by the projection of $\mathbfit{v}$ onto the associated basis vectors\footnote{This is true independently of the coordinate system used to describe the components of $\mathbfit{v}$.},
\begin{align}
\label{Eq:VectorComponents}
v_{\alpha_{1}} &= \mathbfit{v} \cdot \bm{\hat{\alpha}}_{1}\\
v_{\alpha_{2}} &= \mathbfit{v} \cdot \bm{\hat{\alpha}}_{2} \nonumber \\
v_{\beta_{1}} &= \mathbfit{v} \cdot \bm{\hat{\beta}}_{1} \nonumber \\
v_{\beta_{2}} &= \mathbfit{v} \cdot \bm{\hat{\beta}}_{2}  \nonumber \ . 
\end{align}
The components of the coordinate 2-tuple in basis A can be calculated from those in basis B via (e.g. \citealt{2018ApJ...869...79M}),
\begin{align}
v_{\alpha_{1}} &= \mathbfit{v} \cdot \bm{\hat{\alpha}}_{1} = (v_{\beta_{1}}\bm{\hat{\beta}}_{1} +  v_{\beta_{2}}\bm{\hat{\beta}}_{2}) \cdot \bm{\hat{\alpha}}_{1} \\
v_{\alpha_{2}} &= \mathbfit{v} \cdot \bm{\hat{\alpha}}_{2} = (v_{\beta_{1}}\bm{\hat{\beta}}_{1} +  v_{\beta_{2}}\bm{\hat{\beta}}_{2}) \cdot \bm{\hat{\alpha}}_{2} \nonumber 
\end{align}
Thus, the transition matrix encoding the coordinate transformation between the coordinate 2-tuple in basis A and B is given by,
\begin{equation}
\label{Eq:TransitionMatrix}
\mathbfss{T}_{\bm{\hat{\alpha}} - \bm{\hat{\beta}}} = \begin{pmatrix}
\bm{\hat{\alpha}}_{1} \cdot \bm{\hat{\beta}}_{1}  & \bm{\hat{\alpha}}_{1} \cdot \bm{\hat{\beta}}_{2} \\ 
\bm{\hat{\alpha}}_{2} \cdot \bm{\hat{\beta}}_{1}  & \bm{\hat{\alpha}}_{2} \cdot \bm{\hat{\beta}}_{2}
\end{pmatrix} \ ,
\end{equation}
such that,  
\begin{equation}
\label{Eq:ComponentsOfVecBasisTransform}
\begin{pmatrix}
v_{\alpha_{1}}  \\ 
v_{\alpha_{2}}
\end{pmatrix} = 
\begin{pmatrix}
\bm{\hat{\alpha}}_{1} \cdot \bm{\hat{\beta}}_{1}  & \bm{\hat{\alpha}}_{1} \cdot \bm{\hat{\beta}}_{2} \\ 
\bm{\hat{\alpha}}_{2} \cdot \bm{\hat{\beta}}_{1}  & \bm{\hat{\alpha}}_{2} \cdot \bm{\hat{\beta}}_{2}
\end{pmatrix}
\begin{pmatrix}
v_{\beta_{1}}  \\ 
v_{\beta_{2}}
\end{pmatrix} \ .
\end{equation}
Furthermore, \autoref{Eq:TransitionMatrix} is trivially extended to the case that the basis vectors of bases $A$ and $B$ are, themselves, a function of direction on the sky, $\hat{\mathbfit{s}}$, as in \autoref{Eq:TransitionMatrixZaazEquatorialBasisVectors}, in which case we have,
\begin{equation}
\label{Eq:TransitionMatrixS}
\mathbfss{T}_{\bm{\hat{\alpha}} - \bm{\hat{\beta}}}(\hat{\mathbfit{s}}) = \begin{pmatrix}
\bm{\hat{\alpha}}_{1}(\hat{\mathbfit{s}}) \cdot \bm{\hat{\beta}}_{1}(\hat{\mathbfit{s}})  & \bm{\hat{\alpha}}_{1}(\hat{\mathbfit{s}}) \cdot \bm{\hat{\beta}}_{2}(\hat{\mathbfit{s}}) \\ 
\bm{\hat{\alpha}}_{2}(\hat{\mathbfit{s}}) \cdot \bm{\hat{\beta}}_{1}(\hat{\mathbfit{s}})  & \bm{\hat{\alpha}}_{2}(\hat{\mathbfit{s}}) \cdot \bm{\hat{\beta}}_{2}(\hat{\mathbfit{s}})
\end{pmatrix} \ .
\end{equation}

\subsection{Geometry of antenna and sky coordinate frames}
\label{Sec:GeometryOfAntennaAndSkyCoordinateFrames}

\begin{figure*}
	\centerline{
	\includegraphics[width=0.50\textwidth]{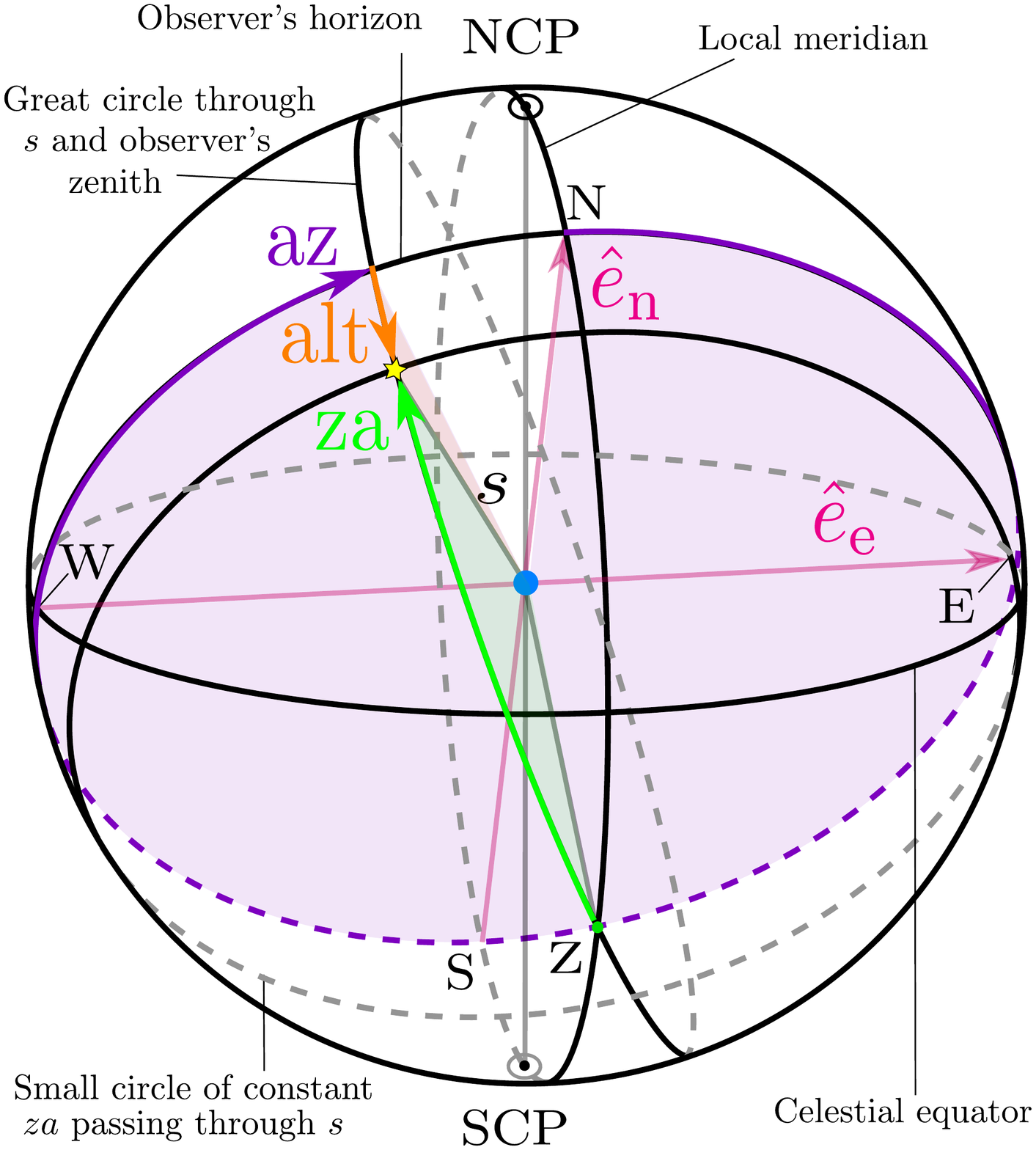} 
	\includegraphics[width=0.50\textwidth]{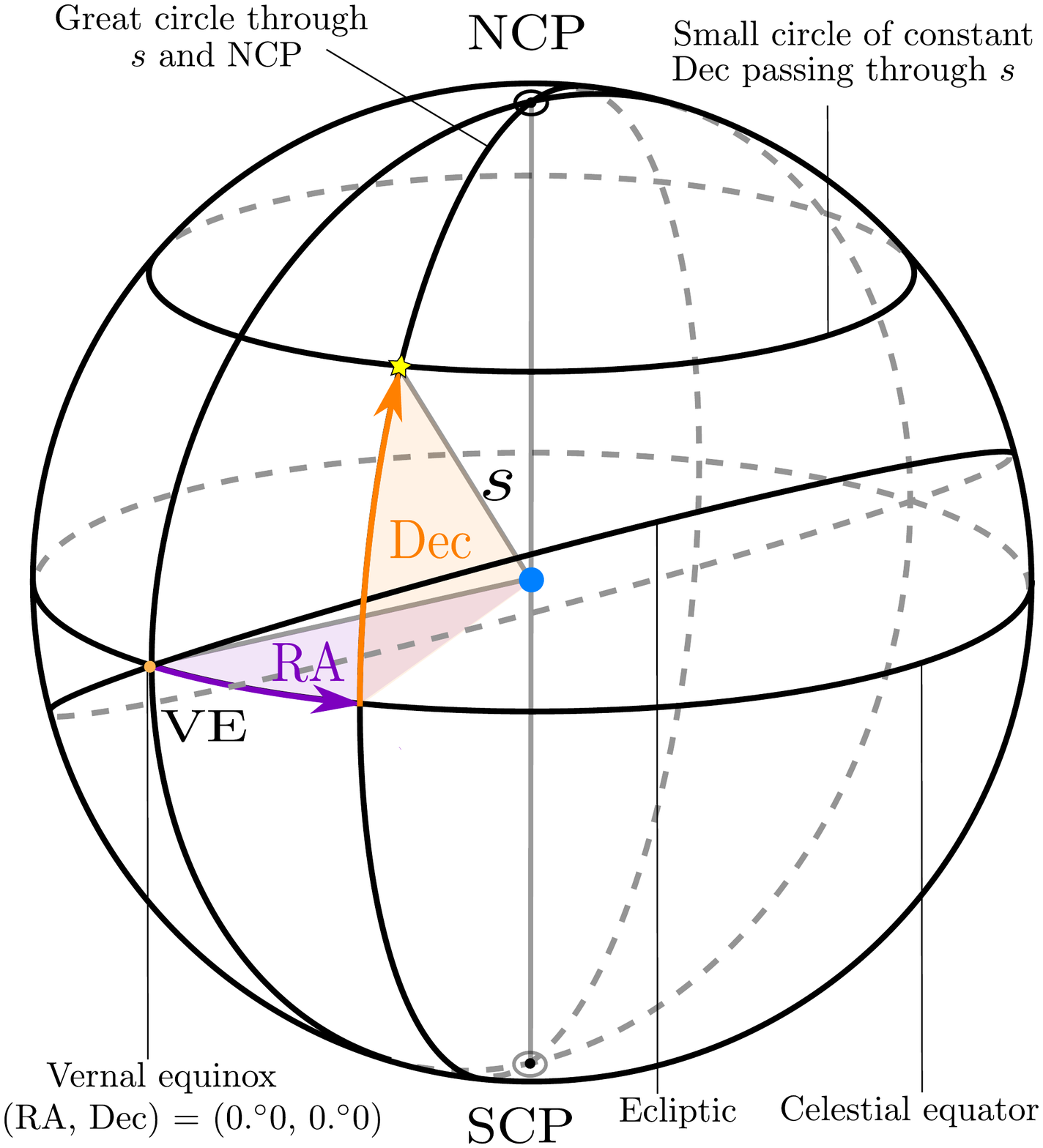}}
\caption{[Left] Azimuth (az), zenith angle (za) and altitude (alt) coordinates of a source on the celestial sphere at position $s$. The zenith pointing, $Z$, and the corresponding plane perpendicular to it and tangent to Earth at the observer, the intersection of which with the celestial sphere describes the `observer's horizon', are shown for an observer at a latitude of $-30\fdg0$. The directions of the geographic cardinal points, in the tangent plane to Earth, at the observer's location, are given by pointings from the origin of the tangent plane to its intersections, on the observer's horizon, with the local meridian and celestial equator. 
Azimuth, $0 \le \mathrm{az} < 2\pi~\mathrm{rad}$, is measured from north (N) towards east (E) in the tangent plane to the Earth at the observer's location. Zenith angle, $0 \le \mathrm{az} < \pi~\mathrm{rad}$, increases from zero at the observer's zenith to $180\fdg0$ at the observer's nadir. 
The east and north basis vectors of an east-north-up (ENU) coordinate system at the location of the observer are labelled as $\hat{\bm{e}}_\mathrm{e}$ and $\hat{\bm{e}}_\mathrm{n}$, respectively. The up-axis, and corresponding basis vector, is not labelled but increases in the zenith pointing direction from zero at the origin of the observer's tangent plane. [Right] Right ascension (RA) and declination (Dec) coordinates of a source on the celestial sphere at the position $s$. 
Right ascension, $0 \le \mathrm{RA} < 2\pi~\mathrm{rad}$, increases eastward from zero at vernal equinox (VE), where VE is defined as the direction of the ascending node of the intersection of the ecliptic and the celestial equator. Declination, $-\pi/2 \le \mathrm{Dec} \le \pi/2~\mathrm{rad}$, increases from zero on the celestial equator to $\pi/2~\mathrm{rad}$ at NCP.
}
\label{Fig:zaazAndRaDecCoordinateFrames} 
\end{figure*}

The transition matrix, $\mathbfss{T}_\mathrm{ENU-equatorial}$, encodes the transform of the components of the electric vector field between an RA, Dec basis in an equatorial coordinate system\footnote{We use the RA, Dec coordinates of the standard equatorial spherical coordinate system, as well as the radial coordinate axis increasing from the origin of the coordinate system.} and an east-north-up (ENU) topocentric coordinate basis. To bridge the gap between the equatorial and the ENU coordinate systems and simplify the derivation of $\mathbfss{T}_\mathrm{ENU-equatorial}$, we introduce a third coordinate system, zaaz\footnote{We use the zenith angle and azimuth coordinates of a standard zaaz coordinate system plus a radial coordinate pointing between the observer and zenith.}.

Since transformations between the equatorial, zaaz and ENU bases are linear, $\mathbfss{T}_\mathrm{ENU-equatorial}$ is the product of two intermediate transition matrices, $\mathbfss{T}_\mathrm{ENU-zaaz}$ and $\mathbfss{T}_\mathrm{zaaz-equatorial}$, between the respective coordinate bases,
\begin{equation}
\label{Eq:TransitionMatrixComb}
\mathbfss{T}_\mathrm{ENU-equatorial} = \mathbfss{T}_\mathrm{ENU-zaaz} \mathbfss{T}_\mathrm{zaaz-equatorial} \ ,
\end{equation}
where $\mathbfss{T}_\mathrm{ENU-zaaz}$ and $\mathbfss{T}_\mathrm{zaaz-equatorial}$ are transition matrices between the Cartesian ENU basis and the corresponding topocentric spherical zaaz coordinate basis, and between the zaaz and equatorial coordinate bases, respectively.
Furthermore, as a transition matrix between the angular basis vectors of two spherical coordinate bases differing only in the poles of their coordinate axes, $\mathbfss{T}_\mathrm{zaaz-equatorial}$ is fully characterised by the relative rotation angle of the respective angular basis vectors (see \autoref{Sec:TzaazEquatorial}).

The coordinate axis definitions and basis vectors relevant to the three coordinate frames used in \autoref{Eq:TransitionMatrixComb} are illustrated in \autoref{Fig:zaazAndRaDecCoordinateFrames}. ENU is a topocentric coordinate system with $\hat{\mathbfit{e}}_\mathrm{n}$ and $\hat{\mathbfit{e}}_\mathrm{e}$ basis vectors lying in the local tangent plane of the observer in the directions of north and east, respectively, and $\hat{\mathbfit{e}}_\mathrm{u}$ points from the observer towards zenith. The zaaz and equatorial coordinate frames are defined on the celestial sphere (i.e. $r=$ constant). Zenith angle ($0 \ge za \ge \pi~\mathrm{rad}$) increases from zero at the observer's zenith to $\pi~\mathrm{rad}$ at the observer's nadir. Azimuth ($0 \ge az \ge 2\pi~\mathrm{rad}$) increases in the direction of east from zero at $\hat{\mathbfit{e}}_\mathrm{n}$. The LST dependence of zaaz coordinates of fixed sources on the celestial sphere follows from the rotation, with the siderial motion of the observer, of the pole of the angular coordinates on the celestial sphere. RA ($-\pi \le \mathrm{RA} \le \pi~\mathrm{rad}$) increases eastward from zero at vernal equinox (VE), where VE is the direction of the ascending node of the intersection of the ecliptic and the celestial equator. Declination ($-\pi/2 \le \mathrm{Dec} \le \pi/2~\mathrm{rad}$) increases from zero on the celestial equator to $\pi/2~\mathrm{rad}$ at NCP. Equatorial coordinates, and, correspondingly, the equatorial vector basis, of a given reference epoch (B1950 or J2000, for example), are stationary on the celestial sphere\footnote{Use of a reference epoch eliminates the time dependence of the coordinate system that is otherwise present due to slow time-dependence of the direction of VE resulting from the precession and nutation of Earth's rotation axis.}.

\subsection{$\mathbfss{T}_\mathrm{ENU-zaaz}$}
\label{Sec:TENUzaaz}

Before deriving the basis vectors of the ENU and zaaz coordinate systems, we note two physical constraints on the electric and voltage vector fields:
\begin{enumerate}
 \item The electric field lies purely in the plane perpendicular to its Poynting vector, thus, in an antenna-centric spherical basis, such as zaaz or equatorial, the radial component of the electric field incident on the antenna is identically zero.
 
 \item A dual-feed antenna is sensitive to the amplitude of the electric field vector in the plane defined by the orientation of the feeds. Without loss of generality, we will assume this is the N-E plane. Thus, in an ENU basis, the $\hat{\mathbfit{e}}_\mathrm{u}$ component of the voltage vector is identically zero.
\end{enumerate}
Here, for generality, we will work with the three dimensional basis vectors at the outset and then apply the above constraints when deriving the transition matrices for the system of interest.

ENU and zaaz coordinates are related as follows (see \autoref{Fig:zaazAndRaDecCoordinateFrames}, left),
\begin{align}
\label{Eq:ENUzaazCoords}
E &= r\sin(za)\sin(az) \\
N &= r\sin(za)\cos(az) \nonumber \\
U &= r\cos(za) \nonumber \ , \\
r &= \sqrt{E^2+N^2+U^2} \\
\cos(za) &= \dfrac{U}{\sqrt{E^2+N^2+U^2}} \nonumber \\
\tan(az) &= \dfrac{E}{N} \nonumber \ .
\end{align}

The ENU basis vectors $\hat{\mathbfit{e}}_\mathrm{e}$, $\hat{\mathbfit{e}}_\mathrm{n}$ and $\hat{\mathbfit{e}}_\mathrm{u}$ are unit vectors in the local east, north and up directions, respectively, and, in an ENU coordinate frame, they have components\footnote{Following \citet{2019ApJ...882...58K}, we use a left-handed ENU coordinate basis.},
\begin{equation}
\label{Eq:ENUbasisVectorsInENU}
\hat{\mathbfit{e}}_\mathrm{n} = 
\begin{pmatrix}
1  \\ 
0  \\ 
0  \\
\end{pmatrix} , \  
\hat{\mathbfit{e}}_\mathrm{e} = 
\begin{pmatrix}
0  \\ 
1  \\ 
0  \\
\end{pmatrix} , \  
\hat{\mathbfit{e}}_\mathrm{u} = 
\begin{pmatrix}
0  \\ 
0  \\ 
1  \\
\end{pmatrix} \ .
\end{equation}
The components of the zaaz unit vectors in the ENU coordinate frame are given by,
\begin{align}
\label{Eq:zaazBasisVectorsInENU}
\hat{\mathbfit{e}}_\mathrm{r}  &= \dfrac{1}{\lvert \nabla r \rvert} \nabla r \\
\hat{\mathbfit{e}}_\mathrm{za} &= \dfrac{1}{\lvert \nabla za \rvert} \nabla za \nonumber \\
\hat{\mathbfit{e}}_\mathrm{az} &= \dfrac{1}{\lvert \nabla az \rvert} \nabla az \nonumber \ ,
\end{align}
where $\nabla$ is the gradient operator. From \autoref{Eq:ENUzaazCoords} and \autoref{Eq:zaazBasisVectorsInENU} we have,
\begin{align}
\label{Eq:zaazBasisVectorsInENUExplicit1}
\hat{\mathbfit{e}}_\mathrm{r}  &= \dfrac{E}{\sqrt{E^2+N^2+U^2}}\, \hat{\mathbfit{e}}_\mathrm{e} + \dfrac{N}{\sqrt{E^2+N^2+U^2}}\, \hat{\mathbfit{e}}_\mathrm{n} + \dfrac{U}{\sqrt{E^2+N^2+U^2}}\, \hat{\mathbfit{e}}_\mathrm{u} \\
\hat{\mathbfit{e}}_\mathrm{za} &= \dfrac{EU}{\sqrt{E^2+N^2}\sqrt{E^2+N^2+U^2}}\, \hat{\mathbfit{e}}_\mathrm{e} + \dfrac{NU}{\sqrt{E^2+N^2}\sqrt{E^2+N^2+U^2}}\, \hat{\mathbfit{e}}_\mathrm{n} \nonumber \\
 &- \dfrac{\sqrt{E^2+N^2}}{\sqrt{E^2+N^2+U^2}}\, \hat{\mathbfit{e}}_\mathrm{u} \nonumber \\
\hat{\mathbfit{e}}_\mathrm{az} &= \dfrac{N}{\sqrt{E^2+N^2}}\, \hat{\mathbfit{e}}_\mathrm{e} - \dfrac{E}{\sqrt{E^2+N^2}}\, \hat{\mathbfit{e}}_\mathrm{n} \nonumber \ ,
\end{align}
or, equivalently, in terms of $r$, $za$ and $az$ coordinates, 
\begin{align}
\label{Eq:zaazBasisVectorsInENUExplicit2}
\hat{\mathbfit{e}}_\mathrm{r}  &= \sin(za)\sin(az)\, \hat{\mathbfit{e}}_\mathrm{e} + \sin(za)\cos(az)\, \hat{\mathbfit{e}}_\mathrm{n} + \cos(za)\, \hat{\mathbfit{e}}_\mathrm{u} \\
\hat{\mathbfit{e}}_\mathrm{za} &= \sin(az)\cos(za)\, \hat{\mathbfit{e}}_\mathrm{e} + \cos(az)\cos(za)\, \hat{\mathbfit{e}}_\mathrm{n} + \sin(za)\, \hat{\mathbfit{e}}_\mathrm{u} \nonumber \\
\hat{\mathbfit{e}}_\mathrm{az} &= \cos(az)\, \hat{\mathbfit{e}}_\mathrm{e} - \sin(az)\, \hat{\mathbfit{e}}_\mathrm{n} \nonumber \ .
\end{align}
From conditions 1 and 2 above, the vector electric field at the antenna is non-zero exclusively in the $\hat{\mathbfit{e}}_\mathrm{za}$ and $\hat{\mathbfit{e}}_\mathrm{az}$ basis vector directions and the component of the electric field vector that will induce a voltage response in the antenna feeds is exclusively in the $\hat{\mathbfit{e}}_\mathrm{e}$, $\hat{\mathbfit{e}}_\mathrm{n}$ basis vector directions. At this stage, we, therefore, restrict calculation of the transition matrix to the projections of basis vectors with these non-zero components. Using \autoref{Eq:ENUbasisVectorsInENU} and \autoref{Eq:zaazBasisVectorsInENUExplicit2}, this yields the transition matrix encoding the conversion of the non-zero and measured components of the electric field vector between zaaz and ENU coordinates,
%
\begin{equation}
\label{Eq:TransitionMatrixENUzaaz}
\mathbfss{T}_\mathrm{ENU-zaaz} = 
\begin{pmatrix}
\hat{\mathbfit{e}}_\mathrm{n} \cdot \hat{\mathbfit{e}}_\mathrm{za}  & \hat{\mathbfit{e}}_\mathrm{n} \cdot \hat{\mathbfit{e}}_\mathrm{az} \\
\hat{\mathbfit{e}}_\mathrm{e} \cdot \hat{\mathbfit{e}}_\mathrm{za}  & \hat{\mathbfit{e}}_\mathrm{e} \cdot \hat{\mathbfit{e}}_\mathrm{az}
\end{pmatrix}
=
\begin{pmatrix}
\cos(az)\cos(za)  & -\sin(az) \\ 
\sin(az)\cos(za)  & \cos(az)
\end{pmatrix} \ .
\end{equation}

\subsection{$\mathbfss{T}_\mathrm{zaaz-equatorial}$}
\label{Sec:TzaazEquatorial}

\begin{figure*}
	\centerline{
	\includegraphics[width=0.50\textwidth]{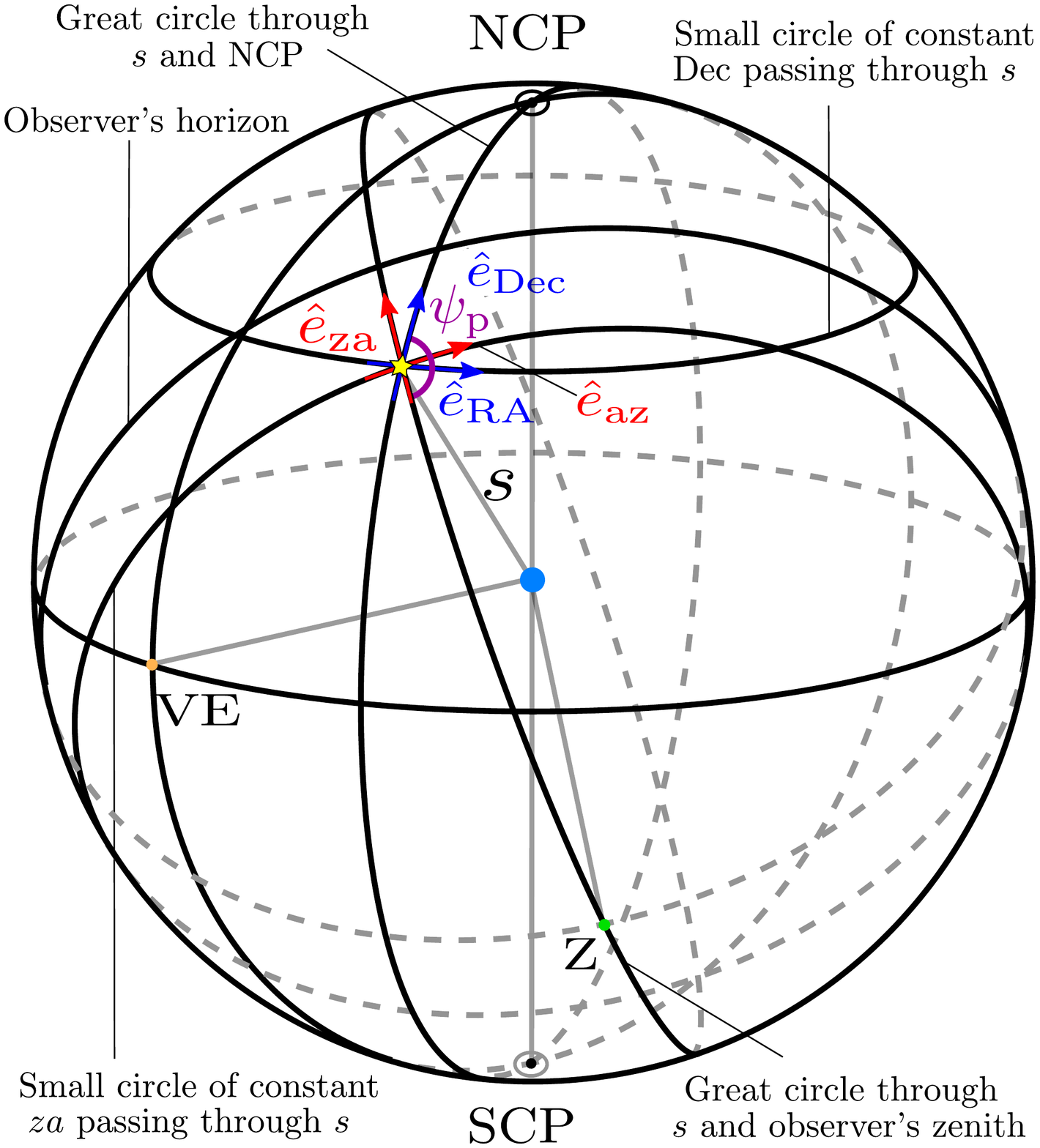}
	\includegraphics[width=0.50\textwidth]{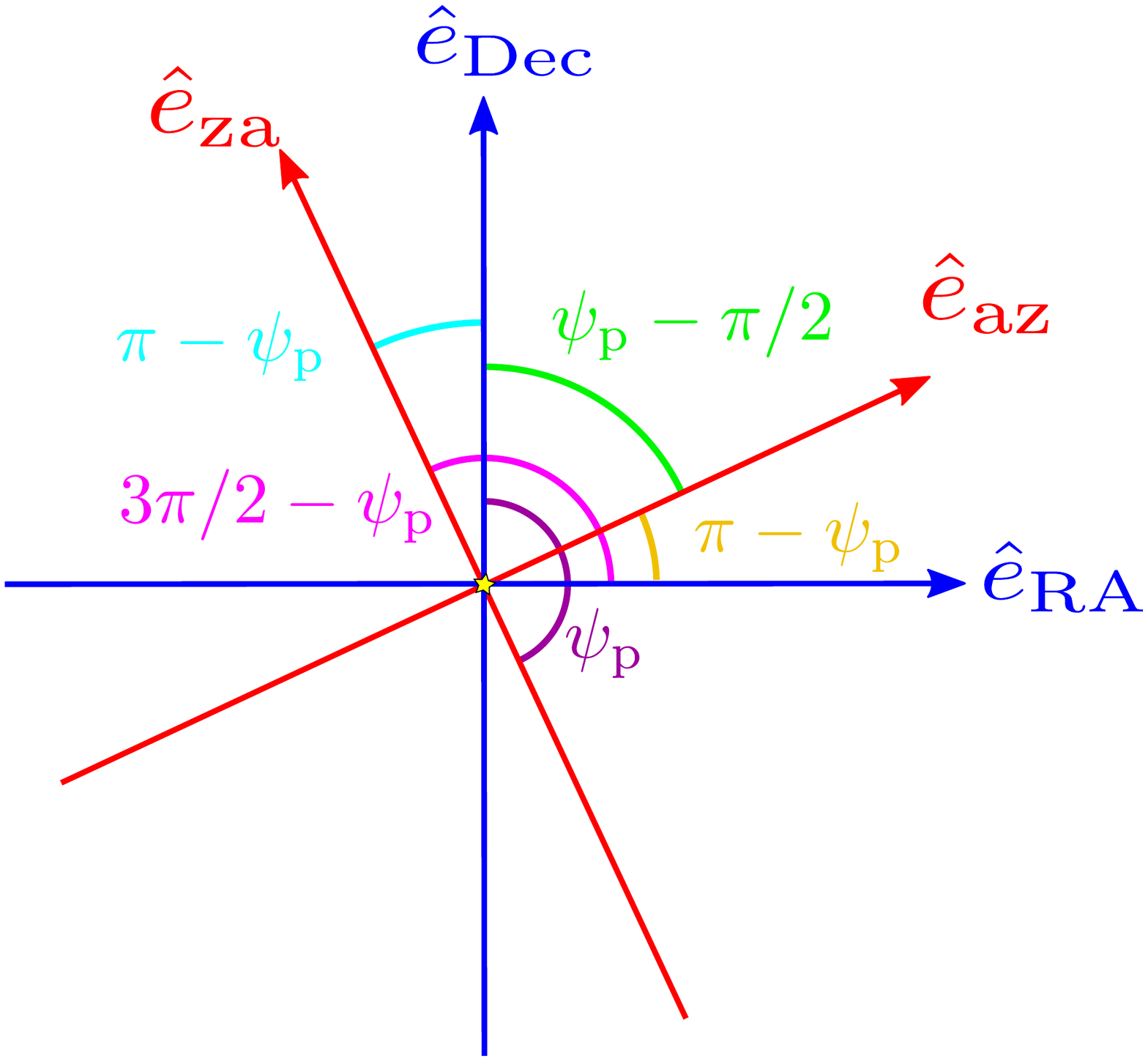}}
\caption{[Left] Basis vectors of right ascension ($\hat{\mathbfit{e}}_\mathrm{Ra}$) and declination ($\hat{\mathbfit{e}}_\mathrm{Dec}$), corresponding to an equatorial coordinate system, and zenith angle ($\hat{\bm{e}}_\mathrm{za}$) and azimuth ($\hat{\bm{e}}_\mathrm{az}$), corresponding to a zaaz coordinate system, are labelled. All four lie in the tangent plane to the celestial sphere at position $s$. $\hat{\mathbfit{e}}_\mathrm{Dec}(\hat{\bm{s}})$ and $\hat{\mathbfit{e}}_\mathrm{Ra}(\hat{\bm{s}})$  point towards the NCP and eastward along the small circle defined by the declination of the source (or great circle for a source on the celestial equator), respectively. $\hat{\bm{e}}_\mathrm{za}(\hat{\bm{s}})$ and $\hat{\bm{e}}_\mathrm{az}(\hat{\bm{s}})$ point away from zenith along the great circle passing though $Z$ and $s$ and from north towards east along the small circle of constant za passing through $s$, respectively. The parallactic angle, $\psi_\mathrm{p}$, between the hour circle through point $s$ and the great circle through $s$ and zenith ($\angle Zs\mathrm{NCP}$) is labelled. Positive parallactic angle is measured eastward from NCP.
[Right] A face-on view of the tangent plane to the celestial sphere at point $s$. Basis vectors of right ascension and declination and zenith angle and azimuth are shown as red and blue arrows, respectively. The parallactic angle, $\psi_\mathrm{p}$, is the angle at the origin subtended between $\hat{\mathbfit{e}}_\mathrm{Dec}$ and -$\hat{\mathbfit{e}}_\mathrm{za}$, with positive $\psi_\mathrm{p}$ measured clockwise from $\hat{\mathbfit{e}}_\mathrm{Dec}$. Both it and the angles between basis vectors, relevant to the $\mathbfss{T}_\mathrm{zaaz-equatorial}$ transition matrix, are labelled.
}
\label{Fig:BasisVectorsOnCelestialSphere} 
\end{figure*}

Since the angular coordinates of the equatorial and zaaz coordinate systems differ only in the locations of their poles, the conversion between their angular bases is given by the $\hat{\mathbfit{s}}$-dependent coordinate transform defined by the rotation between the axes of the respective coordinate systems in the tangent plane to the sphere (see \autoref{Fig:BasisVectorsOnCelestialSphere}). Thus, from \autoref{Fig:BasisVectorsOnCelestialSphere}, right, we have,
\begin{align}
\label{Eq:zaazEquatorialDotProducts}
\hat{\mathbfit{e}}_\mathrm{za} \cdot \hat{\mathbfit{e}}_\mathrm{Dec} &= \cos(\pi -\psi_\mathrm{p}) \\
\hat{\mathbfit{e}}_\mathrm{za} \cdot \hat{\mathbfit{e}}_\mathrm{RA} &= \cos(3\pi/2 - \psi_\mathrm{p}) \nonumber \\
\hat{\mathbfit{e}}_\mathrm{az} \cdot \hat{\mathbfit{e}}_\mathrm{Dec} &= \cos(\psi_\mathrm{p} - \pi/2) \nonumber \\
\hat{\mathbfit{e}}_\mathrm{az} \cdot \hat{\mathbfit{e}}_\mathrm{RA} &= \cos(\pi - \psi_\mathrm{p}) \nonumber \ .
\end{align}
Here, the parallactic angle, $\psi_\mathrm{p}(\hat{\mathbfit{s}})$, is the angle between the hour circle through point $s$ and the great circle through $s$ and zenith, with positive parallactic angle measured eastward from NCP (or equivalently, measured clockwise from $\hat{\mathbfit{e}}_\mathrm{Dec}$ in the tangent plane to the celestial sphere at point $s$). For an antenna located at a latitude $\lambda$, the parallactic angle of a source in the direction $\hat{\mathbfit{s}}$, located at an hour angle $h$, and declination $\delta$, is given by (e.g. \citealt{1999ASPC..180.....T}),
\begin{equation}
\label{Eq:ParallacticAngle}
\tan(\psi_\mathrm{p}) = \dfrac{\cos(\lambda)\sin(h)}{\sin(\lambda)\cos(\delta) - \cos(\lambda)\sin(\delta)\cos(h)} \ .
\end{equation}
Substituting the projections in \autoref{Eq:zaazEquatorialDotProducts} into \autoref{Eq:TransitionMatrix} and simplifying, we have,  
\begin{equation}
\label{Eq:TransitionMatrixZaazEquatorial}
\mathbfss{T}_\mathrm{zaaz-equatorial} = 
\begin{pmatrix}
\hat{\mathbfit{e}}_\mathrm{za} \cdot \hat{\mathbfit{e}}_\mathrm{Dec}  & \hat{\mathbfit{e}}_\mathrm{za} \cdot \hat{\mathbfit{e}}_\mathrm{RA} \\
\hat{\mathbfit{e}}_\mathrm{az} \cdot \hat{\mathbfit{e}}_\mathrm{Dec}  & \hat{\mathbfit{e}}_\mathrm{az} \cdot \hat{\mathbfit{e}}_\mathrm{RA} 
\end{pmatrix}
= -
\begin{pmatrix}
\cos(\psi_\mathrm{p})  & \sin(\psi_\mathrm{p}) \\ 
-\sin(\psi_\mathrm{p})  & \cos(\psi_\mathrm{p})
\end{pmatrix} \ .
\end{equation}

\subsection{$\mathbfss{T}_\mathrm{ENU-equatorial}$}
\label{Sec:TENUequatorial}

Finally, substituting \autoref{Eq:TransitionMatrixENUzaaz} and \autoref{Eq:TransitionMatrixZaazEquatorial} into \autoref{Eq:TransitionMatrixComb}, we arrive at the transition matrix between equatorial and ENU coordinates,
%
\onecolumn
\begin{equation}
\label{Eq:TransitionMatrixCombComponents}
\mathbfss{T}_\mathrm{ENU-equatorial} = \\ 
\begin{pmatrix}
-\sin(az)\sin(\psi_\mathrm{p}) - \cos(az)\cos(za)\cos(\psi_\mathrm{p})  & \sin(az)\cos(\psi_\mathrm{p}) - \cos(az)\cos(za)\sin(\psi_\mathrm{p}) \\ 
\cos(az)\sin(\psi_\mathrm{p}) - \sin(az)\cos(za)\cos(\psi_\mathrm{p})  & -\cos(az)\cos(\psi_\mathrm{p}) - \sin(az)\cos(za)\sin(\psi_\mathrm{p})
\end{pmatrix} \ . 
\end{equation}
\twocolumn

\label{lastpage}

\end{document}